\documentclass[a4paper, 10pt, oneside]{article}
\usepackage[a4paper,inner=2cm,outer=2cm,top=3cm,bottom=2.5cm]{geometry}

\usepackage{amsmath}
\usepackage{amssymb}
\usepackage{amsthm}
\usepackage{dsfont}
\usepackage{slashed}
\usepackage{mathrsfs}
\usepackage{mathtools}
\usepackage{color}

\usepackage{bibgerm}

\usepackage[english]{babel}

\usepackage[utf8x]{inputenc}

\usepackage{graphicx,epsfig}

\usepackage{pstricks,pst-plot,pstricks-add,pst-pdf}

\usepackage{setspace}

\usepackage{booktabs}

\usepackage{float}

\usepackage{nextpage}

\usepackage{newclude}

\usepackage[small,bf]{caption}
\usepackage{subcaption}

\usepackage{url}

\usepackage[colorlinks=true,
	linkcolor=black,
	citecolor=black,
	urlcolor=black,
	filecolor=black
]{hyperref}

\usepackage[T1]{fontenc}

\newcommand{\ChPT}{$\chi$PT}

\renewcommand{\O}{\mathcal{O}}
\newcommand{\p}{\partial}

\renewcommand{\Re}{\mathrm{Re}}
\renewcommand{\Im}{\mathrm{Im}}

\newcommand{\<}{\langle}
\renewcommand{\>}{\rangle}

\newcommand{\dprime}{{\prime\prime}}

\newcommand{\tr}{\mathrm{Tr}}

\newcommand{\mpio}{M_{\pi^0}}
\newcommand{\mpip}{M_{\pi^+}}
\newcommand{\mko}{M_{K^0}}
\newcommand{\mkp}{M_{K^+}}
\newcommand{\meta}{M_\eta}
\newcommand{\ml}{m_\ell}
\newcommand{\mg}{m_\gamma}
\newcommand{\zl}{z_\ell}
\newcommand{\zg}{z_\gamma}
\newcommand{\sg}{s_\gamma}
\newcommand{\tl}{t_\ell}

\newcommand{\dilog}{\mathrm{Li}_2}

\newcommand{\remark}[1]{}

\makeatletter
\renewcommand{\maketag@@@}[1]{\hbox{\m@th\normalsize\normalfont#1}}%
\makeatother

\makeatletter
\renewcommand\paragraph{\@startsection{paragraph}{4}{\z@}%
  {-3.25ex \@plus -1ex \@minus -0.2ex}%
  {0.01pt}%
  {\bfseries}%
}
\makeatother

\usepackage{abstract}

\begin{document}

\mbox{}

\bigskip

\begin{center}
{\LARGE{\bf Isospin Breaking Effects in \boldmath $K_{\ell4}$ Decays}}

\vspace{0.5cm}

Peter Stoffer

\vspace{1em}

\begin{tabular}{c}
Albert Einstein Center for Fundamental Physics, \\
Institute for Theoretical Physics,  University of Bern,\\
Sidlerstr. 5, CH-3012 Bern, Switzerland\\
\end{tabular} 

\end{center}

\vspace{1em}

\hrule

\begin{abstract}
In the framework of chiral perturbation theory with photons and leptons, the one-loop isospin breaking effects in $K_{\ell4}$ decays due to both the photonic contribution and the quark and meson mass differences are computed.

A comparison with the isospin breaking corrections applied by recent high statistics $K_{e4}$ experiments is performed.

The calculation can be used to correct the existing form factor measurements by isospin breaking effects that have not yet been taken into account in the experimental analysis. Based on the present work, possible forthcoming experiments on $K_{e4}$ decays could correct the isospin breaking effects in a more consistent way.
\end{abstract}

\hrule

\vspace{1em}

\setcounter{tocdepth}{3}
\tableofcontents

\clearpage


\section{Introduction}

High-precision hadron physics at low energies pursues mainly two aims: a better understanding of the strong interaction in its non-perturbative regime and the indirect search for new physics beyond the standard model. As perturbative QCD is not applicable, one has to use non-perturbative methods like effective field theories, lattice simulations or dispersion relations. The effective theory describing the strong interaction at low energy is chiral perturbation theory (\ChPT{},  \cite{Weinberg1968, GasserLeutwyler1984, GasserLeutwyler1985}). In order to render it predictive, one has to determine the parameters of the theory, the low-energy constants (LECs), either by comparison with experiments or with the help of lattice calculations.

The $K_{\ell4}$ decay is for several reasons a particularly interesting process. The physical region starts at the $\pi\pi$ threshold, i.e.~at lower energies than $K\pi$ scattering, which gives access to the same low-energy constants. \ChPT{}, being an expansion in the meson masses and momenta, should therefore give a better description of $K_{\ell4}$ than $K\pi$ scattering. Besides providing a very clean access to some of the LECs, $K_{\ell4}$ is, due to its final state, one of the best sources of information on $\pi\pi$ interaction \cite{Shabalin1963,Cabibbo1965,Batley2010}.

The recent high-statistics  $K_{\ell4}$ experiments E865 at BNL \cite{Pislak2003} and NA48/2 at CERN \cite{Batley2010, Batley2012} have achieved an impressive accuracy. The statistical errors at the sub-percent level ask for a consistent treatment of isospin breaking effects. Usually, theoretical calculations are performed in an ideal world with intact isospin, the $SU(2)$ symmetry of up- and down-quarks. The quark mass difference and the electromagnetism break isospin symmetry at the percent level.

Isospin breaking effects in $K_{\ell4}$ have been studied in the previous literature and played a major role concerning the success of standard \ChPT{}. In \cite{Colangelo2009}, the effect of quark and meson mass differences on the phase shifts was studied. The inclusion of this effect brought the NA48/2 measurement of the scattering lengths into perfect agreement with the prediction of the \ChPT{}/Roy equation analysis \cite{Colangelo2000}. For a review of these developments, see \cite{Gasser2009}. The mass effects on the phases at two-loop order have been recently studied in an elaborate dispersive framework \cite{Bernard2013}, which confirms the previous results. In both works, the photonic effects are assumed to be treated consistently in the experimental analysis. The earlier work \cite{Cuplov2003, Cuplov2004} treats both mass and photonic effects. However, the calculation of virtual photon effects is incomplete and real photon corrections are taken into account only in the soft approximation.

The experimental analysis of the latest experiment \cite{Batley2010, Batley2012} treats photonic corrections with the semi-classical Gamow-Sommerfeld factor and PHOTOS Monte Carlo \cite{Barberio1994}, which assumes phase space factorisation.

The need for a theoretical treatment of the full radiative corrections was pointed out in \cite{Colangelo2009} and considered as a long-term project. With the present work, I intend to fill this gap. The obtained results enable a better correction of isospin effects in the data:
\begin{itemize}
	\item as I will explain below, one can improve already now the handling of isospin effects in the data analysis;
	\item in the future, an event generator which incorporates the matrix element calculated here should be written and used to perform the data analysis.
\end{itemize}

The paper is organised as follows. In section~\ref{sec:Kinematics}, I define the kinematics, matrix elements and form factors of $K_{\ell4}$ and the radiative decay $K_{\ell4\gamma}$. In section~\ref{sec:ChPTCalculation}, I calculate the matrix elements within \ChPT{} including leptons and photons \cite{Urech1995, Knecht2000}. In section~\ref{sec:ExtractionOfIsospinCorrections}, I present the strategy of extracting the isospin corrections and perform the phase space integration for the radiative decay. The cancellation of both infrared and mass divergences is demonstrated. In section~\ref{sec:Numerics}, the isospin corrections are evaluated numerically. I compare the full radiative process with the soft photon approximation and with the strategy used in the experimental analysis. The appendices give details on the calculation and explicit results for the matrix elements.

It should be noted that large parts of this work assume a small lepton mass and are therefore not applicable to the muonic mode of the process.


\section{Kinematics and Decay Rate}

\label{sec:Kinematics}

\subsection{The $K_{\ell4}$ Decay}

\subsubsection{Definition of the Decay}

$K_{\ell4}$ is the semileptonic decay of a kaon into two pions, a lepton and a neutrino. Let us consider here the following charged mode:
\begin{align}
	\label{eqn:Kl4Decay}
	K^+(p) \rightarrow \pi^+(p_1) \pi^-(p_2) \ell^+(p_\ell) \nu_\ell(p_\nu),
\end{align}
where $\ell\in\{e,\mu\}$ is either an electron or a muon.

\begin{figure}[ht]
	\centering
	\setlength{\unitlength}{1cm}
	\begin{pspicture}(-1,-1)(13,5)
		\pspolygon[linestyle=none,fillstyle=solid](-1,-1)(13,-1)(13,5)(-1,5)
		\psline[linestyle=dotted](6,2)(7,4)(5,4)
		\psline[linestyle=dashed](1,2)(11,2)
		\psline[linestyle=dashed](2.5,0)(4.5,4)
		\psline[linestyle=dashed](9.5,0)(7.4,4.2)
		\psTextFrame[linestyle=none,fillstyle=solid](5.98,1.7)(6.9,2.3){$K^+$}
		\psline(5,4)(2,4)(0,0)(5,0)(6,2)
		\pspolygon(7,0)(12,0)(9.9,4.2)(4.9,4.2)
		\psline[arrowsize=6pt]{<->}(4,0.5)(3,3.5)
		\rput(4.3,0.5){$\pi^-$}
		\rput(3.3,3.7){$\pi^+$}
		\psline[arrowsize=6pt]{<->}(7,1.5)(10,2.5)
		\rput(10,2.8){$\ell^+$}
		\rput(7,1.2){$\nu_\ell$}
		\psarc{->}(6,2){1}{63.43}{116.57}
		\rput(6,2.7){$\phi$}
		\psarc{<-}(3.5,2){1}{108.43}{180}
		\rput(3,2.4){$\theta_\pi$}
		\psarc{->}(8.5,2){1}{0}{18.43}
		\psline[linecolor=white,linewidth=2pt](8.95,2)(9.45,2)
		\rput(9.25,2){$\theta_\ell$}
		\rput(1,0.5){$\Sigma_{2\pi}$}
		\rput(11,0.5){$\Sigma_{\ell\nu}$}
		\psline[arrowsize=4pt]{->}(3.5,2)(3.875,2.75)
		\rput(4.1,2.6){$\vec c$}
		\psline[arrowsize=4pt]{->}(8.5,2)(8.125,2.75)
		\rput(8.5,2.6){$\vec d$}
		\psline[arrowsize=4pt]{->}(6,2)(5.25,2)
		\rput(5.3,2.3){$\vec v$}
	\end{pspicture}
	\caption{The reference frames and the kinematic variables for the $K_{\ell4}$ decay.}
	\label{img:Kl4Kinematics}
\end{figure}
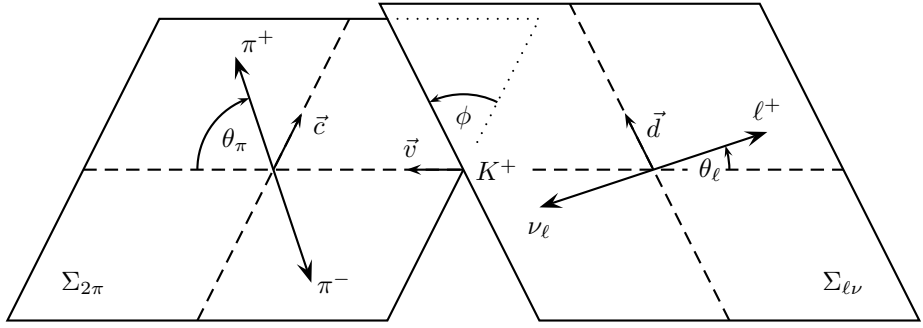

The kinematics of the decay (\ref{eqn:Kl4Decay}) can be described by five variables. The same conventions as in \cite{Bijnens1994} will be used, first introduced by Cabibbo and Maksymowicz \cite{Cabibbo1965}. There appear three different reference frames: the rest frame of the kaon $\Sigma_K$, the $\pi^+\pi^-$ centre-of-mass frame $\Sigma_{2\pi}$ and the $\ell^+\nu$ centre-of-mass frame $\Sigma_{\ell\nu}$. The situation is sketched in figure \ref{img:Kl4Kinematics}.

The five kinematic variables are then:
\begin{itemize}
	\item $s$, the total centre-of-mass squared energy of the two pions,
	\item $s_\ell$, the total centre-of-mass squared energy of the two leptons,
	\item $\theta_\pi$, the angle between the $\pi^+$ in $\Sigma_{2\pi}$ and the dipion line of flight in $\Sigma_K$,
	\item $\theta_\ell$, the angle between the $\ell^+$ in $\Sigma_{\ell\nu}$ and the dilepton line of flight in $\Sigma_K$,
	\item $\phi$, the angle between the dipion plane and the dilepton plane in $\Sigma_K$.
\end{itemize}

The (physical) ranges of these variables are:
\begin{align}
	\begin{alignedat}{2}
		4 \mpip^2 &\le s &&\le (\mkp - m_\ell)^2 , \\
		m_\ell^2 &\le s_\ell &&\le (\mkp - \sqrt{s})^2 , \\
		0 &\le \theta_\pi &&\le \pi , \\
		0 &\le \theta_\ell &&\le \pi , \\
		0 &\le \phi &&\le 2\pi.
	\end{alignedat}
\end{align}

Following \cite{Bijnens1994}, I define the four-momenta:
\begin{align}
	\label{eqn:FourMomenta}
	P := p_1 + p_2 , \quad Q := p_1 - p_2 , \quad L := p_\ell + p_\nu , \quad N := p_\ell - p_\nu .
\end{align}
Total momentum conservation implies $p = P + L$.

I will use the Mandelstam variables
\begin{align}
	s := (p_1 + p_2)^2 , \quad t := (p - p_1)^2 , \quad u := (p - p_2)^2
\end{align}
and the abbreviation
\begin{align}
	\begin{split}
		\label{eqn:DefinitionsXSigmaZ}
		z_\ell &:= m_\ell^2 / s_\ell , \\
		X &:= \frac{1}{2}\lambda^{1/2}_{K\ell}(s) := \frac{1}{2}\lambda^{1/2}(\mkp^2,s,s_\ell) , \quad \lambda(a,b,c) := a^2 + b^2 + c^2 - 2(ab+bc+ca) , \\
		\sigma_\pi &:= \sqrt{1-\frac{4\mpip^2}{s}} , \\
		\nu &:= t - u = -2\sigma_\pi X \cos\theta_\pi , \\
		\Sigma_0 &:= s + t + u = \mkp^2 + 2 \mpip^2 + s_\ell.
	\end{split}
\end{align}
In the appendix \ref{sec:LorentzTransformationsKl4}, the Lorentz transformations between the three reference frames are determined and the Lorentz invariant products of the momenta are computed.

\subsubsection{Matrix Element, Form Factors and Decay Rate}

\paragraph{$K_{\ell4}$ in the Isospin Limit}

After integrating out the $W$ boson in the standard model, we end up with a Fermi type current-current interaction. If we switch off the electromagnetic interaction, the matrix element of $K_{\ell4}$
\begin{align}
	\begin{split}
		{\vphantom{\big\<}}_\mathrm{out}\big\< \pi^+(p_1) \pi^-(p_2) \ell^+(p_\ell) \nu_\ell(p_\nu) \big| K^+(p) \big\>_\mathrm{in} &= \big\< \pi^+(p_1) \pi^-(p_2) \ell^+(p_\ell) \nu_\ell(p_\nu) \big| i T \big| K^+(p)\big\> \\
		&= i (2\pi)^4 \delta^{(4)}(p - P - L ) \; \mathcal{T}
	\end{split}
\end{align}
splits up into a leptonic times a hadronic part:
\begin{align}
	\begin{split}
		\mathcal{T} &= \frac{G_F}{\sqrt{2}} V_{us}^* \bar u(p_\nu) \gamma_\mu (1-\gamma^5)v(p_\ell) \; \big\< \pi^+(p_1) \pi^-(p_2) \big| \bar s \gamma^\mu(1-\gamma^5) u \big| K^+(p) \big\> .
	\end{split}
\end{align}
The hadronic matrix element exhibits the usual $V-A$ structure of weak interaction. Its Lorentz structure allows us to write the two contributions as
\begin{align}
	\big\< \pi^+(p_1) \pi^-(p_2) \big| V_\mu(0) \big| K^+(p)\big\> &= -\frac{H}{\mkp^3} \epsilon_{\mu\nu\rho\sigma} L^\nu P^\rho Q^\sigma , \\
	\big\< \pi^+(p_1) \pi^-(p_2) \big| A_\mu(0) \big| K^+(p) \big\> &= -i \frac{1}{\mkp} \left( P_\mu F + Q_\mu G + L_\mu R \right) ,
\end{align}
where $V_\mu = \bar s \gamma_\mu u$ and $A_\mu = \bar s \gamma_\mu \gamma^5 u$. The form factors $F$, $G$, $R$ and $H$ are functions of $s$, $s_\ell$ and $\cos\theta_\pi$ (or $s$, $t$ and $u$).

In order to write the decay rate in a compact form, it is convenient to introduce new form factors as linear combinations of $F$, $G$, $R$ and $H$ (following \cite{Pais1968, Bijnens1994}) that correspond to definite helicity amplitudes:
\begin{align}
	\label{eqn:Kl4NewFormFactors}
	\begin{split}
		F_1 &:= X F + \sigma_\pi (PL) \cos\theta_\pi G , \\
		F_2 &:= \sigma_\pi \sqrt{s s_\ell} G , \\
		F_3 &:= \sigma_\pi X \sqrt{s s_\ell} \frac{H}{\mkp^2} , \\
		F_4 &:= -(PL) F - s_\ell R - \sigma_\pi X \cos\theta_\pi G.
	\end{split}
\end{align}

The partial decay rate for the $K_{\ell4}$ decay is given by
\begin{align}
	d\Gamma = \frac{1}{2\mkp(2\pi)^8} \sum_\mathrm{spins} | \mathcal{T} |^2 \delta^{(4)}(p - P - L) \frac{d^3 p_1}{2p_1^0} \frac{d^3 p_2}{2p_2^0} \frac{d^3 p_\ell}{2p_\ell^0} \frac{d^3 p_\nu}{2p_\nu^0}.
\end{align}
Since the kinematics is described by five phase space variables, seven integrals can be performed. This leads to
\begin{align}
	\label{eqn:FiveDimensionalPhaseSpace}
	d\Gamma = G_F^2 |V_{us}|^2 \frac{(1-z_\ell) \sigma_\pi(s)X}{2^{13}\pi^6\mkp^5} J_5(s,s_\ell,\theta_\pi,\theta_\ell,\phi) \, ds \, ds_\ell \, d\cos\theta_\pi \, d\cos\theta_\ell \, d\phi .
\end{align}

The explicit expression for $J_5$ is derived in the appendix \ref{sec:DecayRateIsospinLimit}. 

$F_4$ corresponds to the helicity amplitude of the spin 0 or temporal polarisation of the virtual $W$ boson. Its contribution to the decay rate is therefore helicity suppressed by a factor $m_\ell^2$ and invisible in the electron mode. In the chiral expansion, $F_3$ appears due to the chiral anomaly, which is at the level of the Lagrangian an $\mathcal{O}(p^4)$ effect. Thus, the important form factors for the experiment are $F_1$ and $F_2$, or equivalently $F$ and $G$.

\paragraph{$K_{\ell4}$ in the Case of Broken Isospin}

In the presence of electromagnetism, the above factorisation of the $K_{\ell4}$ matrix element into a hadronic and a leptonic part is no longer valid. In addition to the $V-A$ structure, a tensorial form factor has to be taken into account \cite{Cuplov2003,Cuplov2004}:
\begin{align}
	\begin{split}
		\label{eqn:TMatrixBrokenIsospin}
		\mathcal{T} ={}& \frac{G_F}{\sqrt{2}} V_{us}^* \left( \bar u(p_\nu) \gamma_\mu (1-\gamma^5)v(p_\ell) \; ( \mathcal{V}^\mu - \mathcal{A}^\mu ) + \bar u(p_\nu) \sigma_{\mu\nu}(1+\gamma^5) v(p_\ell) \mathcal{T}^{\mu\nu} \right), \\
		\mathcal{V}_\mu :={}& -\frac{H}{\mkp^3} \epsilon_{\mu\nu\rho\sigma} L^\nu P^\rho Q^\sigma , \\
		\mathcal{A}_\mu :={}& -i \frac{1}{\mkp} \left( P_\mu F + Q_\mu G + L_\mu R \right) , \\
		\mathcal{T}^{\mu\nu} :={}& \frac{1}{\mkp^2} p_1^\mu p_2^\nu T ,
	\end{split}
\end{align}
where $\sigma_{\mu\nu} = \frac{i}{2} [ \gamma_\mu, \gamma_\nu ]$. The form factors $F$, $G$, $R$, $H$ and $T$ depend now on all five kinematic variables $s$, $s_\ell$, $\theta_\pi$, $\theta_\ell$ and $\phi$.

I follow \cite{Cuplov2004} and introduce in addition to (\ref{eqn:Kl4NewFormFactors}) the form factor $F_5$ (with a slightly different normalisation):
\begin{align}
	\label{eqn:TensorialFormFactorF5}
	F_5 := \frac{\sigma_\pi(s) s s_\ell}{2 \mkp \ml} \, T .
\end{align}
Still, the phase space is parametrised by the same five kinematic variables and the differential decay rate can be written as in (\ref{eqn:FiveDimensionalPhaseSpace}). In the isospin broken case, the presence of the additional tensorial form factor changes the function $J_5$. We will see that $F_5$ is finite in the limit $\ml\to0$. Its contribution to the decay rate is suppressed by $\ml^2$. Details are given in the appendix \ref{sec:DecayRateIsospinBroken}.

It is convenient to define the following additional Lorentz invariants \cite{Cuplov2004}:
\begin{align}
		t_\ell &:= (p - p_\ell)^2 , \quad u_\ell := (p-p_\nu)^2 , \quad s_{1\ell} := (p_1 + p_\ell)^2, \quad s_{2\ell} := (p_2 + p_\ell)^2 .
\end{align}


\subsection{The Radiative Decay $K_{\ell4\gamma}$}

\subsubsection{Definition of the Decay}

If we consider electromagnetic corrections to $K_{\ell4}$, we have to take into account contributions from both virtual photons and real photon emission, because only an appropriate inclusive observable is free of infrared singularities. As long as we restrict ourselves to $\O(e^2)$ corrections, the radiative process with just one additional final state photon is the only one of interest (each additional photon comes along with a factor $e^2$ in the decay rate).

The radiative process $K_{\ell4\gamma}$ is defined as
\begin{align}
	\begin{split}
		\label{eqn:Kl4gDecay}
		K^+(p) \to \pi^+(p_1) \pi^-(p_2) \ell^+(p_\ell) \nu_\ell(p_\nu) \gamma(q) .
	\end{split}
\end{align}
There are several possibilities to parametrise the phase space. I find it most convenient to replace the dilepton sub-phase space of $K_{\ell4}$ by a convenient three particle phase space.

I describe the kinematics still in three reference frames: the rest frame of the kaon $\Sigma_K$, the dipion centre-of-mass frame $\Sigma_{2\pi}$ and the dilepton-photon centre-of-mass frame $\Sigma_{\ell\nu\gamma}$. In total, we need eight phase space variables:
\begin{itemize}
	\item $s$, the total centre-of-mass squared energy of the two pions,
	\item $s_\ell$, the total centre-of-mass squared energy of the dilepton-photon system,
	\item $\theta_\pi$, the angle between the $\pi^+$ in $\Sigma_{2\pi}$ and the dipion line of flight in $\Sigma_K$,
	\item $\theta_\gamma$, the angle between the photon in $\Sigma_{\ell\nu\gamma}$ and the $\ell\nu\gamma$ line of flight in $\Sigma_K$,
	\item $\phi$, the angle between the dipion plane and the $(\ell\nu)\gamma$ plane in $\Sigma_K$.
	\item $q^0$, the photon energy in $\Sigma_{\ell\nu\gamma}$,
	\item $p_\ell^0$, the lepton energy in $\Sigma_{\ell\nu\gamma}$,
	\item $\phi_\ell$, the angle between the $\ell\nu$ plane in $\Sigma_{\ell\nu\gamma}$ and the $(\ell\nu)\gamma$ plane in $\Sigma_K$.
\end{itemize}
The variables $s$, $s_\ell$, $\theta_\pi$ are defined in analogy to the $K_{\ell4}$ decay. The reason for the chosen parametrisation of the $\ell\nu\gamma$ subsystem is that $p_\ell^0$ and $\phi_\ell$ are of purely kinematic nature, i.e.~the dynamics depends only on the six other variables.

Instead of the $q^0$ and $p_\ell^0$, I will mostly use the dimensionless variables
\begin{align}
	\begin{split}
		x := \frac{2 Lq}{s_\ell} , \quad y := \frac{2 L p_\ell}{s_\ell},
	\end{split}
\end{align}
where $L := p_\ell + p_\nu + q$ and $s_\ell = L^2$. They are related to $q^0$ and $p_\ell^0$ by
\begin{align}
	\begin{split}
		x = \frac{2 q^0}{\sqrt{s_\ell}} , \quad y = \frac{2 p_\ell^0}{\sqrt{s_\ell}} .
	\end{split}
\end{align}

I give the photon an artificial small mass $m_\gamma$ in order to regulate the infrared divergences. The ranges of the phase space variables are:
\begin{align}
	\begin{alignedat}{2}
		4 \mpip^2 &\le s &&\le (\mkp - \ml - m_\gamma)^2 , \\
		(\ml + m_\gamma)^2 &\le s_\ell &&\le (\mkp - \sqrt{s})^2 , \\
		0 &\le \theta_\pi &&\le \pi , \\
		0 &\le \theta_\gamma &&\le \pi , \\
		0 &\le \phi &&\le 2\pi , \\
		0 &\le \phi_\ell &&\le 2\pi .
	\end{alignedat}
\end{align}

Let us determine in the following the ranges of the two variables $x$ and $y$. Introducing the variable $s_{\ell\nu} := (p_\ell+p_\nu)^2$, I find the relations
\begin{align}
	\begin{split}
		q^0 = \frac{s_\ell + \mg^2 - s_{\ell\nu}}{2\sqrt{s_\ell}} , \quad x = 1 + \frac{\mg^2}{s_\ell} - \frac{s_{\ell\nu}}{s_\ell} .
	\end{split}
\end{align}
The range of $s_{\ell\nu}$ is obviously
\begin{align}
	\begin{split}
		\ml^2 \le s_{\ell\nu} \le (\sqrt{s_\ell} - \mg)^2 ,
	\end{split}
\end{align}
which leads to
\begin{align}
	\begin{split}
		2 \sqrt{\zg} \le x \le 1 + \zg - \zl ,
	\end{split}
\end{align}
where I have defined
\begin{align}
	\begin{split}
		\zl := \frac{\ml^2}{s_\ell}, \quad \zg = \frac{\mg^2}{s_\ell} .
	\end{split}
\end{align}

The range of $y$ for a given value of $x$ can be found as follows. Determine the boost from $\Sigma_{\ell\nu\gamma}$ to the $\ell\nu$ centre-of-mass frame $\Sigma_{\ell\nu}$ by considering the vector $p_\ell + p_\nu$ in both frames. Define $z = \cos\hat\theta_\ell$ with $\hat\theta_\ell$ being the angle between the lepton momentum in $\Sigma_{\ell\nu}$ and the dilepton line of flight in $\Sigma_{\ell\nu\gamma}$. Then, with the help of the inverse boost, you will find $y$ in terms of $z$ and $x$:
\begin{align}
	\begin{split}
		y = \frac{z\sqrt{x^2-4\zg}(1+\zg-\zl-x) + (2 - x)(1+\zg+\zl-x)}{2(1+\zg-x)} .
	\end{split}
\end{align}
In the limit $\mg\to0$, I obtain the following range:
\begin{align}
	\begin{split}
		\label{eqn:yRange}
		1 - x + \frac{\zl}{1-x} \le y \le 1 + \zl .
	\end{split}
\end{align}

Similar to $K_{\ell4}$, I introduce for the radiative process the momenta
\begin{align}
	\label{eqn:FourMomentaKl4g}
	P := p_1 + p_2 , \quad Q := p_1 - p_2 , \quad L := q + p_\ell + p_\nu , \quad N := q + p_\ell - p_\nu .
\end{align}
It will be useful to define also the momenta
\begin{align}
	\begin{split}
		\label{eqn:AlternativeFourMomentaKl4g}
		\hat L := p_\ell + p_\nu = L - q , \quad \hat N := p_\ell - p_\nu = N - q .
	\end{split}
\end{align}
Total momentum conservation implies $p = P + L$. I will use the Lorentz invariants
\begin{align}
	s := (p_1 + p_2)^2 , \quad t := (p - p_1)^2 , \quad u := (p - p_2)^2 , \quad \sg := (p_\ell + q)^2 = s_\ell( x + y - 1) .
\end{align}
In the appendix \ref{sec:LorentzTransformationsKl4g}, the Lorentz transformations between the three reference frames are determined and all the Lorentz invariant products are computed.

\subsubsection{Matrix Element, Form Factors and Decay Rate}

\label{sec:Kl4gMatrixElement}

The matrix element of the radiative decay (\ref{eqn:Kl4gDecay}) has the form (in analogy to $K_{\ell3\gamma}$ \cite{Gasser2005})
\begin{align}
	\begin{split}
		\label{eqn:Kl4gTMatrix}
		\mathcal{T}_\gamma &= - \frac{G_F}{\sqrt{2}} e V_{us}^* \epsilon_\mu(q)^*
				\begin{aligned}[t]
					&\bigg[ \mathcal{H}^{\mu\nu} \; \bar u(p_\nu) \gamma_\nu (1-\gamma^5)v(p_\ell) + \mathcal{H}^\nu \; \frac{1}{2 p_\ell q} \bar u(p_\nu) \gamma_\nu (1-\gamma^5)(m_\ell - \slashed p_\ell - \slashed q) \gamma^\mu v(p_\ell) \bigg]
				\end{aligned} \\
			&=: \epsilon_\mu(q)^* \mathcal{M}^\mu ,
	\end{split}
\end{align}
where $\mathcal{H}^\nu = \mathcal{V}^\nu-\mathcal{A}^\nu$ is the hadronic part of the $K_{\ell4}$ matrix element. The second part of the matrix element stems from diagrams where the photon is radiated off the lepton line, the first part contains all the rest. The hadronic tensor $\mathcal{H}^{\mu\nu} = \mathcal{V}^{\mu\nu} - \mathcal{A}^{\mu\nu}$ is defined by
\begin{align}
	\begin{split}
		\mathcal{I}^{\mu\nu} &= i \int d^4x \, e^{iqx} \< \pi^+(p_1) \pi^-(p_2) | T \{ V^\mu_\mathrm{em}(x) I^\nu(0) \} | K^+(p) \> , \\
		\mathcal{I} &= \mathcal{V}, \mathcal{A} , \quad I = V, A,
	\end{split}
\end{align}
and satisfies the Ward identity
\begin{align}
	\begin{split}
		q_\mu \mathcal{H}^{\mu\nu} = \mathcal{H}^\nu ,
	\end{split}
\end{align}
such that the condition $q_\mu \mathcal{M}^\mu = 0$ required by gauge invariance is fulfilled.

If the contributions from the anomalous sector are neglected, the hadronic tensor can be decomposed into dimensionless form factors as (the photon is taken on-shell)
\begin{align}
	\begin{split}
		\mathcal{H}^{\mu\nu} &= \frac{i}{\mkp} g^{\mu\nu} \Pi + \frac{i}{\mkp^2}\left( P^\mu \Pi_0^\nu + Q^\mu \Pi_1^\nu + L^\mu \Pi_2^\nu \right) , \\
		\Pi_i^\nu &= \frac{1}{\mkp} \left( P^\nu \Pi_{i0} + Q^\nu \Pi_{i1} + L^\nu \Pi_{i2} + q^\nu \Pi_{i3}  \right) .
	\end{split}
\end{align}
Gauge invariance requires the following relations:
\begin{align}
	\begin{split}
		\label{eqn:GaugeInvarianceFFRelations}
		\mkp^2 \, F - Pq \, \Pi_{00} - Qq \, \Pi_{10} - Lq \, \Pi_{20} &= 0 , \\
		\mkp^2 \, G - Pq \, \Pi_{01} - Qq \, \Pi_{11} - Lq \, \Pi_{21} &= 0 , \\
		\mkp^2 \, R - Pq \, \Pi_{02} - Qq \, \Pi_{12} - Lq \, \Pi_{22} &= 0 , \\
		\mkp^2 \, \Pi + Pq \, \Pi_{03} + Qq \, \Pi_{13} + Lq \, \Pi_{23} &= 0 , \\
	\end{split}
\end{align}
where $F$, $G$ and $R$ are the $K_{\ell4}$ form factors.

The partial decay rate for $K_{\ell4\gamma}$ is given by
\begin{align}
	d\Gamma_\gamma = \frac{1}{2\mkp(2\pi)^{11}} \sum_{\substack{\mathrm{spins} \\ \mathrm{polar.}}} | \mathcal{T}_\gamma |^2 \delta^{(4)}(p - P - L) \frac{d^3 p_1}{2p_1^0} \frac{d^3 p_2}{2p_2^0} \frac{d^3 p_\ell}{2p_\ell^0} \frac{d^3 p_\nu}{2p_\nu^0} \frac{d^3 q}{2q^0}.
\end{align}
Seven integrals can be performed, leaving the integrals over the eight phase space variables:
\begin{align}
	\label{eqn:EightDimensionalPhaseSpace}
	d\Gamma_\gamma &= G_F^2 |V_{us}|^2 e^2 \frac{s_\ell \, \sigma_\pi(s) X}{2^{20}\pi^9 \mkp^7} J_8(s,s_\ell,\theta_\pi,\theta_\gamma,\phi,x,y,\phi_\ell) \, ds \, ds_\ell \, d\cos\theta_\pi \, d\cos\theta_\gamma \, d\phi \, dx \, dy \, d\phi_\ell .
\end{align}
The procedure how to find the explicit expression for $J_8$ in terms of the form factors is explained in appendix~\ref{sec:RadiativeDecayRate}.


\section{\ChPT{} Calculation of the Amplitudes}

\label{sec:ChPTCalculation}

Isospin symmetry is the symmetry under $SU(2)$ transformations of up- and down-quarks. In nature, this symmetry is realised only approximately. The source of isospin symmetry breaking is twofold: on the one hand, $u$- and $d$-quarks do not have the same mass, on the other hand, their electric charge is different. On the fundamental level of the standard model, we can therefore distinguish between quark mass effects and electromagnetic effects.

Usually, calculations of processes can be simplified substantially if isospin symmetry is assumed to be exact. In order to link such calculations to real word measurements, the effects of isospin breaking have to be known. The aim of this work is to compute such isospin breaking corrections to the $K_{\ell4}$ decay.

As $K_{\ell4}$ is a process that happens at low energies, the hadronic part of the matrix element can not be computed perturbatively in QCD. The low-energy effective theory of QCD, chiral perturbation theory (\ChPT{}) \cite{Weinberg1968, GasserLeutwyler1984, GasserLeutwyler1985}, does not treat quarks and gluons but the Goldstone bosons of the spontaneously broken chiral symmetry of QCD as the degrees of freedom. In this effective theory, the isospin breaking effects show up as differences in the masses of the charged and neutral mesons and in form of photonic corrections. The meson mass differences are due to both isospin breaking sources, the quark mass difference as well as electromagnetism. I compute the isospin breaking effects in $K_{\ell4}$ within \ChPT{} including virtual photons and leptons \cite{Urech1995, Knecht2000}. As this is a well-known framework, I abstain from giving a review but only collect the most important formulae in appendix~\ref{sec:AppendixChPT} in order to settle the conventions.

I take into account only first order corrections in the isospin breaking parameters and effects up to one loop. The leading-order form factors behave as $\O(p)$, i.e.~I consider effects of $\O(p^3)$, $\O(\epsilon \, p^3)$, $\O(e^2 \, p)$, where $e=+|e|$ is the electric unit charge and
\begin{align}
	\begin{split}
		\epsilon := \frac{\sqrt{3}}{4} \frac{m_u - m_d}{\hat m - m_s} , \quad \hat m := \frac{m_u + m_d}{2} .
	\end{split}
\end{align}

Since the chiral anomaly shows up first at next-to-leading chiral order, I do not compute isospin breaking corrections to the form factor $H$.


\subsection{Mass Effects}

In contrast to the photonic effects that appear as $\O(e^2)$ corrections in my calculation, the `non-photonic' electromagnetic effects due to the different meson masses in the loops give corrections of the order $\O(Z e^2)$, where $Z$ is the low-energy constant in $\mathcal{L}_{e^2}$. This allows for a separation of the mass effects from purely photonic corrections (a subtlety concerning the counterterms will be discussed later). Let us thus first discuss the mass effects, i.e.~the isospin corrections in the absence of virtual photons.

These $\O(\epsilon \, p^3)$ and $\O(Z e^2 p)$ effects have been considered in \cite{Cuplov2003, Cuplov2004, Colangelo2009}. The present calculation agrees with the results given in \cite{Cuplov2003, Cuplov2004}. For completeness, I give the explicit expressions in my conventions.

\subsubsection{Leading Order}

\begin{figure}[ht]
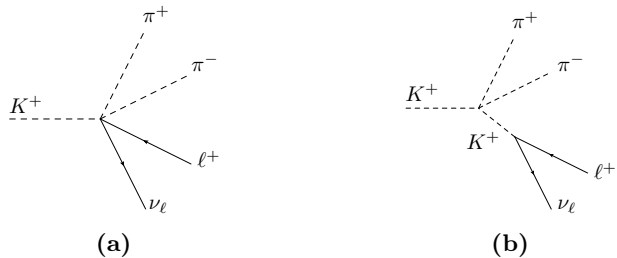

	\centering
	\begin{subfigure}[b]{0.3\textwidth}
		\centering
		\scalebox{0.8}{
			\begin{pspicture}(0,-0.5)(3,3)
				\put(0,1.6){$K^+$}
				\put(2.2,3.1){$\pi^+$}
				\put(3,2.3){$\pi^-$}
				\put(3.1,0.7){$\ell^+$}
				\put(2.3,0){$\nu_\ell$}
				\includegraphics[width=3cm]{images/LO1}
			\end{pspicture}
			}
		\caption{}
		\label{img:Kl4_LO1}
	\end{subfigure}
	\begin{subfigure}[b]{0.3\textwidth}
		\centering
		\scalebox{0.8}{
			\begin{pspicture}(0,-0.5)(3,3)
				\put(0,1.8){$K^+$}
				\put(1.75,3.0){$\pi^+$}
				\put(2.5,2.3){$\pi^-$}
				\put(3.1,0.5){$\ell^+$}
				\put(2.5,0){$\nu_\ell$}
				\put(1.0,1.0){$K^+$}
				\includegraphics[width=3cm]{images/LO2}
			\end{pspicture}
			}
		\caption{}
		\label{img:Kl4_LO2}
	\end{subfigure}
	\caption{Tree-level diagrams for the $K_{\ell4}$ decay.}
	\label{img:Kl4LO}
\end{figure}

At leading order, we have to compute two tree-level diagrams, shown in figure~\ref{img:Kl4LO}.

Diagram~\ref{img:Kl4_LO1} contributes to the form factors $F$, $G$ and $R$, whereas diagram~\ref{img:Kl4_LO2} only contributes to the form factor $R$. This is true for all diagrams with an intermediate kaon pole, also at one-loop level.

The leading-order results for the form factors are:
\begin{align}
	\begin{split}
		F^\mathrm{LO}_\mathrm{ME} &= G^\mathrm{LO}_\mathrm{ME} = \frac{\mkp}{\sqrt{2} F_0} , \\
		R^\mathrm{LO}_\mathrm{ME} &= \frac{\mkp}{2 \sqrt{2}F_0} \frac{\mkp^2 - s -s_\ell - \nu - 4 \Delta_\pi}{\mkp^2 - s_\ell} , \\
		T^\mathrm{LO}_\mathrm{ME} &= 0.
	\end{split}
\end{align}
Only the form factor $R$ gets at leading order an isospin correction due to the mass effects.

\subsubsection{Next-to-Leading Order}

Since the contributions of both $R$ and $T$ to the decay rate are suppressed by $\ml^2$ and experimentally inaccessible in the electron mode, I will calculate only corrections to the form factors $F$ and $G$. Thus, I neglect at next-to-leading order all diagrams that contribute only to the form factor $R$, i.e.~diagrams with a kaon pole in the $s_\ell$-channel. It is convenient to write the NLO expressions for the form factors as
\begin{align}
	\begin{split}
		\label{eqn:NLOFormFactorsMassEffects}
		F^\mathrm{NLO}_\mathrm{ME} &= F^\mathrm{LO}_\mathrm{ME} \left( 1 + \delta F^\mathrm{NLO}_\mathrm{ME} \right) , \\
		G^\mathrm{NLO}_\mathrm{ME} &= G^\mathrm{LO}_\mathrm{ME} \left( 1 + \delta G^\mathrm{NLO}_\mathrm{ME} \right) .
	\end{split}
\end{align}
Since the LO contribution is of $\O(p)$, the order of the NLO corrections considered here is
\begin{align}
	\begin{split}
		\delta F^\mathrm{NLO}_\mathrm{ME}, \delta G^\mathrm{NLO}_\mathrm{ME} = \O(p^2) + \O(\epsilon \, p^2) + \O(Z e^2) .
	\end{split}
\end{align}

Of course, the loop integrals appearing at NLO are UV-divergent. I will regularise them dimensionally and renormalise the theory as usual in the $\overline{MS}$ scheme. The divergent parts of the loop integrals are cancelled by the divergent parts of the LECs.

The explicit NLO results are rather lengthy and can be found in appendix~\ref{sec:AppendixDiagramsMassEffects}.

\paragraph{Loop Diagrams}

\begin{figure}[ht]
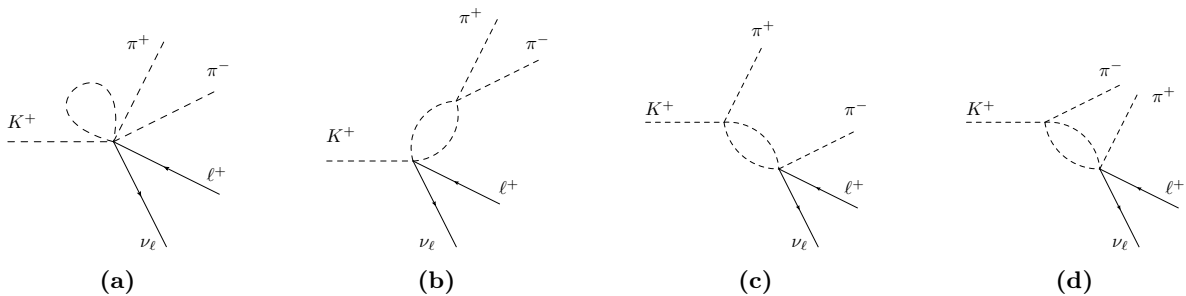

	\centering
	\begin{subfigure}[b]{0.24\textwidth}
		\centering
		\scalebox{0.7}{
			\begin{pspicture}(0,-0.5)(4,5)
				\put(0,2.25){$K^+$}
				\put(2.25,3.75){$\pi^+$}
				\put(3.75,3.25){$\pi^-$}
				\put(3.75,1.25){$\ell^+$}
				\put(2.5,0){$\nu_\ell$}
				\includegraphics[width=4cm]{images/NLO_Tadpole}
			\end{pspicture}
			}
		\caption{}
		\label{img:Kl4_NLOTadpole}
	\end{subfigure}
	\begin{subfigure}[b]{0.24\textwidth}
		\centering
		\scalebox{0.7}{
			\begin{pspicture}(0,-0.5)(4,5)
				\put(0,2.0){$K^+$}
				\put(2.5,4.25){$\pi^+$}
				\put(3.75,3.75){$\pi^-$}
				\put(3.25,1.0){$\ell^+$}
				\put(1.75,0){$\nu_\ell$}
				\includegraphics[width=4cm]{images/NLO_SLoop}
			\end{pspicture}
			}
		\caption{}
		\label{img:Kl4_NLOSloop}
	\end{subfigure}
	\begin{subfigure}[b]{0.24\textwidth}
		\centering
		\scalebox{0.7}{
			\begin{pspicture}(0,-0.5)(4,5)
				\put(0,2.5){$K^+$}
				\put(2.0,4.0){$\pi^+$}
				\put(3.75,2.5){$\pi^-$}
				\put(3.75,1.0){$\ell^+$}
				\put(2.75,0){$\nu_\ell$}
				\includegraphics[width=4cm]{images/NLO_TLoop}
			\end{pspicture}
			}
		\caption{}
		\label{img:Kl4_NLOTloop}
	\end{subfigure}
	\begin{subfigure}[b]{0.24\textwidth}
		\centering
		\scalebox{0.7}{
			\begin{pspicture}(0,-0.5)(4,5)
				\put(0,2.5){$K^+$}
				\put(3.5,2.75){$\pi^+$}
				\put(2.5,3.25){$\pi^-$}
				\put(3.75,1.0){$\ell^+$}
				\put(2.75,0){$\nu_\ell$}
				\includegraphics[width=4cm]{images/NLO_ULoop}
			\end{pspicture}
			}
		\caption{}
		\label{img:Kl4_NLOUloop}
	\end{subfigure}
	\caption{One-loop diagrams contributing to the $K_{\ell4}$ form factors $F$ and $G$.}
	\label{img:Kl4_Loops}
\end{figure}

At NLO, we have to compute the tadpole diagram~\ref{img:Kl4_NLOTadpole} with all possible mesons ($\pi^0$, $\pi^+$, $K^0$, $K^+$ and $\eta$) in the loop as well as the diagrams~\ref{img:Kl4_NLOSloop}-\ref{img:Kl4_NLOUloop} with two-meson intermediate states in the $s$-, $t$- and $u$-channel.

The contributions of the meson loop diagrams can be expressed in terms of the scalar loop functions $A_0$ and $B_0$ (which should not be confused with the low-energy constant $B_0$).

\paragraph{Counterterms}

\begin{figure}[ht]
	\centering
		\scalebox{0.7}{
			\begin{pspicture}(0,-0.5)(4,5)
				\put(0,2.25){$K^+$}
				\put(2.25,3.75){$\pi^+$}
				\put(3.75,3.25){$\pi^-$}
				\put(3.75,1.25){$\ell^+$}
				\put(2.5,0){$\nu_\ell$}
				\includegraphics[width=4cm]{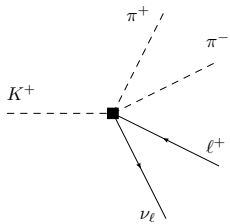}
			\end{pspicture}
			}
	\caption{Counterterm diagram contributing to the $K_{\ell4}$ form factors $F$ and $G$.}
	\label{img:Kl4_CT1}
\end{figure}

I express the one-loop corrections in terms of the scalar loop functions $A_0$ and $B_0$. These loop functions contain UV divergences that have to cancel against the UV divergences in the counterterms and the field strength renormalisation. The only relevant counterterm diagram is shown in figure~\ref{img:Kl4_CT1}. It contains a vertex from the NLO Lagrangian. Now, a subtlety arises here. As we are interested in the mass effects, we have neglected pure $\O(e^2)$ loop corrections, but kept $\O(Z e^2)$ contributions. If we used the same prescription for the counterterms, the UV divergences would not cancel. The reason is that some of the electromagnetic LECs $K_i$ contain a factor $Z$ in their beta function, hence their divergent part is multiplied by $Z$ and contributes at $\O(Z e^2)$. We therefore have to assign also these LECs to the mass effects.

\paragraph{External Leg Corrections}

\begin{figure}[H]
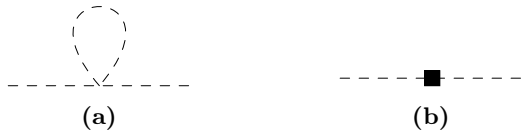

	\centering
	\begin{subfigure}[b]{0.25\textwidth}
		\centering
		\scalebox{0.8}{
			\begin{pspicture}(0,0)(4,3)
				\includegraphics[width=3cm]{images/SE_Loop}
			\end{pspicture}
			}
		\caption{}
		\label{img:Kl4_SELoop}
	\end{subfigure}
	\begin{subfigure}[b]{0.25\textwidth}
		\centering
		\scalebox{0.8}{
			\begin{pspicture}(0,0)(4,3)
				\includegraphics[width=3cm]{images/SE_CT}
			\end{pspicture}
			}
		\caption{}
		\label{img:Kl4_SECT}
	\end{subfigure}
	\caption{Meson self-energy diagrams.}
	\label{img:Kl4_ExternalLegs}
\end{figure}

The last contribution at NLO are the external leg corrections. We have to compute only the field strength renormalisation of the mesons (the lepton propagators get no corrections). For the self-energy of the mesons at NLO, the corrections to the propagator shown in figure~\ref{img:Kl4_ExternalLegs} have to be taken into account. All the Goldstone bosons $\pi^+$, $\pi^0$, $K^+$, $K^0$ and $\eta$ have to be inserted in the tadpole diagram.

\paragraph{Renormalisation}

The complete expressions for the form factors at NLO including the mass effects are
\begin{align}
	\begin{split}
		X^\mathrm{NLO}_\mathrm{ME} &= X^\mathrm{LO}_\mathrm{ME} \left( 1 + \delta X^\mathrm{NLO}_\mathrm{ME} \right) , \\
	\end{split}		\tag{\ref{eqn:NLOFormFactorsMassEffects}}
\end{align}
with
\begin{align}
	\begin{split}
		\delta X^\mathrm{NLO}_\mathrm{ME} &= \delta X^\mathrm{NLO}_\mathrm{tadpole} + \delta X^\mathrm{NLO}_\text{$s$-loop} + \delta X^\mathrm{NLO}_\text{$t$-loop} + \delta X^\mathrm{NLO}_\text{$u$-loop} + \delta X^\mathrm{NLO}_\mathrm{ct} + \delta X_Z^\mathrm{NLO} ,
	\end{split}
\end{align}
where $X \in \{ F, G \}$. The explicit expressions for the individual contributions can be found in the appendix~\ref{sec:AppendixDiagramsMassEffects}. The form factors have to be UV-finite, hence, we should check that in the above sum, all the UV divergences cancel. If I replace the LECs with help of (\ref{eqn:RenormalisedLECs}) and the loop functions with (\ref{eqn:RenormalisedLoopFunctions}), I find indeed that all the terms proportional to the UV divergence $\lambda$ (\ref{eqn:UVDivergenceLambda}) cancel.


\subsection{Photonic Effects}

In a next step, I calculate in the effective theory the effects due to the presence of photons. I include virtual photon corrections up to NLO, i.e.~I have to compute again one-loop diagrams, counterterms and external leg corrections. The sum of these contributions will be UV-finite but contain IR and collinear (in the limit $m_\ell\to0$) singularities. As it is well known, the IR divergences will cancel in the sum of the decay rates of $K_{\ell4}$ and the soft real photon emission process $K^+ \to \pi^+ \pi^- \ell^+ \nu_\ell \gamma_\mathrm{soft}$. The collinear divergence is in the physical case regulated by the lepton mass, which plays the role of a natural cut-off. It cancels in the sum of the decay rates of $K_{\ell4}$ and the (soft and hard) collinear real photon emission process. (Note that at $\O(e^2)$, the emission of only one photon has to be taken into account.)
The fully inclusive decay rate $K^+ \to \pi^+ \pi^- \ell^+ \nu_\ell (\gamma)$ is free of IR and mass divergences and does not depend on a cut-off, in accordance with the KLN theorem \cite{KinoshitaSirlin1959,Kinoshita1962,Lee1964}.

As in the case of the mass effects, also the photonic effects have already been computed in \cite{Cuplov2003, Cuplov2004}. However, in these works a whole gauge invariant class of diagrams appearing at NLO has been overlooked\footnote{I thank V.~Cuplov for confirming this.}. The present calculation confirms the results for the diagrams calculated in \cite{Cuplov2004} (in \cite{Cuplov2003}, eq.~(72) gives a wrong result for one of the diagrams) and completes it with the missing class of diagrams.

For the calculation of the photonic effects, I take into account NLO corrections of $\O(e^2)$, but I neglect contributions of $\O(Z e^2)$ as well as $\O(\epsilon \, p^2)$, since they are treated by the mass effects.

\subsubsection{Leading Order}

\begin{figure}[ht]
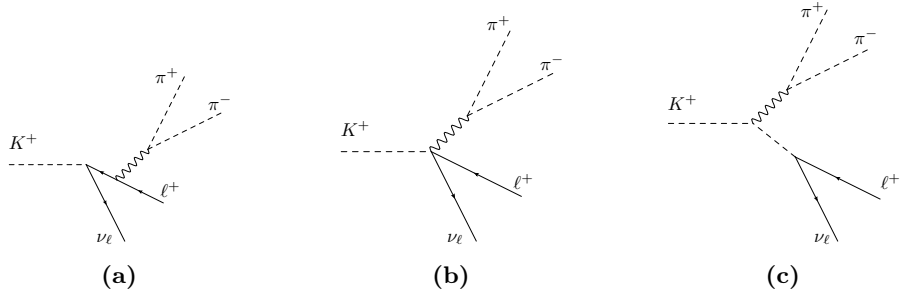

	\centering
	\begin{subfigure}[b]{0.25\textwidth}
		\centering
		\scalebox{0.7}{
			\begin{pspicture}(0,-0.5)(4,5)
				\put(0,1.75){$K^+$}
				\put(2.75,3.0){$\pi^+$}
				\put(3.75,2.5){$\pi^-$}
				\put(2.85,0.85){$\ell^+$}
				\put(1.65,0){$\nu_\ell$}
				\includegraphics[width=4cm]{images/LO3}
			\end{pspicture}
			}
		\caption{}
		\label{img:Kl4_LO3}
	\end{subfigure}
	\begin{subfigure}[b]{0.25\textwidth}
		\centering
		\scalebox{0.7}{
			\begin{pspicture}(0,-0.5)(4,5)
				\put(0,2.0){$K^+$}
				\put(2.75,4.0){$\pi^+$}
				\put(3.75,3.25){$\pi^-$}
				\put(3.25,1.0){$\ell^+$}
				\put(2.0,0){$\nu_\ell$}
				\includegraphics[width=4cm]{images/LO4}
			\end{pspicture}
			}
		\caption{}
		\label{img:Kl4_LO4}
	\end{subfigure}
	\begin{subfigure}[b]{0.25\textwidth}
		\centering
		\scalebox{0.7}{
			\begin{pspicture}(0,-0.5)(4,5)
				\put(0,2.5){$K^+$}
				\put(2.5,4.25){$\pi^+$}
				\put(3.5,3.75){$\pi^-$}
				\put(4.0,1.0){$\ell^+$}
				\put(2.75,0){$\nu_\ell$}
				\includegraphics[width=4cm]{images/LO5}
			\end{pspicture}
			}
		\caption{}
		\label{img:Kl4_LO5}
	\end{subfigure}
	\caption{Tree-level diagrams for the $K_{\ell4}$ decay with a virtual photon.}
	\label{img:Kl4LO_Photons}
\end{figure}

Photonic effects appear already at leading order in the effective theory, i.e.~at $\O(e^2 p^{-1})$, as diagrams with a virtual photon splitting into two pions. In addition to the $\O(e^0 p)$ tree-level diagrams in figure~\ref{img:Kl4LO}, the diagrams shown in figure~\ref{img:Kl4LO_Photons} have to be calculated.

The diagram~\ref{img:Kl4_LO3} contributes to the form factors $G$, $R$ and the tensorial form factor $T$. However, the contribution to $G$ gets exactly cancelled by the diagram~\ref{img:Kl4_LO4}. Diagram~\ref{img:Kl4_LO5} only contributes to $R$.

Therefore, the contribution of the diagrams in figure~\ref{img:Kl4LO_Photons} does not alter the form factors $F$ and $G$:
\begin{align}
	\begin{split}
		F^\mathrm{LO}_{\mathrm{virt.}\gamma} &= \frac{\mkp}{\sqrt{2} F_0} , \quad G^\mathrm{LO}_{\mathrm{virt.}\gamma} =  \frac{\mkp}{\sqrt{2} F_0} .
	\end{split}
\end{align}
The other form factors read (in agreement with \cite{Cuplov2003})
\begin{align}
	\begin{split}
		R^\mathrm{LO}_{\mathrm{virt.}\gamma} &=  \frac{\mkp}{2 \sqrt{2}F_0} \left( \frac{\mkp^2 - s -s_\ell - \nu}{\mkp^2 - s_\ell} + \frac{4 e^2 F_0}{s} \left( \frac{s_{1\ell} - s_{2\ell}}{u_\ell - \ml^2} + \frac{\nu}{\mkp^2 - s_\ell} \right) \right) , \\
		T^\mathrm{LO}_{\mathrm{virt.}\gamma} &= 2 \sqrt{2} e^2 F_0 \frac{\mkp^2 \ml}{s(u_\ell - \ml^2)} .
	\end{split}
\end{align}
We see that the tensorial form factor $F_5$, which was defined above,
\begin{align}
	\begin{split}
		F_5 = \frac{\sigma_\pi(s) s s_\ell}{2 \mkp \ml} \, T ,
	\end{split} \tag{\ref{eqn:TensorialFormFactorF5}}
\end{align}
stays finite in the limit $\ml\to0$. This shows that its contribution to the decay rate (see (\ref{eqn:DecayRateFormFactorsIsoBrokenTensorial}) and (\ref{eqn:DecayRateFormFactorsIsoBrokenInterference}) in the appendix) is suppressed by $\ml^2$. In the following, I will therefore only consider the form factors $F$ and $G$.

\subsubsection{Next-to-Leading Order}

In order to regularise the IR divergence of loop diagrams with virtual photons, I introduce an artificial photon mass $\mg$ and use the propagator of a massive vector field:
\begin{align}
	\begin{split}
		\tilde G^{\mu\nu}(k) = \frac{-i}{k^2 - \mg^2 + i\epsilon} \left( g^{\mu\nu} - \frac{k^\mu k^\nu}{ \mg^2} \right) .
	\end{split}
\end{align}
The same regulator has to be used in the calculation of the radiative process. In the end, one has to take the limit $\mg\to0$, which restores gauge invariance. Terms that do not contribute in this limit are neglected.

For the NLO calculation of photonic effects, I consider all contributions to the form factors $F$ and $G$ of $\O(e^2 p)$. They consist of loop diagrams with virtual photons, counter\-terms and external leg corrections for $K_{\ell4}$. On the other hand, tree diagrams for the radiative process $K_{\ell4\gamma}$ have to be included at the level of the decay rate.

It is again convenient to write the NLO contribution in the form
\begin{align}
	\begin{split}
		F^\mathrm{NLO}_{\mathrm{virt.}\gamma} &= F^\mathrm{LO}_{\mathrm{virt.}\gamma} \left( 1 + \delta F^\mathrm{NLO}_{\mathrm{virt.}\gamma} \right) , \\
		G^\mathrm{NLO}_{\mathrm{virt.}\gamma} &= G^\mathrm{LO}_{\mathrm{virt.}\gamma} \left( 1 + \delta G^\mathrm{NLO}_{\mathrm{virt.}\gamma} \right) .
	\end{split}
\end{align}
The explicit results are collected in the appendix~\ref{sec:AppendixDiagramsPhotonicEffects}.

\paragraph{Loop Diagrams}

\begin{figure}[H]
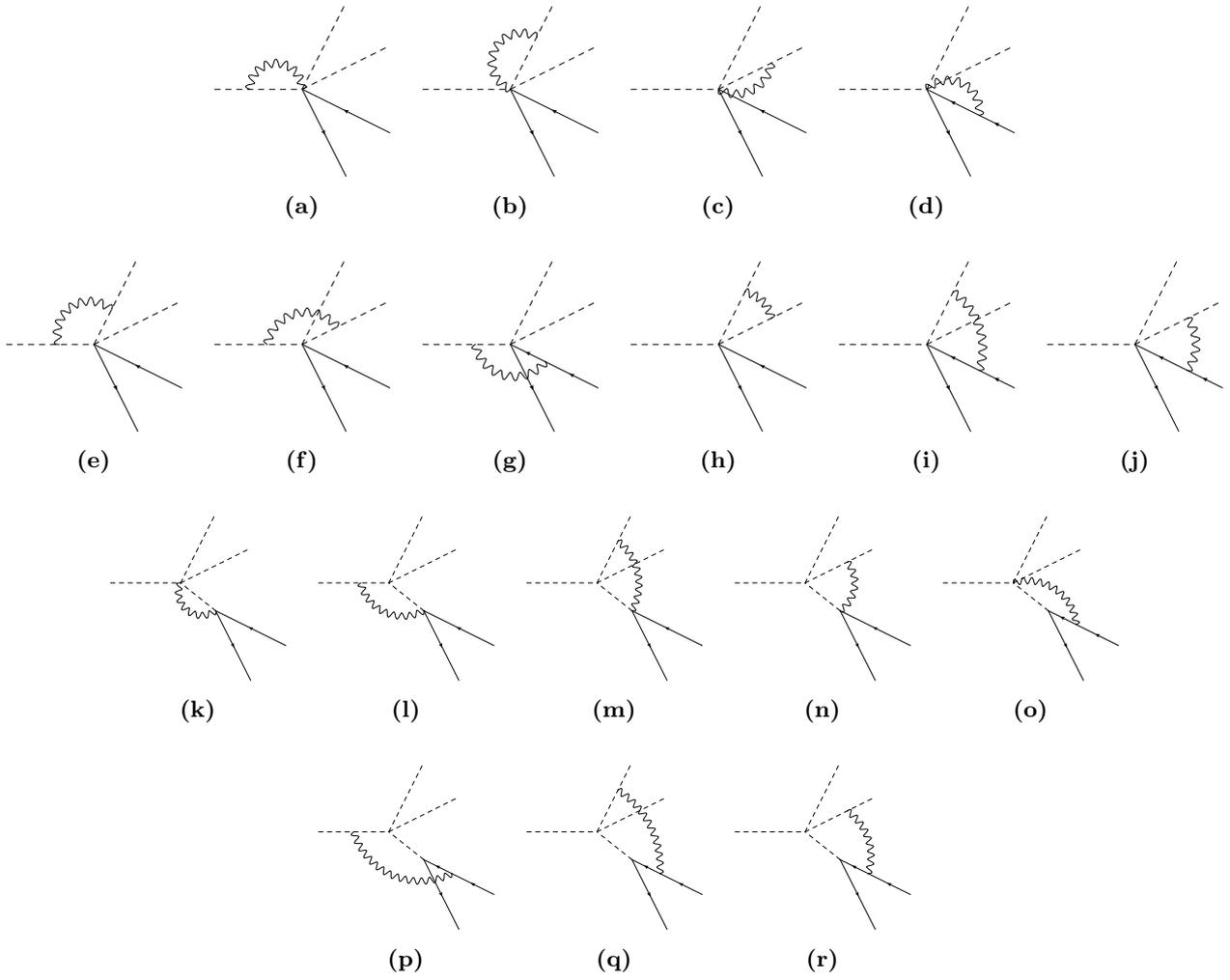

	\centering
	\begin{subfigure}[b]{0.16\textwidth}
		\centering
		\scalebox{0.8}{
			\begin{pspicture}(0,-0.5)(4,5)
				\includegraphics[width=3cm]{images/NLO_gLoop1}
			\end{pspicture}
			}
		\caption{}
		\label{img:Kl4_NLOgLoop1}
	\end{subfigure}
	\begin{subfigure}[b]{0.16\textwidth}
		\centering
		\scalebox{0.8}{
			\begin{pspicture}(0,-0.5)(4,5)
				\includegraphics[width=3cm]{images/NLO_gLoop2}
			\end{pspicture}
			}
		\caption{}
		\label{img:Kl4_NLOgLoop2}
	\end{subfigure}
	\begin{subfigure}[b]{0.16\textwidth}
		\centering
		\scalebox{0.8}{
			\begin{pspicture}(0,-0.5)(4,5)
				\includegraphics[width=3cm]{images/NLO_gLoop3}
			\end{pspicture}
			}
		\caption{}
		\label{img:Kl4_NLOgLoop3}
	\end{subfigure}
	\begin{subfigure}[b]{0.16\textwidth}
		\centering
		\scalebox{0.8}{
			\begin{pspicture}(0,-0.5)(4,5)
				\includegraphics[width=3cm]{images/NLO_gLoop4}
			\end{pspicture}
			}
		\caption{}
		\label{img:Kl4_NLOgLoop4}
	\end{subfigure}
	
	\vspace{0.5cm}

	\begin{subfigure}[b]{0.16\textwidth}
		\centering
		\scalebox{0.8}{
			\begin{pspicture}(0,-0.5)(4,5)
				\includegraphics[width=3cm]{images/NLO_gLoop5}
			\end{pspicture}
			}
		\caption{}
		\label{img:Kl4_NLOgLoop5}
	\end{subfigure}
	\begin{subfigure}[b]{0.16\textwidth}
		\centering
		\scalebox{0.8}{
			\begin{pspicture}(0,-0.5)(4,5)
				\includegraphics[width=3cm]{images/NLO_gLoop6}
			\end{pspicture}
			}
		\caption{}
		\label{img:Kl4_NLOgLoop6}
	\end{subfigure}
	\begin{subfigure}[b]{0.16\textwidth}
		\centering
		\scalebox{0.8}{
			\begin{pspicture}(0,-0.5)(4,5)
				\includegraphics[width=3cm]{images/NLO_gLoop7}
			\end{pspicture}
			}
		\caption{}
		\label{img:Kl4_NLOgLoop7}
	\end{subfigure}
	\begin{subfigure}[b]{0.16\textwidth}
		\centering
		\scalebox{0.8}{
			\begin{pspicture}(0,-0.5)(4,5)
				\includegraphics[width=3cm]{images/NLO_gLoop8}
			\end{pspicture}
			}
		\caption{}
		\label{img:Kl4_NLOgLoop8}
	\end{subfigure}
	\begin{subfigure}[b]{0.16\textwidth}
		\centering
		\scalebox{0.8}{
			\begin{pspicture}(0,-0.5)(4,5)
				\includegraphics[width=3cm]{images/NLO_gLoop9}
			\end{pspicture}
			}
		\caption{}
		\label{img:Kl4_NLOgLoop9}
	\end{subfigure}
	\begin{subfigure}[b]{0.16\textwidth}
		\centering
		\scalebox{0.8}{
			\begin{pspicture}(0,-0.5)(4,5)
				\includegraphics[width=3cm]{images/NLO_gLoop10}
			\end{pspicture}
			}
		\caption{}
		\label{img:Kl4_NLOgLoop10}
	\end{subfigure}

	\vspace{0.5cm}

	\begin{subfigure}[b]{0.16\textwidth}
		\centering
		\scalebox{0.8}{
			\begin{pspicture}(0,-0.5)(4,5)
				\includegraphics[width=3cm]{images/NLO_gLoop11}
			\end{pspicture}
			}
		\caption{}
		\label{img:Kl4_NLOgLoop11}
	\end{subfigure}
	\begin{subfigure}[b]{0.16\textwidth}
		\centering
		\scalebox{0.8}{
			\begin{pspicture}(0,-0.5)(4,5)
				\includegraphics[width=3cm]{images/NLO_gLoop12}
			\end{pspicture}
			}
		\caption{}
		\label{img:Kl4_NLOgLoop12}
	\end{subfigure}
	\begin{subfigure}[b]{0.16\textwidth}
		\centering
		\scalebox{0.8}{
			\begin{pspicture}(0,-0.5)(4,5)
				\includegraphics[width=3cm]{images/NLO_gLoop13}
			\end{pspicture}
			}
		\caption{}
		\label{img:Kl4_NLOgLoop13}
	\end{subfigure}
	\begin{subfigure}[b]{0.16\textwidth}
		\centering
		\scalebox{0.8}{
			\begin{pspicture}(0,-0.5)(4,5)
				\includegraphics[width=3cm]{images/NLO_gLoop14}
			\end{pspicture}
			}
		\caption{}
		\label{img:Kl4_NLOgLoop14}
	\end{subfigure}
	\begin{subfigure}[b]{0.16\textwidth}
		\centering
		\scalebox{0.8}{
			\begin{pspicture}(0,-0.5)(4,5)
				\includegraphics[width=3cm]{images/NLO_gLoop15}
			\end{pspicture}
			}
		\caption{}
		\label{img:Kl4_NLOgLoop15}
	\end{subfigure}

	\vspace{0.5cm}

	\begin{subfigure}[b]{0.16\textwidth}
		\centering
		\scalebox{0.8}{
			\begin{pspicture}(0,-0.5)(4,5)
				\includegraphics[width=3cm]{images/NLO_gLoop16}
			\end{pspicture}
			}
		\caption{}
		\label{img:Kl4_NLOgLoop16}
	\end{subfigure}
	\begin{subfigure}[b]{0.16\textwidth}
		\centering
		\scalebox{0.8}{
			\begin{pspicture}(0,-0.5)(4,5)
				\includegraphics[width=3cm]{images/NLO_gLoop17}
			\end{pspicture}
			}
		\caption{}
		\label{img:Kl4_NLOgLoop17}
	\end{subfigure}
	\begin{subfigure}[b]{0.16\textwidth}
		\centering
		\scalebox{0.8}{
			\begin{pspicture}(0,-0.5)(4,5)
				\includegraphics[width=3cm]{images/NLO_gLoop18}
			\end{pspicture}
			}
		\caption{}
		\label{img:Kl4_NLOgLoop18}
	\end{subfigure}

	\caption{First set of one-loop diagrams with virtual photons: they are obtained by a virtual photon insertion in the tree diagrams in figure~\ref{img:Kl4LO} (I drop the labels for the external particles as they are always the same). Diagrams contributing only to the form factor $R$ are omitted.}
	\label{img:Kl4_gLoops}
\end{figure}

A first class of loop diagrams is obtained by adding a virtual photon to the tree diagrams in figure~\ref{img:Kl4LO}. All diagrams contributing to $F$ and $G$ are shown in figure~\ref{img:Kl4_gLoops}. Again, diagrams with a virtual kaon pole are omitted, as they contribute only to $R$.

I choose to express most of the results in terms of the basic scalar loop functions $A_0$, $B_0$, $C_0$ and $D_0$. 

\begin{figure}[H]
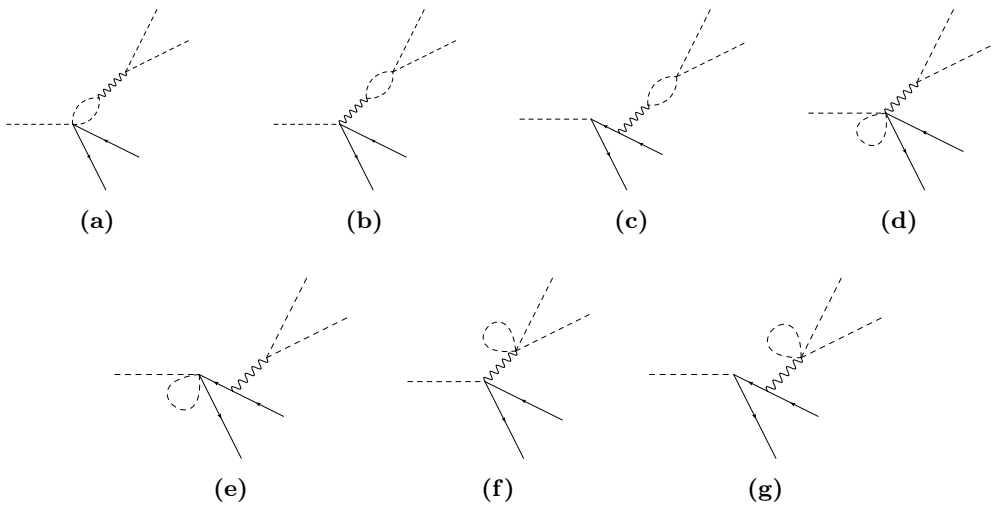

	\centering
	\begin{subfigure}[b]{0.2\textwidth}
		\centering
		\scalebox{0.8}{
			\begin{pspicture}(0,-0.5)(4,5)
				\includegraphics[height=3cm]{images/NLO_mLoop1}
			\end{pspicture}
			}
		\caption{}
		\label{img:Kl4_NLOmLoop1}
	\end{subfigure}
	\begin{subfigure}[b]{0.2\textwidth}
		\centering
		\scalebox{0.8}{
			\begin{pspicture}(0,-0.5)(4,5)
				\includegraphics[height=3cm]{images/NLO_mLoop2}
			\end{pspicture}
			}
		\caption{}
		\label{img:Kl4_NLOmLoop2}
	\end{subfigure}
	\begin{subfigure}[b]{0.2\textwidth}
		\centering
		\scalebox{0.8}{
			\begin{pspicture}(0,-0.5)(4,5)
				\includegraphics[height=3cm]{images/NLO_mLoop3}
			\end{pspicture}
			}
		\caption{}
		\label{img:Kl4_NLOmLoop3}
	\end{subfigure}
	\begin{subfigure}[b]{0.2\textwidth}
		\centering
		\scalebox{0.8}{
			\begin{pspicture}(0,-0.5)(4,5)
				\includegraphics[height=3cm]{images/NLO_mLoop4}
			\end{pspicture}
			}
		\caption{}
		\label{img:Kl4_NLOmLoop4}
	\end{subfigure}
	
	\vspace{0.5cm}

	\begin{subfigure}[b]{0.2\textwidth}
		\centering
		\scalebox{0.8}{
			\begin{pspicture}(0,-0.5)(4,5)
				\includegraphics[height=3cm]{images/NLO_mLoop5}
			\end{pspicture}
			}
		\caption{}
		\label{img:Kl4_NLOmLoop5}
	\end{subfigure}
	\begin{subfigure}[b]{0.2\textwidth}
		\centering
		\scalebox{0.8}{
			\begin{pspicture}(0,-0.5)(4,5)
				\includegraphics[height=3cm]{images/NLO_mLoop6}
			\end{pspicture}
			}
		\caption{}
		\label{img:Kl4_NLOmLoop6}
	\end{subfigure}
	\begin{subfigure}[b]{0.2\textwidth}
		\centering
		\scalebox{0.8}{
			\begin{pspicture}(0,-0.5)(4,5)
				\includegraphics[height=3cm]{images/NLO_mLoop7}
			\end{pspicture}
			}
		\caption{}
		\label{img:Kl4_NLOmLoop7}
	\end{subfigure}

	\caption{Second set of one-loop diagrams with virtual photons: they are obtained by a meson loop insertion in the tree diagrams in figure~\ref{img:Kl4LO_Photons}. Diagrams contributing only to the form factor $R$ are omitted.}
	\label{img:Kl4_mLoops}
\end{figure}

\clearpage

The contributions of the diagrams~\ref{img:Kl4_NLOgLoop1} - \ref{img:Kl4_NLOgLoop4}, where one end of the photon line is attached to a charged external line and the other end to the vertex, are all IR-finite.

The next six (triangle) diagrams, obtained by attaching a virtual photon to two external lines, generate an IR divergence. My results differ from \cite{Cuplov2004} only by the contribution of the additional term in the propagator for the massive vector boson. This contribution will cancel in the sum with the external leg corrections.

The remaining eight diagrams in this first set consist of one bulb, four triangle and finally three box diagrams that are obtained by an insertion of a virtual photon into diagram~\ref{img:Kl4_LO2}.

A second set of loop diagrams, shown in figure~\ref{img:Kl4_mLoops}, is obtained by inserting virtual mesons into the tree-level diagrams in figure~\ref{img:Kl4LO_Photons}. Although the contributions of the LO diagrams in figure~\ref{img:Kl4LO_Photons} to the form factors $F$ and $G$ vanish, the NLO diagrams give a finite contribution to $G$. To my knowledge, they have not been considered in the previous literature.

In diagrams \ref{img:Kl4_NLOmLoop1} - \ref{img:Kl4_NLOmLoop3}, we have to insert charged mesons in the loop. In the tadpole loops, all octet mesons have to be included.

\paragraph{Counterterms}

\begin{figure}[ht]
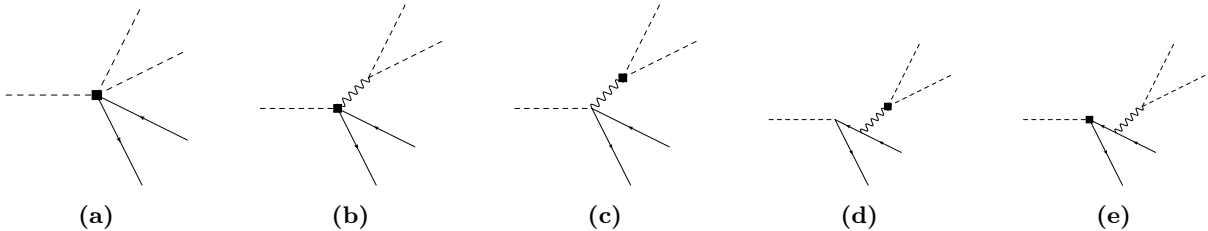

	\centering
	\begin{subfigure}[b]{0.19\textwidth}
		\centering
		\scalebox{0.8}{
			\begin{pspicture}(0,-0.5)(4,5)
				\includegraphics[width=3cm]{images/NLO_CT1}
			\end{pspicture}
			}
		\caption{}
		\label{img:Kl4_NLOCT1}
	\end{subfigure}
	\begin{subfigure}[b]{0.19\textwidth}
		\centering
		\scalebox{0.8}{
			\begin{pspicture}(0,-0.5)(4,5)
				\includegraphics[width=3cm]{images/NLO_CT2}
			\end{pspicture}
			}
		\caption{}
		\label{img:Kl4_NLOCT2}
	\end{subfigure}
	\begin{subfigure}[b]{0.19\textwidth}
		\centering
		\scalebox{0.8}{
			\begin{pspicture}(0,-0.5)(4,5)
				\includegraphics[width=3cm]{images/NLO_CT3}
			\end{pspicture}
			}
		\caption{}
		\label{img:Kl4_NLOCT3}
	\end{subfigure}
	\begin{subfigure}[b]{0.19\textwidth}
		\centering
		\scalebox{0.8}{
			\begin{pspicture}(0,-0.5)(4,5)
				\includegraphics[width=3cm]{images/NLO_CT4}
			\end{pspicture}
			}
		\caption{}
		\label{img:Kl4_NLOCT4}
	\end{subfigure}
	\begin{subfigure}[b]{0.19\textwidth}
		\centering
		\scalebox{0.8}{
			\begin{pspicture}(0,-0.5)(4,5)
				\includegraphics[width=3cm]{images/NLO_CT5}
			\end{pspicture}
			}
		\caption{}
		\label{img:Kl4_NLOCT5}
	\end{subfigure}

	\caption{Counterterms needed to renormalise the loops with virtual photons.}
	\label{img:Kl4_gCT}
\end{figure}

In order to renormalise the UV divergences in the loop functions, we have to compute the counterterm contribution, i.e.~tree-level diagrams with one vertex from $\mathcal{L}_{p^4}$, $\mathcal{L}_{e^2p^2}$ or $\mathcal{L}_\mathrm{lept}$. These diagrams are shown in figure~\ref{img:Kl4_gCT}. The loop diagrams of the first class, figure~\ref{img:Kl4_gLoops}, need only the counterterm~\ref{img:Kl4_NLOCT1}, the remaining four counterterm diagrams renormalise the meson loops of the second class, figure~\ref{img:Kl4_mLoops}.

\paragraph{External Leg Corrections}

In order to complete the NLO calculation, we need the external leg corrections at $\O(e^2 p)$. At this order, the corrections for both charged mesons and lepton have to be taken into account.

\begin{figure}[H]
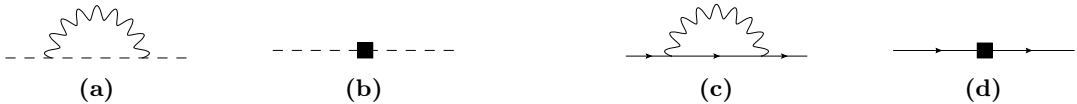

	\centering
	\begin{subfigure}[b]{0.2\textwidth}
		\centering
		\scalebox{0.8}{
			\begin{pspicture}(0,0)(4,3)
				\includegraphics[width=3cm]{images/mSE_gLoop}
			\end{pspicture}
			}
		\caption{}
		\label{img:Kl4_mSEgLoop}
	\end{subfigure}
	\begin{subfigure}[b]{0.2\textwidth}
		\centering
		\scalebox{0.8}{
			\begin{pspicture}(0,0)(4,3)
				\includegraphics[width=3cm]{images/SE_CT}
			\end{pspicture}
			}
		\caption{}
		\label{img:Kl4_mSECT}
	\end{subfigure}
	\hspace{1cm}
	\begin{subfigure}[b]{0.2\textwidth}
		\centering
		\scalebox{0.8}{
			\begin{pspicture}(0,0)(4,3)
				\includegraphics[width=3cm]{images/lSE_gLoop}
			\end{pspicture}
			}
		\caption{}
		\label{img:Kl4_lSEgLoop}
	\end{subfigure}
	\begin{subfigure}[b]{0.2\textwidth}
		\centering
		\scalebox{0.8}{
			\begin{pspicture}(0,0)(4,3)
				\includegraphics[width=3cm]{images/lSE_CT}
			\end{pspicture}
			}
		\caption{}
		\label{img:Kl4_lSECT}
	\end{subfigure}
	\caption{Meson and lepton self-energy diagrams.}
	\label{img:Kl4_gExternalLegs}
\end{figure}

The calculation of the field strength renormalisation and its contribution to the form factors can be found in the appendix~\ref{sec:AppendixExternalLegCorrectionsPhotonicEffects}.

\paragraph{Renormalisation}

The form factors at $\O(e^2 p)$ are given by
\begin{align}
	\begin{split}
		X^\mathrm{NLO}_{\mathrm{virt.}\gamma} &= X^\mathrm{LO}_{\mathrm{virt.}\gamma} \left( 1 + \delta X^\mathrm{NLO}_{\mathrm{virt.}\gamma}  \right) , \quad X\in\{F,G\} ,
	\end{split}
\end{align}
where the NLO corrections consists of
\begin{align}
	\begin{split}
		\delta X^\mathrm{NLO}_{\mathrm{virt.}\gamma} &= \delta X^\mathrm{NLO}_{\gamma-\mathrm{loop}} + \delta X^\mathrm{NLO}_{\gamma-\mathrm{pole}} + \delta X^\mathrm{NLO}_{\gamma-\mathrm{ct}} + \delta X^\mathrm{NLO}_{\gamma-Z} .
	\end{split}
\end{align}
The individual contributions are all given explicitly in the appendix~\ref{sec:AppendixDiagramsPhotonicEffects}. With these expressions, it can be easily verified that the contributions stemming from the additional term $k^\mu k^\nu / \mg^2$ in the propagator for a massive gauge boson (with respect to a massless propagator in Feynman gauge) cancel in the above sum (in the limit $\mg\to0$). In appendix~\ref{sec:RadiativeDecayRate}, I show that the radiative decay rate only gets $\O(\mg^2)$ contributions from the additional term in the propagator. Hence, in the limit $\mg\to0$, the longitudinal modes decouple and gauge invariance is restored~\cite{Ticciati1999}.

Next, let us check that the UV-divergent parts cancel in the sum of all NLO contributions. Working in the $\overline{MS}$ scheme, I replace the LECs according to (\ref{eqn:RenormalisedLECs}) with their renormalised counterparts. Introducing also the renormalised loop functions (\ref{eqn:RenormalisedLoopFunctions}) and tensor coefficient functions (\ref{eqn:RenormalisedTensorCoefficients}), I find that all the terms proportional to the UV divergence $\lambda$ cancel.

\subsubsection{Real Photon Emission}

\label{sec:MatrixElementRealPhotonEmission}

\begin{figure}[ht]
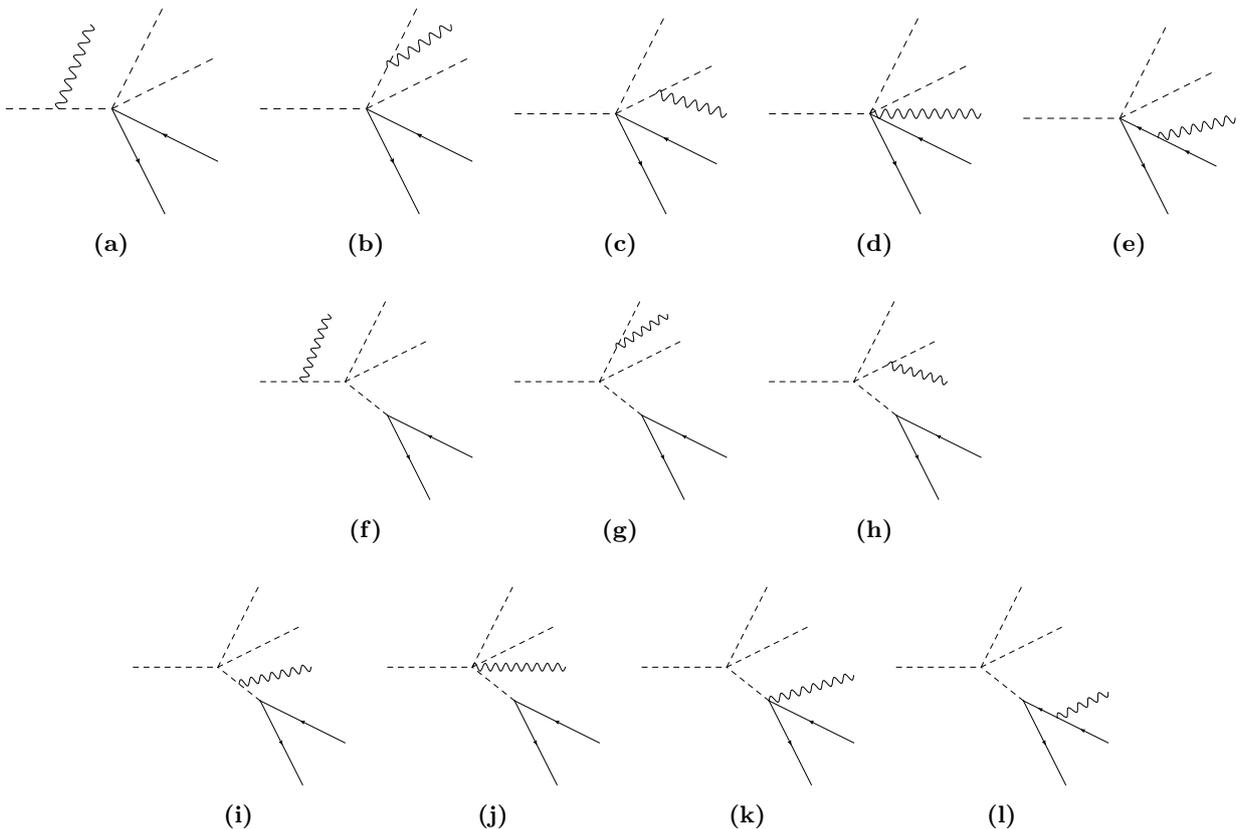

	\centering
	\begin{subfigure}[b]{0.19\textwidth}
		\centering
		\scalebox{0.8}{
			\begin{pspicture}(0,0)(4,3)
				\includegraphics[width=3.5cm]{images/RE1}
			\end{pspicture}
			}
		\caption{}
		\label{img:Kl4g1}
	\end{subfigure}
	\begin{subfigure}[b]{0.19\textwidth}
		\centering
		\scalebox{0.8}{
			\begin{pspicture}(0,0)(4,3)
				\includegraphics[width=3.5cm]{images/RE2}
			\end{pspicture}
			}
		\caption{}
		\label{img:Kl4g2}
	\end{subfigure}
	\begin{subfigure}[b]{0.19\textwidth}
		\centering
		\scalebox{0.8}{
			\begin{pspicture}(0,0)(4,3)
				\includegraphics[width=3.5cm]{images/RE3}
			\end{pspicture}
			}
		\caption{}
		\label{img:Kl4g3}
	\end{subfigure}
	\begin{subfigure}[b]{0.19\textwidth}
		\centering
		\scalebox{0.8}{
			\begin{pspicture}(0,0)(4,3)
				\includegraphics[width=3.5cm]{images/RE4}
			\end{pspicture}
			}
		\caption{}
		\label{img:Kl4g4}
	\end{subfigure}
	\begin{subfigure}[b]{0.19\textwidth}
		\centering
		\scalebox{0.8}{
			\begin{pspicture}(0,0)(4,3)
				\includegraphics[width=3.5cm]{images/RE5}
			\end{pspicture}
			}
		\caption{}
		\label{img:Kl4g5}
	\end{subfigure}
	
	\vspace{0.5cm}
	
	\begin{subfigure}[b]{0.19\textwidth}
		\centering
		\scalebox{0.8}{
			\begin{pspicture}(0,0)(4,3)
				\includegraphics[width=3.5cm]{images/RE6}
			\end{pspicture}
			}
		\caption{}
		\label{img:Kl4g6}
	\end{subfigure}
	\begin{subfigure}[b]{0.19\textwidth}
		\centering
		\scalebox{0.8}{
			\begin{pspicture}(0,0)(4,3)
				\includegraphics[width=3.5cm]{images/RE7}
			\end{pspicture}
			}
		\caption{}
		\label{img:Kl4g7}
	\end{subfigure}
	\begin{subfigure}[b]{0.19\textwidth}
		\centering
		\scalebox{0.8}{
			\begin{pspicture}(0,0)(4,3)
				\includegraphics[width=3.5cm]{images/RE8}
			\end{pspicture}
			}
		\caption{}
		\label{img:Kl4g8}
	\end{subfigure}
	
	\vspace{0.5cm}
	
	\begin{subfigure}[b]{0.19\textwidth}
		\centering
		\scalebox{0.8}{
			\begin{pspicture}(0,0)(4,3)
				\includegraphics[width=3.5cm]{images/RE9}
			\end{pspicture}
			}
		\caption{}
		\label{img:Kl4g9}
	\end{subfigure}
	\begin{subfigure}[b]{0.19\textwidth}
		\centering
		\scalebox{0.8}{
			\begin{pspicture}(0,0)(4,3)
				\includegraphics[width=3.5cm]{images/RE10}
			\end{pspicture}
			}
		\caption{}
		\label{img:Kl4g10}
	\end{subfigure}
	\begin{subfigure}[b]{0.19\textwidth}
		\centering
		\scalebox{0.8}{
			\begin{pspicture}(0,0)(4,3)
				\includegraphics[width=3.5cm]{images/RE11}
			\end{pspicture}
			}
		\caption{}
		\label{img:Kl4g11}
	\end{subfigure}
	\begin{subfigure}[b]{0.19\textwidth}
		\centering
		\scalebox{0.8}{
			\begin{pspicture}(0,0)(4,3)
				\includegraphics[width=3.5cm]{images/RE12}
			\end{pspicture}
			}
		\caption{}
		\label{img:Kl4g12}
	\end{subfigure}
	\caption{Tree-level diagrams for the decay $K_{\ell4\gamma}$.}
	\label{img:Kl4g}
\end{figure}

As explained before, an IR-finite result can only be obtained for a sufficiently inclusive observable. In the present case, we have to add the $\O(e^2)$ contribution of the radiative process at the decay rate level. Let us therefore compute the $\O(e)$ tree-level amplitude for $K_{\ell4\gamma}$.

The relevant diagrams are shown in figure~\ref{img:Kl4g}. If we use the decomposition of the matrix element defined in section~\ref{sec:Kl4gMatrixElement}, the diagrams~\ref{img:Kl4g5} and \ref{img:Kl4g12} just reproduce the second term in (\ref{eqn:Kl4gTMatrix}), where the hadronic part is given by the LO $K_{\ell4}$ form factors in the isospin limit.

The diagrams~\ref{img:Kl4g4} and \ref{img:Kl4g11}, where the photon is emitted off the vertex, correspond to the form factor $\Pi$:
\begin{align}
	\begin{split}
		\Pi =  \frac{\mkp}{2\sqrt{2} F_0} \left( 5 - \frac{ s + \nu }{\mkp^2 - s_\ell} \right) ,
	\end{split}
\end{align}
where $\nu = t-u$.

The form factors $\Pi_{ij}$ correspond to the remaining 8 diagrams, where the photon is emitted off a meson line or a mesonic vertex. The form factors multiplying $\bar u(p_\nu) \slashed P (1 - \gamma^5) v(p_\ell)$ or $\bar u(p_\nu) \slashed Q (1 - \gamma^5) v(p_\ell)$ have a simple form:
\begin{align}
	\begin{split}
		\Pi_{00} &= \Pi_{01} = - \frac{\mkp^3}{\sqrt{2} F_0} \left( \frac{2}{\mg^2 - 2 pq} + \frac{1}{\mg^2 + 2 p_1q} - \frac{1}{\mg^2 + 2 p_2q} \right) , \\
		\Pi_{10} &= \Pi_{11} = - \frac{\mkp^3}{\sqrt{2} F_0} \left( \frac{1}{\mg^2 + 2 p_1q} + \frac{1}{\mg^2 + 2 p_2q} \right) , \\
		\Pi_{20} &= \Pi_{21} = - \frac{\mkp^3}{\sqrt{2} F_0} \frac{2}{\mg^2 - 2 pq} .
	\end{split}
\end{align}
The remaining form factors are a bit more complicated. They satisfy the relations
\begin{align}
	\begin{split}
		\Pi_{03} &= - \Pi_{02} - \frac{\mkp^3}{\sqrt{2} F_0} \frac{2}{\mg^2 + 2 p_1q} , \\
		\Pi_{13} &= - \Pi_{12} - \frac{\mkp^3}{\sqrt{2} F_0} \frac{2}{\mg^2 + 2 p_1q} , \\
		\Pi_{23} &= - \Pi_{22} .
	\end{split}
\end{align}
In the limit $\mg\to0$, I find
\begin{align}
	\begin{split}
		\Pi_{02} &= \frac{\mkp^3}{\sqrt{2} F_0} \frac{1}{\mkp^2 - s_\ell + 2 Lq} \begin{aligned}[t]
				&\Bigg( \frac{\mkp^2 - s - t + u - s_\ell}{4} \left( \frac{2}{pq} - \frac{1}{p_1q} + \frac{1}{p_2q} \right) + \frac{Lq -  Qq}{pq} - \frac{Lq}{p_1q}  - \frac{Qq}{p_2q} \Bigg) ,
				\end{aligned} \\
		\Pi_{12} &= \frac{\mkp^3}{\sqrt{2} F_0} \frac{1}{\mkp^2 - s_\ell + 2 Lq} \begin{aligned}[t]
				&\Bigg( \frac{\mkp^2 - s - t + u - s_\ell}{4} \left( - \frac{1}{p_1q} - \frac{1}{p_2q} \right) - \frac{Lq}{p_1q} + \frac{Qq}{p_2q} + 3 \Bigg) ,
				\end{aligned} \\
		\Pi_{22} &= \frac{\mkp^3}{\sqrt{2} F_0} \frac{1}{\mkp^2 - s_\ell + 2 Lq} \begin{aligned}[t]
				& \Bigg( \frac{\mkp^2 - s - t + u - s_\ell}{2} \left( \frac{1}{pq} + \frac{2}{\mkp^2 - s_\ell} \right) + \frac{Lq - Qq}{pq} + 1 \Bigg) .
				\end{aligned}
	\end{split}
\end{align}
These expressions fulfil the relations (\ref{eqn:GaugeInvarianceFFRelations}) as required by gauge invariance.

The contribution of the diagrams~\ref{img:Kl4g6}-\ref{img:Kl4g10} to the decay rate is helicity suppressed by a factor of $\ml^2$. The suppression at leading chiral order also works for the diagrams~\ref{img:Kl4g11} and \ref{img:Kl4g12}. One could therefore omit all diagrams with a kaon pole in the limit $\ml\to0$. However, from a technical point of view, this barely reduces the complexity of the calculation. Hence, I have given here the results for the form factors using the complete set of diagrams. Moreover, at higher chiral order, one has to expect structure dependent contributions not suppressed by $\ml^2$.


\section{Extraction of the Isospin Corrections}

\label{sec:ExtractionOfIsospinCorrections}

This section discusses the extraction of the isospin breaking corrections to the $K_{\ell4}$ form factors and decay rate. While the experiments are performed in our real world, where isospin is broken, it is useful to translate measured quantities into the context of an ideal world with conserved isospin, i.e.~a world with no electromagnetism and identical up- and down-quark masses. The motivation for doing such a transformation is that in an isospin symmetric world, calculations may become much easier. The isospin breaking corrections for $K_{\ell4}$ will be used in a forthcoming dispersive treatment of this decay \cite{Colangelo2012, Stoffer2013, Colangelo2013}, which is performed in the isospin limit.

Correcting the isospin breaking effects in existing experimental data on the $K_{\ell4}$ form factors is a delicate matter: the $K_{\ell4}$ form factors are in the real world themselves not observable quantities, because they are not infrared-safe. As explained above, any experiment will measure a semi-inclusive decay rate of $K_{\ell4}$ and $K_{\ell4+n\gamma}$, typically with some cuts on the real photons. These cuts are detector specific and naturally defined in the lab frame. It is almost impossible to handle such cuts in an analytic way. Therefore, one must rely on a Monte Carlo simulation of the semi-inclusive decay that models the detector geometry and all the applied cuts in order to extract isospin corrected quantities. I suggest for future experiments the inclusion of the here presented amplitudes for $K_{\ell4}$ and $K_{\ell4\gamma}$ in a Monte Carlo simulation like PHOTOS \cite{Barberio1994}.

The isospin corrections due to the mass effects can be extracted directly for the form factors. For the photonic effects, I calculate the radiative corrections to the (semi-)inclusive decay rate.

\subsection{Mass Effects}

I define the isospin breaking corrections to the form factors as follows.

The measured semi-inclusive differential decay rate $d\Gamma_{(\gamma, \mathrm{cut})}^\mathrm{exp}$ (neglecting experimental uncertainties) equals the result from the presented NLO calculation up to higher order in the chiral expansion or the isospin breaking parameters:
\begin{align}
	\begin{split}
		d\Gamma_{(\gamma, \mathrm{cut})}^\mathrm{exp} = d\Gamma_{(\gamma, \mathrm{cut})}^\mathrm{NLO} + h.o. &= d\Gamma^\mathrm{NLO}_\mathrm{iso} + \Delta d\Gamma^\mathrm{NLO}_\mathrm{ME} + \Delta d\Gamma^\mathrm{NLO}_{\mathrm{virt.}\gamma} + \int_\mathrm{cut} d\Gamma_{\gamma} \\
			& + \O(p^6, \epsilon \, p^6, Ze^2 p^4, e^2 p^4) + \O(\epsilon^2, \epsilon \, e^2, e^4),
	\end{split}
\end{align}
where the real photon in the radiative decay rate is integrated using the same cuts as in the experiment. I expect the contribution of higher order in the breaking parameters to be negligible, while the $\O(p^6)$ contribution is certainly not. The different terms are of the following order:
\begin{align}
	\begin{split}
		d\Gamma^\mathrm{NLO}_\mathrm{iso} &= \O( p^4 ), \quad \Delta d\Gamma^\mathrm{NLO}_\mathrm{ME} = \O( \epsilon \, p^4, Z e^2 p^2 ) , \\
		\Delta d\Gamma^\mathrm{NLO}_{\mathrm{virt.}\gamma} &= \O( e^2 p^2 ) , \quad \int_\mathrm{cut} d\Gamma_{\gamma} = \O( e^2 p^2 ) .
	\end{split}
\end{align}
The NA48/2 analysis assumes the following isospin breaking effects:
\begin{align}
	\begin{split}
		d\Gamma^\mathrm{exp}_{(\gamma,\mathrm{cut})} &= d\Gamma^\mathrm{exp} + \Delta d\Gamma_\mathrm{Coulomb} + \Delta d\Gamma_\mathrm{PHOTOS}^\mathrm{cut} .
	\end{split}
\end{align}
If I assume
\begin{align}
	\begin{split}
		\Delta d\Gamma_\mathrm{Coulomb} + \Delta d\Gamma_\mathrm{PHOTOS}^\mathrm{cut} \approx \Delta d\Gamma^\mathrm{NLO}_{\mathrm{virt.}\gamma} + \int_\mathrm{cut} d\Gamma_{\gamma} + \O(e^2 p^4) ,
	\end{split}
\end{align}
(an approximation that I will test later), the form factors given by the experiment contain only the isospin breaking mass effects (note that $X^\mathrm{LO} = \O(p)$):
\begin{align}
	\begin{split}
		X^\mathrm{exp} &= X^\mathrm{NLO}_\mathrm{ME} + \O(p^5, \epsilon \, p^5, Z e^2 p^3).
	\end{split}
\end{align}
The relative isospin corrections to the form factors due to the mass effects are
\begin{align}
	\begin{split}
		\delta_\mathrm{ME} X &:=  1 - \frac{X_\mathrm{iso}}{X_\mathrm{ME}} =  1 - \frac{X_\mathrm{iso}^\mathrm{NLO}}{X_\mathrm{ME}^\mathrm{NLO}}  + \O(\epsilon \, p^4, Z e^2 p^2) .
	\end{split}
\end{align}
The uncertainty can be naïvely estimated to be $\O(\epsilon \, p^4 , Z e^2 p^2)  \approx 0.2\%$. The mass effects at NNLO in the chiral expansion have been studied in a dispersive treatment \cite{Bernard2013} and found to be small given the present experimental accuracy.

The definition of the isospin limit is a convention. I choose here
\begin{align}
	\begin{split}
		X^\mathrm{NLO}_\mathrm{iso} := \lim_{\substack{\epsilon\to0, \\ e^2\to0}} \; \lim_{\substack{\mpio\to\mpip^\mathrm{exp}, \\ \mko\to\mkp^\mathrm{exp}}} X^\mathrm{NLO}_\mathrm{ME} .
	\end{split}
\end{align}

\subsection{Photonic Effects}

In this section, I calculate the (semi-)inclusive decay rate for $K_{\ell4(\gamma)}$. This will allow on the one hand for a more precise treatment of photonic corrections in future experiments (compared to previous treatments that do not make use of the matrix elements). On the other hand, I will be able to study the approximation
\begin{align}
	\begin{split}
		\Delta d\Gamma_\mathrm{Coulomb} + \Delta d\Gamma_\mathrm{PHOTOS}^\mathrm{cut} \approx \Delta d\Gamma^\mathrm{NLO}_{\mathrm{virt.}\gamma} + \int_\mathrm{cut} d\Gamma_{\gamma} + \O(e^2 p^4) ,
	\end{split}
\end{align}
although not for the experimental cuts, but for a simplified cut that can be handled analytically.

\subsubsection{Strategy for the Phase Space Integration}

I have introduced a finite photon mass as a regulator and will eventually send this regulator to zero (in the inclusive decay rate). We are not interested in the full dependence of the decay rate on the photon mass, but only in terms that do not vanish in the limit $m_\gamma\to0$, i.e.~in the IR-singular and finite pieces.

I use this fact to simplify the calculation as follows. I split the phase space of the radiative decay into a soft photon and a hard photon region. In the soft region, I use the soft photon approximation (SPA) to simplify the amplitude. This region will produce the IR singularity, which cancels against the divergence in the virtual corrections. The hard region gives an IR-finite result. Here, the limit $m_\gamma\to0$ can be taken immediately. The dependence on the photon energy cut $\Delta\varepsilon$ that separates the soft from the hard region must cancel in the sum of the two contributions. The hard region either covers the whole hard photon phase space, or alternatively, an additional cut on the maximum photon energy in the dilepton-photon system can be introduced rather easily.

\subsubsection{Soft Region}

Let us calculate the soft photon amplitude. In the real emission amplitude
\begin{align}
	\begin{split}
		\mathcal{T}_\gamma &= - \frac{G_F}{\sqrt{2}} e V_{us}^* \epsilon_\mu(q)^* \bigg[ \mathcal{H}^{\mu\nu} \; \mathcal{L}_\nu +  \mathcal{H}_\nu \; \mathcal{\tilde L}^{\mu\nu} \bigg]
	\end{split}
\end{align}
where
\begin{align}
	\begin{split}
		\mathcal{L}_\nu &:=  \bar u(p_\nu) \gamma_\nu (1-\gamma^5)v(p_\ell) , \\
		\mathcal{\tilde L}^{\mu\nu} &:= \frac{1}{2 p_\ell q} \bar u(p_\nu) \gamma^\nu (1-\gamma^5)(m_\ell - \slashed p_\ell - \slashed q) \gamma^\mu v(p_\ell) , \\
		\mathcal{H}_\nu &:= \frac{i}{\mkp} \left( P_\nu F + Q_\nu G + L_\nu R \right) , \\
		\mathcal{H}^{\mu\nu} &:= \frac{i}{\mkp} g^{\mu\nu} \Pi + \frac{i}{\mkp^2}\left( P^\mu \Pi_0^\nu + Q^\mu \Pi_1^\nu + L^\mu \Pi_2^\nu \right) , \\
		\Pi_i^\nu &:= \frac{1}{\mkp} \left( P^\nu \Pi_{i0} + Q^\nu \Pi_{i1} + L^\nu \Pi_{i2} + q^\nu \Pi_{i3}  \right) ,
	\end{split}
\end{align}
I neglect according to the SPA terms with a $q$ in the numerator, i.e.~the $\slashed q$ in $\mathcal{\tilde L}^{\mu\nu}$ and the $q^\nu$ in $\Pi_i^\nu$. If I insert the tree-level expressions for the form factors and consistently keep only terms that diverge as $q^{-1}$, I find that the soft photon amplitude factorises as
\begin{align}
	\label{eqn:SoftPhotonFactorisation}
	\begin{split}
		\mathcal{T}_\gamma^\mathrm{soft} &= e \mathcal{T}^\mathrm{LO}_\mathrm{iso} \left( - \frac{p \epsilon^*(q)}{p q} + \frac{p_\ell \epsilon^*(q)}{p_\ell q} + \frac{p_1 \epsilon^*(q)}{p_1 q} - \frac{p_2 \epsilon^*(q)}{p_2 q} \right) ,
	\end{split}
\end{align}
where $\mathcal{T}^\mathrm{LO}_\mathrm{iso}$ is the tree-level $K_{\ell4}$ matrix element in the isospin limit.

In the SPA, also the photon momentum in the delta function of the phase space measure is neglected. This means that we can essentially use $K_{\ell4}$ kinematics to describe the other momenta:
\begin{align}
	\label{eqn:DecayRateSoftRegion}
	\begin{split}
		d\Gamma_\gamma^\mathrm{soft} &= \frac{1}{2\mkp} \widetilde{dp_1} \widetilde{dp_2} \widetilde{dp_\ell} \widetilde{dp_\nu} \widetilde{dq_{\;}} \delta^{(4)}(p - p_1 - p_2 - p_\ell - p_\nu) \sum_{\substack{\mathrm{spins},\\ \mathrm{polar.}}} \left| \mathcal{T}_\gamma^\mathrm{soft} \right|^2 \\
			&= e^2 d\Gamma_\mathrm{iso}^\mathrm{LO} \int\limits_{|\vec q| \le \Delta \varepsilon} \widetilde{dq_{\;}} \sum_\mathrm{polar.} \left| - \frac{p \epsilon^*(q)}{p q} + \frac{p_\ell \epsilon^*(q)}{p_\ell q} + \frac{p_1 \epsilon^*(q)}{p_1 q} - \frac{p_2 \epsilon^*(q)}{p_2 q} \right|^2 \\
			&= - e^2 d\Gamma_\mathrm{iso}^\mathrm{LO} \int\limits_{|\vec q| \le \Delta \varepsilon} \widetilde{dq_{\;}} \begin{aligned}[t]
				&\bigg[ \frac{\mkp^2}{(pq)^2} + \frac{\ml^2}{(p_\ell q)^2} + \frac{\mpip^2}{(p_1 q)^2} + \frac{\mpip^2}{(p_2 q)^2} \\
				 & - \frac{2 p p_\ell}{(pq)(p_\ell q)} - \frac{2 p p_1}{(pq)(p_1 q)} + \frac{2 p p_2}{(pq)(p_2 q)} \\
				 & + \frac{2 p_1 p_\ell}{(p_1 q)(p_\ell q)} - \frac{2 p_2 p_\ell}{(p_2 q)(p_\ell q)} - \frac{2 p_1 p_2}{(p_1 q)(p_2 q)} \bigg] , \end{aligned}
	\end{split}
\end{align}
where I use the abbreviation
\begin{align}
	\begin{split}
		\widetilde{dk} := \frac{d^3 k}{(2\pi)^3 2k^0} .
	\end{split}
\end{align}
These are standard bremsstrahlung integrals, which have been computed in \cite{Hooft1979} (see also \cite{Itzykson1980}). They amount to
\begin{align}
	\begin{split}
		I_1(k) :={}& \int\limits_{|\vec q| \le \Delta\varepsilon} \widetilde{dq_{\;}} \frac{1}{(k q)^2} = \frac{1}{8\pi^2} \frac{1}{k^2} \Bigg[ 2 \ln\left( \frac{2 \Delta\varepsilon}{\mg} \right) - \frac{k^0}{|\vec k|} \ln\left( \frac{k^0 + |\vec k|}{k^0 - |\vec k|} \right) \Bigg] + \O(\mg^2) .
	\end{split}
\end{align}
The integrals with two different momenta are more complicated:
\begin{align}
	\begin{split}
		I_2(k_1,k_2) :={}& \int\limits_{|\vec q| \le \Delta\varepsilon} \widetilde{dq_{\;}} \frac{1}{(k_1 q)(k_2 q)} = \frac{\alpha}{8\pi^2} \left[ \frac{2}{k_1^2 - k^2} \ln\left( \frac{k_1^2}{k^2} \right) \ln\left( \frac{2\Delta\varepsilon}{\mg} \right) + \tilde I_2(k_1,k_2) \right] + \O(\mg^2) , \\
		\tilde I_2(k_1,k_2) ={}& \frac{1}{k_1^0 - k^0} \frac{1}{v} \begin{aligned}[t]
			& \Bigg[ \frac{1}{4} \ln^2\left( \frac{u^0 - |\vec u|}{u^0 + |\vec u|} \right) + \dilog\left( \frac{v-u^0 + |\vec u|}{v} \right) + \dilog\left( \frac{v-u^0-|\vec u|}{v} \right) \Bigg] \Bigg|_{u=k}^{u=k_1} , \end{aligned}
	\end{split}
\end{align}
where $k = \alpha k_2$ and $\alpha$ is the solution of $(k_1 - \alpha k_2)^2=0$ such that $\alpha k_2^0 - k_1^0$ has the same sign as $k_1^0$. Further, $v$ is defined as
\begin{align}
	\begin{split}
		v := \frac{k_1^2 - k^2}{2(k_1^0 - k^0)} .
	\end{split}
\end{align}

I find it most convenient to evaluate the soft photon contribution in the rest frame of the two leptons and the photon, $\Sigma_{\ell\nu\gamma}$. The particle momenta in this frame are given by
\begin{align}
	\begin{split}
		\begin{aligned}[t]
			p^0 &= \frac{\mkp^2 - s + s_\ell}{2\sqrt{s_\ell}} , \quad & |\vec p| &= \frac{\lambda_{K\ell}^{1/2}(s)}{2 \sqrt{s_\ell}} , \\
			p_\ell^0 &= \frac{\sqrt{s_\ell}}{2}(1+\zl) , \quad & |\vec p_\ell| &= \frac{\sqrt{s_\ell}}{2}(1-\zl) , \\
			p_1^0 &= \frac{PL + \sigma_\pi X \cos\theta_\pi}{2\sqrt{s_\ell}} , \quad & |\vec p_1| &= \sqrt{(p_1^0)^2 - \mpip^2} , \\
			p_2^0 &= \frac{PL - \sigma_\pi X \cos\theta_\pi}{2\sqrt{s_\ell}} , \quad & |\vec p_2| &= \sqrt{(p_2^0)^2 - \mpip^2} .
		\end{aligned}
	\end{split}
\end{align}

The bremsstrahlung integrals become
\begin{align}
	\begin{split}
		I_1(p) &= \frac{1}{8\pi^2} \frac{1}{\mkp^2} \Bigg[ 2 \ln\left( \frac{2 \Delta\varepsilon}{\mg} \right) - \frac{\mkp^2-s+s_\ell}{\lambda_{K\ell}^{1/2}(s)} \ln\left( \frac{\mkp^2 - s + s_\ell + \lambda_{K\ell}^{1/2}(s)}{\mkp^2 - s + s_\ell - \lambda_{K\ell}^{1/2}(s)} \right) \Bigg] , \\
		I_1(p_\ell) &= \frac{1}{8\pi^2} \frac{1}{\ml^2} \Bigg[ 2 \ln\left( \frac{2 \Delta\varepsilon}{\mg} \right) + \frac{1+\zl}{1-\zl} \ln(\zl) \Bigg] , \\
		I_1(p_1) &= \frac{1}{8\pi^2} \frac{1}{\mpip^2} \Bigg[ 2 \ln\left( \frac{2 \Delta\varepsilon}{\mg} \right) - \frac{p_1^0}{|\vec p_1|} \ln\left( \frac{p_1^0 + |\vec p_1|}{p_1^0 - |\vec p_1|} \right) \Bigg], \\
		I_1(p_2) &= \frac{1}{8\pi^2} \frac{1}{\mpip^2} \Bigg[ 2 \ln\left( \frac{2 \Delta\varepsilon}{\mg} \right) - \frac{p_2^0}{|\vec p_2|} \ln\left( \frac{p_2^0 + |\vec p_2|}{p_2^0 - |\vec p_2|} \right) \Bigg] .
	\end{split}
\end{align}

The evaluation of the integrals with two momenta is straightforward but a bit tedious. I give here the respective values of $\alpha(k_1,k_2)$:
\begin{align}
	\begin{alignedat}{2}
		\alpha(p, p_\ell) &= \frac{\lambda^{1/2}(\tl,\mkp^2,\ml^2) + \mkp^2 + \ml^2 - \tl}{2 \ml^2}, \quad & \alpha(p, p_1) &= \frac{\lambda_{K\pi}^{1/2}(t) + \mkp^2 + \mpip^2 - t}{2\mpip^2} , \\
		\alpha(p_1, p_\ell) &= \frac{\lambda^{1/2}(s_{1\ell},\mpip^2, \ml^2) -\ml^2 - \mpip^2 + s_{1\ell}}{2\ml^2} , \quad & \alpha(p, p_2) &= \frac{\lambda_{K\pi}^{1/2}(u) + \mkp^2 + \mpip^2 - u}{2\mpip^2} , \\
		\alpha(p_2, p_\ell) &= \frac{\lambda^{1/2}(s_{2\ell},\mpip^2, \ml^2) -\ml^2 - \mpip^2 + s_{2\ell}}{2\ml^2} , \quad & \alpha(p_1, p_2) &= \frac{s \sigma_\pi + s - 2 \mpip^2}{2\mpip^2} .
	\end{alignedat}
\end{align}

\subsubsection{Hard Region}

The hard region is defined as the phase space region where $|\vec q| > \Delta \varepsilon$, i.e.
\begin{align}
	\begin{split}
		x > x_\mathrm{min} = \frac{2\Delta\varepsilon}{\sqrt{s_\ell}}  =: \tilde x_\mathrm{min} (1-\zl) ,
	\end{split}
\end{align}
where the variable $\tilde x_\mathrm{min}$ is independent of $s_\ell$.

Here, the full $K_{\ell4\gamma}$ kinematics has to be applied. However, as the hard region does not produce any IR singularity, the limit $\mg\to0$ can be taken at the very beginning.

In the appendix~\ref{sec:RadiativeDecayRate}, I have derived the expression for the decay rate
\begin{align}
	\begin{split}
		d\Gamma_\gamma^\mathrm{hard} &= G_F^2 |V_{us}|^2 e^2 \frac{s_\ell \, \sigma_\pi(s) X}{2^{20}\pi^9 \mkp^7} J_8 \, ds \, ds_\ell \, d\cos\theta_\pi \, d\cos\theta_\gamma \, d\phi \, dx \, dy \, d\phi_\ell ,
	\end{split}
\end{align}
where
\begin{align}
	\begin{split}
		J_8 &= \mkp^4 \sum_\mathrm{polar.} \epsilon_\mu(q)^* \epsilon_\rho(q)
			\begin{aligned}[t]
				&\Bigg[ \mathcal{H}_\nu \mathcal{H}^*_\sigma \sum_\mathrm{spins} \mathcal{\tilde L}^{\mu\nu} \mathcal{\tilde L}^{*\rho\sigma} + \mathcal{H}^{\mu\nu} \mathcal{H}^{*\rho\sigma} \sum_\mathrm{spins} \mathcal{L}_\nu \mathcal{L}^*_\sigma + 2 \Re\bigg( \mathcal{H}^{\mu\nu} \mathcal{H}^{*\sigma} \sum_\mathrm{spins} \mathcal{L}_\nu \mathcal{\tilde L}^{*\rho}{}_{\sigma} \bigg) \Bigg] .
			\end{aligned}
	\end{split}
\end{align}
Since the form factors only depend on the first six phase space variables, the integrals over $y$ and $\phi_\ell$ can be performed without knowledge of the dynamics. The $K_{\ell4}$ form factors and the form factor $\Pi$ depend on $s$, $s_\ell$ and $\cos\theta_\pi$ only (at the order we consider). I therefore split the hadronic tensor into two pieces
\begin{align}
	\begin{split}
		\mathcal{H}^{\mu\nu} = \frac{i}{\mkp} g^{\mu\nu} \Pi + \frac{i}{\mkp^2} \mathcal{\tilde H}^{\mu\nu} , \quad \mathcal{\tilde H}^{\mu\nu} = P^\mu \Pi_0^\nu + Q^\mu \Pi_1^\nu + L^\mu \Pi_2^\nu
	\end{split}
\end{align}
and write $J_8$ as follows:
\begin{align}
	\begin{split}
		J_8 &= J_8^{\ell\ell} + J_8^{hh} + J_8^\mathrm{int} , \\
		J_8^{\ell\ell} &= \mkp^4 \sum_\mathrm{polar.} \epsilon_\mu(q)^* \epsilon_\rho(q) 
			\begin{aligned}[t]
				& \Bigg[ \mathcal{H}_\nu \mathcal{H}^*_\sigma \sum_\mathrm{spins} \mathcal{\tilde L}^{\mu\nu} \mathcal{\tilde L}^{*\rho\sigma} + \frac{1}{\mkp^2} g^{\mu\nu} g^{\rho\sigma} | \Pi |^2 \sum_\mathrm{spins} \mathcal{L}_\nu \mathcal{L}^*_\sigma \\
				&+ \frac{i}{\mkp} \bigg( g^{\mu\nu} \Pi \, \mathcal{H}^{*\sigma} \sum_\mathrm{spins} \mathcal{L}_\nu \mathcal{\tilde L}^{*\rho}{}_{\sigma} - g^{\mu\nu} \Pi^* \, \mathcal{H}^{\sigma} \sum_\mathrm{spins} \mathcal{L}_\nu^* \mathcal{\tilde L}^{\rho}{}_{\sigma} \bigg) \Bigg] ,
			\end{aligned} \\
		J_8^{hh} &= \sum_\mathrm{polar.} \epsilon_\mu(q)^* \epsilon_\rho(q) \Bigg[ \mathcal{\tilde H}^{\mu\nu} \mathcal{\tilde H}^{*\rho\sigma} \sum_\mathrm{spins} \mathcal{L}_\nu \mathcal{L}^*_\sigma \Bigg] , \\
		J_8^\mathrm{int} &= \mkp^2 \sum_\mathrm{polar.} \epsilon_\mu(q)^* \epsilon_\rho(q)
			\begin{aligned}[t]
				&\Bigg[ \frac{1}{\mkp} \left( g^{\mu\nu} \mathcal{\tilde H}^{*\rho\sigma} \Pi + \mathcal{\tilde H}^{\mu\nu} g^{\rho\sigma} \Pi^*\right) \sum_\mathrm{spins} \mathcal{L}_\nu \mathcal{L}^*_\sigma \\
				&+ i \bigg( \mathcal{\tilde H}^{\mu\nu} \mathcal{H}^{*\sigma} \sum_\mathrm{spins} \mathcal{L}_\nu \mathcal{\tilde L}^{*\rho}{}_{\sigma} - \mathcal{\tilde H}^{*\mu\nu} \mathcal{H}^{\sigma} \sum_\mathrm{spins} \mathcal{L}_\nu^* \mathcal{\tilde L}^{\rho}{}_{\sigma} \bigg) \Bigg] .
			\end{aligned}
	\end{split}
\end{align}
The first term, $J_8^{\ell\ell}$, denotes the absolute square of the contributions where the photon is attached to the lepton line (either the external line or the vertex). Here, the hadronic part is described by the $K_{\ell4}$ form factors and $\Pi$. I can therefore integrate directly over the five phase space variables $\cos\theta_\gamma$, $\phi$, $x$, $y$ and $\phi_\ell$.

The second term, $J_8^{hh}$, is the absolute square of the contributions with the photon emitted off the hadrons. The form factors $\Pi_{ij}$ describe here the hadronic part. As they depend on six phase space variables, I perform first the integral over $\phi_\ell$ and $y$, then insert the explicit tree-level expressions for the form factors $\Pi_{ij}$, given in section~\ref{sec:MatrixElementRealPhotonEmission}. I further integrate the decay rate and keep it differential only with respect to $s$, $s_\ell$ and $\cos\theta_\pi$. The same strategy applies to the third term, $J_8^\mathrm{int}$, the interference of off-lepton and off-hadron emission.

It is important to note that for a vanishing lepton mass $\ml$, the phase space integrals containing $\mathcal{H}^\mu$ produce a singularity for collinear photons. The lepton mass plays the role of a natural cut-off for this collinear divergence, which emerges as a $\ln\ml^2$ mass singularity. In those integrals, the limit $\ml\to0$ must not be taken before the integration.

Let us now consider the three parts separately.

I perform the five phase space integrals in the $\ell\ell$-part and apply an expansion for small values of $\tilde x_\mathrm{min}$, keeping only the logarithmic term. Only after the integration, it is safe to expand the result for small values of $\ml$:
\begin{align}
	\begin{split}
		\frac{d\Gamma_\gamma^{\mathrm{hard},\ell\ell}}{ds ds_\ell d\cos\theta_\pi} &= e^2 G_F^2 |V_{us}|^2 \frac{\sigma_\pi(s) X}{9 \cdot 2^{15} \, \pi^7 \mkp^5} \begin{aligned}[t]
			& \bigg( 2 \left( |F_1|^2 + \sin^2\theta_\pi |F_2|^2 \right) \left( 12 \ln \tilde x_\mathrm{min} - 3 \ln\zl + 5 \right) \\
			& + 3 \left| F_4 + s_\ell \Pi \right|^2 \bigg) + \O(\zl \ln \zl) . \end{aligned}
	\end{split}
\end{align}
The soft photon contribution corresponding to the square of the off-lepton emission amplitude is given by $I_1(p_\ell)$. In the sum of the soft and the hard photon emission, the dependence on $\Delta\varepsilon$ drops out:
\begin{align}
	\label{eqn:RadiativeDecayRateLL}
	\begin{split}
		\frac{d\Gamma_\gamma^{\ell\ell}}{ds ds_\ell d\cos\theta_\pi} &= e^2 G_F^2 |V_{us}|^2 \frac{\sigma_\pi(s) X}{9 \cdot 2^{15} \, \pi^7 \mkp^5} \begin{aligned}[t]
			& \bigg( 2 \left( |F_1|^2 + \sin^2\theta_\pi |F_2|^2 \right) \left( 5 + 6 \ln \zg  - 9 \ln\zl \right) \\
			& + 3 \left|F_4+s_\ell \Pi \right|^2  \bigg) + \O(\zl \ln \zl) . \end{aligned}
	\end{split}
\end{align}
I can introduce an additional cut on the photon energy in $\Sigma_{\ell\nu\gamma}$ by integrating $x$ only over a part of the hard region:
\begin{align}
	\begin{split}
		\tilde x_\mathrm{min} (1-\zl) < x < \tilde x_\mathrm{max}(1 - \zl) .
	\end{split}
\end{align}
Instead of (\ref{eqn:RadiativeDecayRateLL}), I find then
\begin{align}
	\label{eqn:RadiativeDecayRateLLCut}
	\begin{split}
		\frac{d\Gamma_{\gamma,\mathrm{cut}}^{\ell\ell}}{ds ds_\ell d\cos\theta_\pi} &= e^2 G_F^2 |V_{us}|^2 \frac{\sigma_\pi(s) X}{9 \cdot 2^{15} \, \pi^7 \mkp^5} \begin{aligned}[t]
			& \bigg( 2  (|F_1|^2 + \sin^2\theta_\pi |F_2|^2) \\
			& \cdot \begin{aligned}[t] & \Big( \tilde x_\mathrm{max} (9 - \tilde x_\mathrm{max}(3 + \tilde x_\mathrm{max}))  + 6 \ln\zg  - 3 (2 + \tilde x_\mathrm{max}^2) \ln\zl \\
				& - 3 (1 - \tilde x_\mathrm{max}^2) \ln(1 - \tilde x_\mathrm{max}) - 12 \ln(\tilde x_\mathrm{max}) \Big) \end{aligned} \\
			& + 3 \tilde x_\mathrm{max}^2 (3 - 2 \tilde x_\mathrm{max}) |F_4 + s_\ell \Pi|^2
  \bigg) + \O(\zl \ln \zl) . \end{aligned}
	\end{split}
\end{align}

The integration of the $hh$-part is more involved. I perform the integrals over $\phi_\ell$ and $y$ analytically, insert the explicit form factors $\Pi_{ij}$ and integrate over $x$ analytically, too (either with or without the energy cut $\tilde x_\mathrm{max}$). Although, with some effort, the integrals over $\phi$ and $\cos\theta_\gamma$ could be performed analytically, I choose to integrate these two angles numerically: since they only describe the orientation of the dilepton-photon three-body system with respect to the pions, these two integrals contain nothing delicate. The dependence on the cuts $\tilde x_\mathrm{min}$ and $\tilde x_\mathrm{max}$ is manifest after the integration over $x$ and collinear singularities cannot show up in the remaining integrals. I therefore write the $hh$-part as
\begin{align}
	\begin{split}
		\frac{d\Gamma_{\gamma,\mathrm{cut}}^{\mathrm{hard},hh}}{ds ds_\ell d\cos\theta_\pi} &= e^2 G_F^2 |V_{us}|^2 \frac{s_\ell \sigma_\pi(s) X}{2^{20} \pi^9 \mkp^7}  \begin{aligned}[t]
				& \Bigg( \ln\left(\frac{\tilde x_\mathrm{min}}{\tilde x_\mathrm{max}}\right) \int_{-1}^1  d\cos\theta_\gamma \int_0^{2\pi} d\phi \, j_1^{hh}(s,s_\ell,\cos\theta_\pi,\cos\theta_\gamma,\phi) \\
				& + \int_{-1}^1  d\cos\theta_\gamma \int_0^{2\pi} d\phi \, j_{2,\mathrm{cut}}^{hh}(s,s_\ell,\cos\theta_\pi,\cos\theta_\gamma,\phi) \Bigg) .
			\end{aligned}
	\end{split}
\end{align}
The function $j_1^{hh}$ is given by
\begin{align}
	\begin{split}
		j_1^{hh}(s,s_\ell,\cos\theta_\pi,\cos\theta_\gamma,\phi) &= \frac{32\pi \mkp^4}{3 F_0^2} \left( (PL + X \sigma_\pi \cos\theta_\pi)^2 - 4 s_\ell \mpip^2 \right) \\
			& \quad \cdot \begin{aligned}[t]
				&\Bigg( \frac{s}{A_1^2} + \frac{s}{A_2^2} + \frac{2 PL + s + s_\ell}{(PL + s_\ell + \cos\theta_\gamma X)^2} + \frac{2 (PL + s)}{A_1 (PL + s_\ell + \cos\theta_\gamma X)} \\
				& - \frac{2 (PL + s)}{A_2 (PL + s_\ell + \cos\theta_\gamma X)}  - \frac{2s + 4 \cos\theta_\pi X \sigma_\pi}{A_1 A_2} \\
				& + \frac{ 4 \cos\theta_\pi X s_\ell \sigma_\pi}{A_1 A_2 (PL + s_\ell + \cos\theta_\gamma X)}  - \frac{4 s  \sigma_\pi^2 (PL + \cos\theta_\gamma X)^2}{A_1^2 A_2^2} \Bigg) , \end{aligned}
	\end{split}
\end{align}
where the $\phi$-dependence is hidden in
\begin{align}
	\begin{split}
		A_1 &= PL + \cos\theta_\gamma X - \cos\theta_\gamma \cos\theta_\pi PL \sigma_\pi - \cos\theta_\pi X \sigma_\pi \\
			& \quad + \cos\phi \, \sigma_\pi \sqrt{(1 - \cos\theta_\gamma^2) (1 - \cos\theta_\pi^2) s s_\ell} , \\
		A_2 &= PL + \cos\theta_\gamma X + \cos\theta_\gamma \cos\theta_\pi PL \sigma_\pi + \cos\theta_\pi X \sigma_\pi \\
			& \quad - \cos\phi \, \sigma_\pi \sqrt{(1 - \cos\theta_\gamma^2) (1 - \cos\theta_\pi^2) s s_\ell} .
	\end{split}
\end{align}

The integrand $j_{2,\mathrm{cut}}^{hh}$ of the second numerical integral is a lengthy expression that I do not state here explicitly.

The soft photon contribution to this second part contains the six bremsstrahlung integrals $I_1(p)$, $I_1(p_1)$, $I_1(p_2)$, $I_2(p,p_1)$, $I_2(p,p_2)$ and $I_2(p_1,p_2)$. It is easy to verify numerically that in the sum of the contributions from soft and hard region, the dependence on $\Delta\varepsilon$ (i.e.~on $\tilde x_\mathrm{min}$) again drops out. The analytic result of the integral over $j_1^{hh}$ can therefore be inferred from the soft photon $hh$-part (note that these bremsstrahlung integrals do not depend on $\phi$ or $\cos\theta_\ell$).

The interference term of off-lepton and off-hadron photon emission is the last and most intricate part of the phase space integral calculation. On the one hand, the explicit form factors $\Pi_{ij}$ have to be inserted after the $\phi_\ell$- and $y$-integration. On the other hand, while the part of the interference term containing $\Pi$ is free of collinear singularities and independent of $\tilde x_\mathrm{min}$, the contrary is true for the part involving the $K_{\ell4}$ form factors. I again integrate over $\phi_\ell$, $y$ and $x$ analytically, expand the result for small $\ml^2$ and obtain the structure
\begin{align}
	\label{eqn:HardPhotonInterferenceTerms}
	\begin{split}
		\frac{d\Gamma_{\gamma,\mathrm{cut}}^{\mathrm{hard,int}}}{ds ds_\ell d\cos\theta_\pi} &= e^2 G_F^2 |V_{us}|^2 \frac{s_\ell \sigma_\pi(s) X}{2^{20} \pi^9 \mkp^7} \\
			& \quad \cdot \begin{aligned}[t]
				& \Bigg( \ln \zl \left(\tilde x_\mathrm{max} + \ln\left(\frac{ \tilde x_\mathrm{min}}{ \tilde x_\mathrm{max}}\right) \right) \int_{-1}^1 d\cos\theta_\gamma \int_0^{2\pi} d\phi \, j_1^\mathrm{int}(s,s_\ell,\cos\theta_\pi,\cos\theta_\gamma,\phi) \\
				& + \ln \left( \frac{\tilde x_\mathrm{min}}{\tilde x_\mathrm{max}}\right) \int_{-1}^1 d\cos\theta_\gamma \int_0^{2\pi} d\phi \, j_2^\mathrm{int}(s,s_\ell,\cos\theta_\pi,\cos\theta_\gamma,\phi) \\
				& + \int_{-1}^1  d\cos\theta_\gamma \int_0^{2\pi} d\phi \, j_{3,\mathrm{cut}}^\mathrm{int}(s,s_\ell,\cos\theta_\pi,\cos\theta_\gamma,\phi) \Bigg) .
			\end{aligned}
	\end{split}
\end{align}
I perform the integrals over $\phi$ and $\cos\theta_\gamma$ numerically. The expressions for the integrands $j_i^\mathrm{int}$ are too lengthy to be given explicitly. $j_{3,\mathrm{cut}}^\mathrm{int}$ depends on the cut $\tilde x_\mathrm{max}$.

Again, the sum of the soft and hard photon contribution must not depend on $\Delta\varepsilon$. I expand the soft contribution, given by the remaining bremsstrahlung integrals $I_2(p,p_\ell)$, $I_2(p_1,p_\ell)$ and $I_2(p_2,p_\ell)$, in $\ml$ and neglect terms that vanish for $\ml\to0$:
\begin{align}
	\begin{split}
		\frac{d\Gamma_\gamma^\mathrm{soft,int}}{ds ds_\ell d\cos\theta_\pi} &= - e^2  \int_{-1}^1  d\cos\theta_\ell \int_0^{2\pi} d\phi \; d\Gamma_\mathrm{iso}^\mathrm{LO} \frac{1}{16\pi^2} \begin{aligned}[t]
				&\Bigg( \ln\left( \frac{2\Delta\varepsilon}{\mg} \right) \begin{aligned}[t]
					& \bigg[ 4 \ln \zl + b_1^\mathrm{int}(s,s_\ell,\cos\theta_\pi,\cos\theta_\ell,\phi) \bigg] \end{aligned} \\
				& + \ln^2 \zl + b_2^\mathrm{int}(s,s_\ell,\cos\theta_\pi,\cos\theta_\ell,\phi) \Bigg) , \end{aligned} 
	\end{split}
\end{align}
where the $b_i^\mathrm{int}$ are again rather lengthy expressions.

I perform the integrals over $\cos\theta_\ell$ and $\phi$ numerically and find that the dependence on $\Delta\varepsilon$ drops out indeed in the sum of soft and hard photon contribution.

\subsubsection{Cancellation of Divergences}

Both the virtual corrections and the real emission contain infrared divergences. These divergences, which are regulated by the artificial photon mass $m_\gamma$, must vanish in the inclusive decay rate. In the radiative process, the IR divergence is generated in the soft region, which I have treated in the soft photon approximation.

Furthermore, collinear (or mass) divergences arise in the virtual corrections and in the soft and hard region of the radiative process. They are regulated by the lepton mass $m_\ell$ that acts as a natural cutoff. According to the KLN theorem \cite{KinoshitaSirlin1959,Kinoshita1962,Lee1964}, there must not be any divergences in the fully inclusive decay rate. Since the limit $\ml\to0$ is usually taken in experimental analyses, I apply the same approximation to the inclusive decay rate. Here, however, it is crucial that the collinear divergences indeed cancel.

Note that I use everywhere the physical lepton mass, which can be identified (up to higher order effects) with the renormalised mass. A necessary condition for the KLN theorem to hold in this representation is that the mass renormalisation does not diverge in the limit $\ml\to0$. This condition is fulfilled by (\ref{eqn:LeptonMassRenormalisation}).

\paragraph{Infrared Singularities}

In the virtual corrections, the six triangle diagrams~\ref{img:Kl4_NLOgLoop5}-\ref{img:Kl4_NLOgLoop10} and the external leg corrections are IR-divergent. The relevant loop functions are given in appendix~\ref{sec:IRdivergentLoopFunctions}.

\begin{figure}[ht]
	\centering
	\scalebox{0.8}{
		\begin{pspicture}(0,0)(8,6)
			\put(4.5,5.35){(a)}
			\put(2.5,5.35){(b)}
			\put(1.75,4.75){(c)}
			\put(5.75,5.2){(d)}
			\includegraphics[width=8cm]{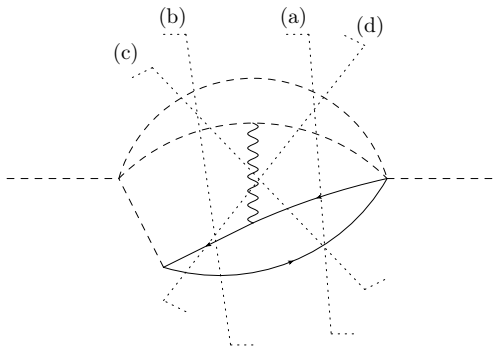}
		\end{pspicture}
		}
	\caption{Four-loop kaon self-energy diagram with four- or five-particle cuts.}
	\label{img:KaonSEFourLoop}
\end{figure}

\begin{figure}[ht]
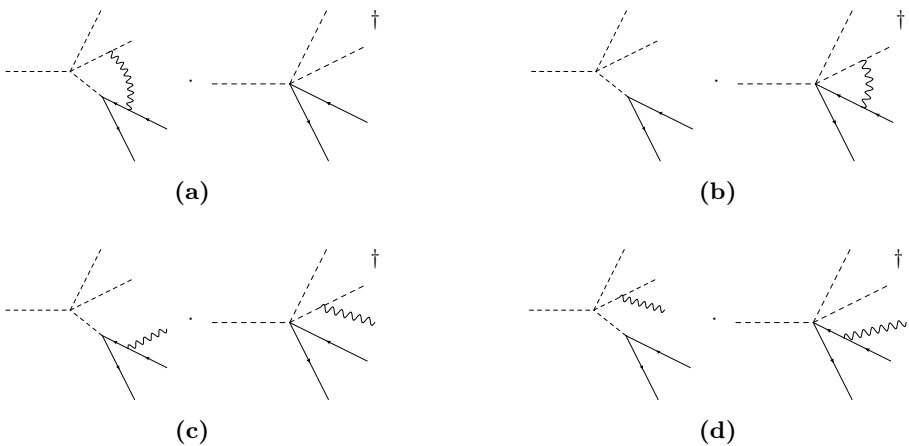

	\centering
	\begin{subfigure}[b]{0.4\textwidth}
		\centering
		\scalebox{0.8}{
			\begin{pspicture}(0,0)(6,3)
				\put(3,1.25){$\cdot$}
				\put(6,2.25){$\dagger$}
				\includegraphics[height=2.5cm]{images/NLO_gLoop18}
				\hspace{0.5cm}
				\includegraphics[height=2.5cm]{images/LO1}
			\end{pspicture}
			}
		\caption{}
		\label{img:PhaseSpaceProductA}
	\end{subfigure}
	\begin{subfigure}[b]{0.4\textwidth}
		\centering
		\scalebox{0.8}{
			\begin{pspicture}(0,0)(6,3)
				\put(3,1.25){$\cdot$}
				\put(6,2.25){$\dagger$}
				\includegraphics[height=2.5cm]{images/LO2}
				\hspace{0.5cm}
				\includegraphics[height=2.5cm]{images/NLO_gLoop10}
			\end{pspicture}
			}
		\caption{}
		\label{img:PhaseSpaceProductB}
	\end{subfigure}
	
	\vspace{0.5cm}

	\begin{subfigure}[b]{0.4\textwidth}
		\centering
		\scalebox{0.8}{
			\begin{pspicture}(0,0)(6,3)
				\put(3,1.25){$\cdot$}
				\put(6,2.25){$\dagger$}
				\includegraphics[height=2.5cm]{images/RE12}
				\hspace{0.5cm}
				\includegraphics[height=2.5cm]{images/RE3}
			\end{pspicture}
			}
		\caption{}
		\label{img:PhaseSpaceProductC}
	\end{subfigure}
	\begin{subfigure}[b]{0.4\textwidth}
		\centering
		\scalebox{0.8}{
			\begin{pspicture}(0,0)(6,3)
				\put(3,1.25){$\cdot$}
				\put(6,2.25){$\dagger$}
				\includegraphics[height=2.5cm]{images/RE8}
				\hspace{0.5cm}
				\includegraphics[height=2.5cm]{images/RE5}
			\end{pspicture}
			}
		\caption{}
		\label{img:PhaseSpaceProductD}
	\end{subfigure}
	\caption{Phase space products corresponding to the four cuts of the kaon self-energy diagram.}
	\label{img:PhaseSpaceProducts}
\end{figure}

A priori, one would expect that the box diagrams~\ref{img:Kl4_NLOgLoop16}-\ref{img:Kl4_NLOgLoop18} also give rise to an IR singularity, because the scalar four point loop function $D_0$ is IR-divergent as well. However, as can be shown with Passarino-Veltman reduction techniques  \cite{Hooft1979, Passarino1979} and the explicit expressions for the IR-divergent scalar box integral \cite{Beenakker1990}, the contribution of the box diagrams to the form factors $F$ and $G$ are IR-finite. This can be understood rather easily: consider the four-loop kaon self-energy diagram in figure~\ref{img:KaonSEFourLoop}. This diagram is an IR-finite quantity and so must be the sum of its four- and five-particle cuts. Each of the four cuts corresponds to a phase space integral of the product of two diagrams, shown in figure~\ref{img:PhaseSpaceProducts}. Now, as the IR divergence in the radiative process is generated in the soft region, where the matrix element factorises into the LO non-radiative process times the soft photon factor (\ref{eqn:SoftPhotonFactorisation}), the IR divergence has to drop out already in the differential inclusive decay rate, where the photon is integrated. The phase space products~\ref{img:PhaseSpaceProductB}-\ref{img:PhaseSpaceProductD} can only contribute to the term $R F^*$, $R G^*$ and $|R|^2$. Therefore, the phase space product~\ref{img:PhaseSpaceProductA} cannot give an IR-divergent contribution to $|F|^2$ or $|G|^2$. Hence, the box diagram on the left-hand side of the product can only give IR-divergent contributions to $R$. An analogous argument works for the two other box diagrams.

Let us now turn our attention to the IR divergences of the virtual corrections. Summing all the IR-divergent contributions (after UV renormalisation), I find
\begin{align}
	\begin{split}
		\delta F^\mathrm{NLO,IR}_{\mathrm{virt.}\gamma} &= \delta F^\mathrm{NLO,IR}_{\gamma-\mathrm{loop},e-j} + \delta F_{\gamma-Z}^\mathrm{NLO,IR} = \delta G^\mathrm{NLO,IR}_{\mathrm{virt.}\gamma} = \delta G^\mathrm{NLO,IR}_{\gamma-\mathrm{loop},e-j} + \delta G_{\gamma-Z}^\mathrm{NLO,IR} \\
		&= 2 e^2 \begin{aligned}[t]
			& \Bigg( (\mkp^2+\mpip^2-t) C_0^\mathrm{IR}(\mpip^2,t, \mkp^2,\mg^2,\mpip^2,\mkp^2) \\
			& - (\mkp^2+\mpip^2-u) C_0^\mathrm{IR}(\mpip^2,u,\mkp^2,\mg^2,\mpip^2,\mkp^2) \\
			& + (\mkp^2+\ml^2-\tl) C_0^\mathrm{IR}(\ml^2,\tl,\mkp^2,\mg^2,\ml^2,\mkp^2) \\
			& + (2\mpip^2-s) C_0^\mathrm{IR}(\mpip^2,s,\mpip^2,\mg^2,\mpip^2,\mpip^2) \\
			& - (\mpip^2 + \ml^2 - s_{1\ell} ) C_0^\mathrm{IR}(\ml^2, s_{1\ell}, \mpip^2, \mg^2, \ml^2, \mpip^2) \\
			& + (\mpip^2 + \ml^2 - s_{2\ell}) C_0^\mathrm{IR}(\ml^2, s_{2\ell}, \mpip^2, \mg^2, \ml^2, \mpip^2) \\
			& - \frac{1}{8\pi^2} \ln\zg \Bigg) =: \delta X^\mathrm{NLO,IR}_{\mathrm{virt.}\gamma} , \end{aligned} \\
	\end{split}
\end{align}
where
\begin{align}
	\begin{split}
		C_0^\mathrm{IR}(m^2, s, M^2, \mg^2, m^2, M^2) &= - \frac{1}{16\pi^2} \frac{x_s}{m M(1-x_s^2)} \ln x_s \ln \zg , \\
			x_s &=  -\frac{1 - \sqrt{1 - \frac{4 m M}{s - (m-M)^2}}}{1 + \sqrt{1 - \frac{4 m M}{s - (m-M)^2}}} .
	\end{split}
\end{align}

The infrared-divergent part of the NLO decay rate is given by
\begin{align}
	\begin{split}
		d\Gamma^\mathrm{NLO,IR} = d\Gamma^\mathrm{LO}_\mathrm{iso} \; 2 \Re(\delta X^\mathrm{NLO,IR}_{\mathrm{virt.}\gamma}) + \O(\zl \ln\zl) .
	\end{split}
\end{align}
By extracting the IR divergence (terms proportional to $\ln\zg$) out of the soft photon contribution to the radiative decay rate (\ref{eqn:DecayRateSoftRegion}), it is now easy to verify that the sum of virtual corrections and soft bremsstrahlung (where the photon is integrated) and hence the inclusive decay rate is free of infrared divergences:
\begin{align}
	\begin{split}
		d\Gamma^\mathrm{NLO,IR} + d\Gamma_\gamma^\mathrm{soft,IR} = 0 .
	\end{split}
\end{align}

\paragraph{Collinear Singularities}

Both the soft and the hard region of the radiative process give rise to collinear singularities, terms proportional to $\ln\zl$. Let us now check that these mass divergences cancel in the fully inclusive decay rate (the cut on the photon energy must be removed for this purpose, i.e.~I take the limit $\tilde x_\mathrm{max} \to 1$). Virtual photon corrections can produce a collinear divergence if one end of the photon line is attached to the lepton line. Since the mass divergence in the radiative process is produced in the collinear region of the phase space (soft and hard), where the matrix element could be factorised similarly to the soft region \cite{Bohm2001}, one can argue in an analogous way as for the IR divergences that the contribution of the box diagrams to the form factors $F$ and $G$ has no mass divergence. This is confirmed by the explicit expressions for the diagrams. The only collinear divergent contributions stem from the external leg correction for the lepton and the three diagrams~\ref{img:Kl4_NLOgLoop7}, \ref{img:Kl4_NLOgLoop9} and \ref{img:Kl4_NLOgLoop10}.

The external leg correction for the lepton contains the following collinear divergence:
\begin{align}
	\begin{split}
		\delta F_{\gamma-Z}^{\mathrm{NLO,coll}} &= \delta G_{\gamma-Z}^{\mathrm{NLO,coll}} = \frac{3 e^2}{32\pi^2} \ln\zl ,
	\end{split}
\end{align}
contributing to the decay rate as
\begin{align}
	\begin{split}
		d\Gamma^\mathrm{NLO,coll}_{Z} &= d\Gamma^{\mathrm{LO}}_\mathrm{iso} \; \frac{3e^2}{16\pi^2} \ln\zl .
	\end{split}
\end{align}
This cancels exactly the mass divergence in the $\ell\ell$-part of the real photon corrections (\ref{eqn:RadiativeDecayRateLL}).

Next, I collect the mass divergent terms contained in the three relevant loop diagrams:
\begin{align}
	\begin{split}
		\delta F^\mathrm{NLO,coll}_{\gamma-\mathrm{loop}} &= \delta G^\mathrm{NLO,coll}_{\gamma-\mathrm{loop}} = \frac{e^2}{16\pi^2} \ln\zl \left( \frac{1}{2} \ln \zl - \ln \zg - 2 \right) , 
	\end{split}
\end{align}
resulting in a collinear divergence in the decay rate of
\begin{align}
	\begin{split}
		d\Gamma^\mathrm{NLO,coll}_{\mathrm{loop}} &= d\Gamma^{\mathrm{LO}}_\mathrm{iso} \; \frac{e^2}{16\pi^2} \ln\zl \left( \ln \zl - 2 \ln \zg - 4 \right) .
	\end{split}
\end{align}
This singularity must cancel with the mass divergence in the interference term of the radiative decay rate. The divergent contribution from the soft photon region is given by
\begin{align}
	\begin{split}
		d\Gamma^\mathrm{soft,int}_{\gamma,\mathrm{coll}} &= - d\Gamma^\mathrm{LO}_\mathrm{iso} \frac{e^2}{16\pi^2} \ln\zl \left( \ln\zl + 4 \ln\left( \frac{2\Delta\varepsilon}{\mg} \right) \right) \\
			&=  - d\Gamma^\mathrm{LO}_\mathrm{iso} \frac{e^2}{16\pi^2} \ln\zl \left( \ln\zl - 2 \ln\zg + 4 \ln\left( \frac{2\Delta\varepsilon}{\sqrt{s_\ell}} \right) \right) .
	\end{split}
\end{align}
In the sum of virtual and soft real corrections, the double divergences (double collinear and soft-collinear) cancel:
\begin{align}
	\begin{split}
		d\Gamma^\mathrm{NLO,coll}_{\mathrm{loop}} + d\Gamma^\mathrm{soft,int}_{\gamma,\mathrm{coll}} &= - d\Gamma^\mathrm{LO}_\mathrm{iso} \frac{e^2}{4\pi^2} \ln\zl \left( 1 + \ln \tilde x_\mathrm{min} \right) .
	\end{split}
\end{align}
This single divergence must cancel against the one in the hard real corrections (\ref{eqn:HardPhotonInterferenceTerms}). By evaluating numerically the integral over $j_1^\mathrm{int}$, I have checked that this cancellation takes place.

I have now verified that the fully inclusive decay rate
\begin{align}
	\begin{split}
		\frac{d\Gamma_{(\gamma)}}{ds ds_\ell d\cos\theta_\pi} &= \frac{d\Gamma_{\mathrm{virt.}\gamma}^\mathrm{NLO}}{ds ds_\ell d\cos\theta_\pi} + \frac{d\Gamma_{\gamma}^\mathrm{soft}}{ds ds_\ell d\cos\theta_\pi} + \frac{d\Gamma_{\gamma}^\mathrm{hard}}{ds ds_\ell d\cos\theta_\pi}
	\end{split}
\end{align}
does not depend on the energy cut separating the soft from the hard region and contains neither infrared nor collinear (mass) singularities. The calculation is therefore in accordance with the KLN theorem. Note that this is a necessary but highly non-trivial consistency check, since the two regions of the radiative phase space are parametrised differently.


\section{Numerical Evaluation}

\label{sec:Numerics}

The existing high statistics experiments on $K_{\ell4}$, E865 \cite{Pislak2003, Pislak2010} and NA48/2 \cite{Batley2010, Batley2012}, have applied isospin corrections to a certain extent and with different approximations. In the NA48/2 experiment, the data was corrected by the semi-classical Gamow-Sommerfeld (or Coulomb) factor and with help of PHOTOS \cite{Barberio1994}. The E865 experiment used the same analytic prescription by Diamant-Berger \cite{DiamantBerger1976} as the older Geneva-Saclay experiment \cite{Rosselet1977}. Both treatments did not make use of the full matrix element and relied on factorisation of the tree-level amplitude as it happens in a soft and collinear photon approximation. The isospin breaking due to the mass effects was not taken into account.

Unfortunately, in the case of NA48/2, an analysis without the effect of PHOTOS is not available. Hence, it seems almost impossible to make use of the here calculated photonic effects for a full a posteriori correction of the form factors. Nevertheless, I have a program at hand that calculates the effect of PHOTOS on the (partially) inclusive decay rate\footnote{I am very grateful to B.~Bloch-Devaux for providing me with this program.}. This enables me to perform a comparison of the here presented calculation with the effect of PHOTOS, using the simple photon energy cut in $\Sigma_{\ell\nu\gamma}$ described in the previous section.

I pursue therefore two aims in the following sections. First, the isospin corrections due to the mass effects can be extracted directly for the form factors. Second, for the photonic effects, I calculate the radiative corrections to the (semi-)inclusive decay rate. These isospin breaking effects are then compared with the correction applied by NA48/2.

\subsection{Corrections due to the Mass Effects}

As explained in the previous chapter, the isospin breaking effects due to the quark and meson mass differences can be extracted on the level of the amplitude or form factors. I now evaluate these corrections numerically.

The form factors have the partial wave expansions \cite{Bijnens1994}
\begin{align}
	\begin{split}
		F + \frac{\sigma_\pi PL}{X} \cos\theta_\pi G &= \sum_{l=0}^\infty P_l(\cos\theta_\pi) f_l(s,s_\ell) , \\
		G &= \sum_{l=1}^\infty P_l^\prime(\cos\theta_\pi) g_l(s,s_\ell) ,
	\end{split}
\end{align}
where $P_l$ are the Legendre polynomials. The NA48/2 experiment \cite{Batley2012} uses the expansion
\begin{align}
	\begin{split}
		F &= F_s e^{i\delta_s} + F_p e^{i\delta_p} \cos\theta_\pi + \ldots, \\
		G &= G_p e^{i\delta_p} + \ldots
	\end{split}
\end{align}
and defines
\begin{align}
	\begin{split}
		\tilde G_p &= G_p + \frac{X}{\sigma_\pi PL} F_p .
	\end{split}
\end{align}
Hence, I identify
\begin{align}
	\begin{split}
		F_s &= | f_0 |, \quad \tilde G_p = \frac{X}{\sigma_\pi PL} | f_1 |, \quad G_p = | g_1 |
	\end{split}
\end{align}
and calculate the partial wave projections
\begin{align}
	\begin{split}
		f_l &= \frac{2l+1}{2} \int_{-1}^1 d\cos\theta_\pi P_l(\cos\theta_\pi) \left( F + \frac{\sigma_\pi PL}{X} \cos\theta_\pi G \right) , \\
		g_l &= \int_{-1}^1 d\cos\theta_\pi \frac{P_{l-1}(\cos\theta_\pi) - P_{l+1}(\cos\theta_\pi)}{2} \, G .
	\end{split}
\end{align}
At the order that I consider, the isospin correction due to the mass effects to the norms and phases of the partial waves is then given by
\begin{align}
	\begin{split}
		\delta_\mathrm{ME} F_s :={}& 1 - \frac{1}{|f_0|}\lim\limits_{\mathrm{isospin}}  |f_0| = 1 - \frac{1}{| \Re(f_0) |} \lim_\mathrm{isospin} |\Re(f_0)| + \O(p^4) , \\
		\delta_\mathrm{ME} \tilde G_p :={}& 1 - \frac{1}{|f_1|} \lim_{\mathrm{isospin}}  |f_1| = 1 - \frac{1}{| \Re(f_1) |} \lim_\mathrm{isospin} |\Re(f_1)| + \O(p^4) , \\
		\delta_\mathrm{ME} G_p :={}& 1 - \frac{1}{ |g_1|} \lim_{\mathrm{isospin}}  |g_1| = 1 - \frac{1}{| \Re(g_1) |} \lim_\mathrm{isospin} |\Re(g_1) | + \O(p^4) , \\
		\Delta_\mathrm{ME} \delta_0^0 :={}& \arg(f_0) - \lim_\mathrm{isospin} \arg(f_0) = \frac{\Im(f_0)}{f_0^\mathrm{LO}} - \lim_\mathrm{isospin} \frac{\Im(f_0)}{f_0^\mathrm{LO}} + \O(p^4) , \\
		\Delta_\mathrm{ME} \delta_1^1 :={}& \arg(f_1) - \lim_\mathrm{isospin} \arg(f_1) = \frac{\Im(f_1)}{f_1^\mathrm{LO}} - \lim_\mathrm{isospin} \frac{\Im(f_1)}{f_1^\mathrm{LO}} + \O(p^4) \\
			={}& \arg(g_1) - \lim_\mathrm{isospin} \arg(g_1) = \frac{\Im(g_1)}{g_1^\mathrm{LO}} - \lim_\mathrm{isospin} \frac{\Im(g_1)}{g_1^\mathrm{LO}} + \O(p^4) .
	\end{split}
\end{align}
The isospin correction to the $P$-wave phase shift vanishes at this order. Using the inputs described in \cite{Colangelo2009,Bernard2013}, I reproduce their NLO results for the $S$-wave phase shift.

The correction to the phase depends on the pion decay constant and the breaking parameters. In the correction to the norm of the partial waves, also the low-energy constants $L_4^r$, $K_2^r$, $K_4^r$ and $K_6^r$ appear ($K_4^r$ only appears in the correction to the $P$-wave).

I have presented the analytic results of the loop calculation in terms of the decay constant in the chiral limit $F_0$. Unfortunately, different lattice determinations do not yet agree on its value \cite{Aoki2013}. For the numerics, I convert the results to an expansion in $1/F_\pi$ using the relation between $F_0$ and $F_\pi$ in pure QCD at $\O(p^4, \epsilon p^4)$ \cite{Neufeld1996},
\begin{align}
	\begin{split}
		F_\pi &= F_0 \begin{aligned}[t]
			& \Bigg[ 1 + \frac{4}{F_0^2} \Big( L_4^r(\mu) (M_\pi^2 + 2 M_K^2) + L_5^r(\mu) M_\pi^2 \Big) \\
			& - \frac{1}{2(4\pi)^2 F_0^2} \left( 2 M_\pi^2 \ln\left( \frac{M_\pi^2}{\mu^2}\right) + M_K^2 \ln\left( \frac{M_K^2}{\mu^2} \right) \right)  \Bigg], \end{aligned}
	\end{split}
\end{align}
where $M_{\pi,K}$ denote the masses in the isospin limit, defined as
\begin{align}
	\begin{split}
		M_\pi^2 = \mpio^2 , \quad M_K^2 = \frac{1}{2}\left( \mkp^2 + \mko^2 - \mpip^2 + \mpio^2 \right) .
	\end{split}
\end{align}
For $F_\pi$ and the meson masses, I use the current PDG values \cite{Beringer2012}.

Another strategy would be to work directly with $F_0$ and assign a large error that covers the different determinations, as done in~\cite{Bernard2013}. I use the solution based on the expansion in $1/F_0$ with a central value of $F_0 = 75 \; \mathrm{MeV}$ for a very rough estimate of higher order corrections.

The correction to the norms of the partial waves depends rather strongly on the value of $L_4^r$. The $\O(p^4)$ fits in \cite{BijnensTalavera2002, Bijnens2012} give the large value $L_4^r = 1.5 \cdot 10^{-3}$. I decide however, to rely on the lattice estimate of \cite{MILC2009}, recommended in \cite{Aoki2013}, but to use a more conservative uncertainty of $\pm 0.5\cdot 10^{-3}$ (see table~\ref{tab:InputsMassEffects}).

For the NLO constants of the electromagnetic sector, I use the estimates of \cite{Ananthanarayan2004} and assign a 100\% error. For the isospin breaking parameter $\epsilon$, I take the latest recommendation in the FLAG report~\cite{Aoki2013},
\begin{align}
	\begin{split}
		\epsilon = \frac{\sqrt{3}}{4 R} , \quad R = 35.8 \pm 2.6 ,
	\end{split}
\end{align}
where I added the lattice and electromagnetic errors in quadrature.

I fix the electromagnetic low-energy constant $Z$ with the LO relation to the pion mass difference (\ref{eqn:LOMassDifferences}).

\begin{table}[H]
	\centering
	\begin{tabular}{c c r}
		\toprule
		$10^3 \cdot L_4^r(\mu)$ & $0.04 \pm 0.50$ & \cite{Aoki2013} \\
		$10^3 \cdot L_5^r(\mu)$ & $0.84 \pm 0.50$ & \cite{Aoki2013} \\[0.05cm]
		\hline \\[-0.35cm]
		$10^3 \cdot K_2^r(\mu)$ & $0.69 \pm 0.69$ & \cite{Ananthanarayan2004} \\
		$10^3 \cdot K_4^r(\mu)$ & $1.38 \pm 1.38$ & \cite{Ananthanarayan2004} \\
		$10^3 \cdot K_6^r(\mu)$ & $2.77 \pm 2.77$ & \cite{Ananthanarayan2004} \\[0.05cm]
		\hline \\[-0.35cm]
		$F_\pi$ & $(92.21 \pm 0.14)$~MeV & \cite{Beringer2012} \\[0.05cm]
		\hline \\[-0.35cm]
		$R$ & $35.8 \pm 2.6\hphantom{0}$ & \cite{Aoki2013} \\
		\bottomrule
	\end{tabular}
	\caption{Input parameters for the evaluation of the mass effects ($\mu = 770$~MeV).}
	\label{tab:InputsMassEffects}
\end{table}

\begin{figure}[ht]
	\centering
	\large
	\scalebox{0.63}{
		\input{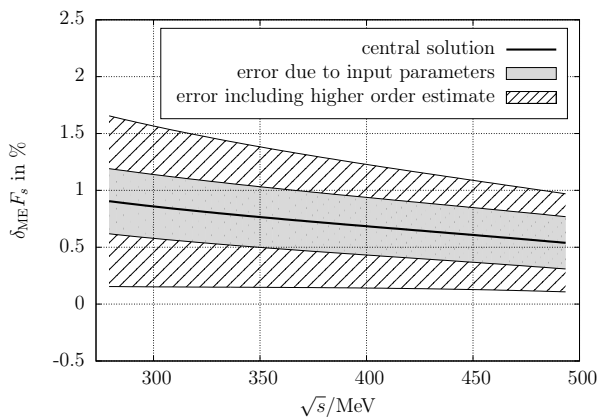}
		}
	\caption{Relative value of the mass effect correction to the $S$-wave $F_s$ for $s_\ell = 0$. The exact meaning of the error bands is explained in the text.}
	\label{plot:SWave}
\end{figure}

\begin{figure}[ht]
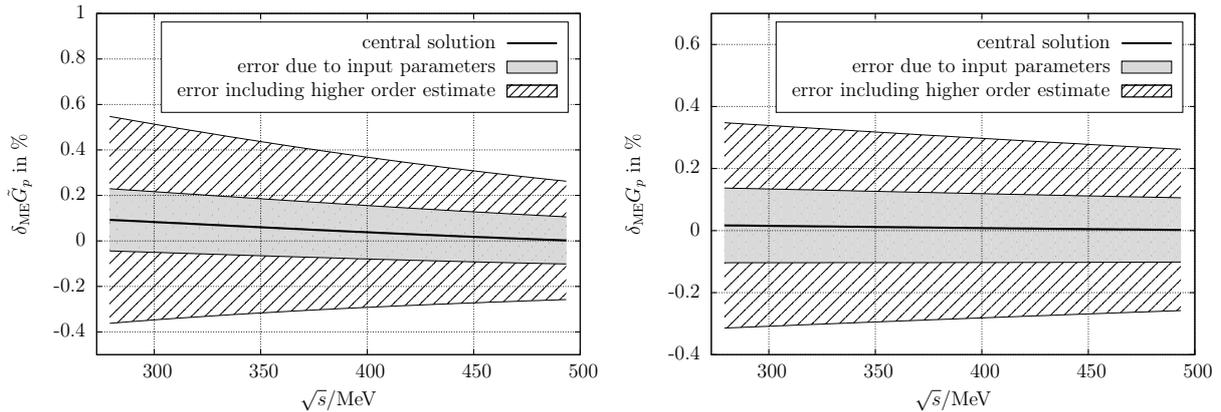

	\centering
	\large
	\scalebox{0.63}{
		\input{plots/gpt.tex}
		\input{plots/gp.tex}
		}
	\caption{Relative value of the mass effect corrections to the $P$-waves $\tilde G_p$ and $G_p$ for $s_\ell = 0$.}
	\label{plot:PWaves}
\end{figure}

The plots in figures~\ref{plot:SWave} and \ref{plot:PWaves} show the relative isospin correction due to the mass effects for the norm of the partial waves. I separately show the error band due to the variation of the input parameters and the error band that also includes the estimate of higher order corrections, given by the difference between the $F_\pi$- and the $F_0$-solution, added in quadrature. The error due to the input parameters is dominated by the uncertainty of the low-energy constant $L_4^r$. The LECs of the electromagnetic sector and the isospin breaking parameter $R$ play a minor role.

In contrast to the $S$-wave, where the isospin corrections are at the percent level, the effect in the two $P$-waves is within the uncertainty compatible with zero. The dependence on $s_\ell$ is rather weak and covered by the error bands.

To conclude this section, I suggest to apply the additional isospin breaking corrections to the NA48/2 measurement \cite{Batley2012} shown in table~\ref{tab:IsospinCorrectionsMassEffectsNA48}. In order to obtain the partial waves of the form factors in the isospin limit, one has to subtract the given corrections. The corrections to the $P$-waves are certainly negligible. However, for the $S$-wave, the isospin correction (and also its uncertainty, unfortunately) is much larger than the experimental errors.

\begin{table}[H]
	\footnotesize
	\centering
	\begin{tabular}{c c c c c c c c}
		\toprule
		$\sqrt{s}/$MeV & $\sqrt{s_\ell}/$MeV & $F_s$ \cite{Batley2010,Batley2012} & $\delta_\mathrm{ME} F_s \cdot F_s$ & $\tilde G_p$ \cite{Batley2010,Batley2012} & $\delta_\mathrm{ME}\tilde G_p \cdot \tilde G_p$ & $G_p$ \cite{Batley2010,Batley2012} & $\delta_\mathrm{ME} G_p \cdot G_p$ \\[0.1cm]
		\hline \\[-0.3cm]
		286.06 &	126.44 &				5.7195(122) &				$0.050(16)(38)$ &	4.334(76) &	$0.003(6)(16)$	&	5.053(266) 				& $0.001(6)(15)$ \\
		295.95 &	142.60 &				5.8123(101) &				$0.050(16)(37)$ &	4.422(61) &	$0.002(6)(16)$	&	5.186(165) 				& $0.001(6)(16)$ \\
		304.88 &	141.31 &				5.8647(102) &				$0.049(16)(36)$ &	4.550(52) &	$0.002(6)(16)$	&	4.941(123) 				& $0.001(6)(15)$ \\
		313.48 &	137.47 &				5.9134(104) &				$0.048(16)(36)$ &	4.645(47) &	$0.002(6)(16)$	&	4.896(104) 				& $0.001(6)(14)$ \\
		322.02 &	130.92 &				5.9496(\hphantom{0}95) &	$0.048(16)(35)$ &	4.711(47) &	$0.002(6)(16)$	&	5.245(\hphantom{0}99) 		& $0.001(6)(15)$ \\
		330.80 &	124.14 &				5.9769(103) &				$0.047(16)(34)$ &	4.767(44) &	$0.002(6)(15)$	&	5.283(\hphantom{0}92) 		& $0.001(6)(15)$ \\
		340.17 &	116.91 &				6.0119(\hphantom{0}98) &	$0.046(16)(34)$ &	4.780(45) &	$0.002(6)(15)$	&	5.054(\hphantom{0}90) 		& $0.001(6)(14)$ \\
		350.94 &	108.19 &				6.0354(\hphantom{0}96) &	$0.046(16)(33)$ &	4.907(39) &	$0.002(6)(15)$	&	5.264(\hphantom{0}72) 		& $0.001(6)(15)$ \\
		364.57 &	\hphantom{0}98.53 &	6.0532(\hphantom{0}96) &	$0.044(16)(32)$ &	5.019(40) &	$0.002(6)(15)$	&	5.357(\hphantom{0}64) 		& $0.001(6)(15)$ \\
		389.95 &	\hphantom{0}80.62 &	6.1314(184) &				$0.043(16)(30)$ &	5.163(42) &	$0.001(6)(15)$	&	5.418(\hphantom{0}64) 		& $0.001(6)(15)$ \\
		\bottomrule
	\end{tabular}
	\caption[Isospin breaking corrections due to the mass effects.]{Isospin breaking corrections due to the mass effects, calculated for the bins of the NA48/2 measurement \cite{Batley2010,Batley2012}. For comparison, I quote the values of the partial waves with their uncertainties (statistical and systematic errors added in quadrature) without including the dominant error of the normalisation. Note that the uncertainties of $F_s$ are taken from \cite{Batley2010}, as the values displayed in \cite{Batley2012} are not correct\footnotemark{}. The first error to the isospin correction is due to the input parameters, the second is a rough estimate of higher order corrections.}
	\label{tab:IsospinCorrectionsMassEffectsNA48}
\end{table}
\footnotetext{I thank B.~Bloch-Devaux for the confirmation thereof.}

\subsection{Discussion of the Photonic Effects}

For the numerical evaluation of the photonic effects, I compute the (semi-)inclusive decay rate, differential with respect to $s$, $s_\ell$ and $\cos\theta_\pi$. After some general considerations and tests, I compare the resulting $\O(e^2)$ correction to the one applied in the NA48/2 experiment \cite{Batley2012}, i.e.~the Gamow-Sommerfeld factor combined with PHOTOS~\cite{Barberio1994}.

For the numerical evaluation of the inclusive decay rate $d\Gamma_{(\gamma)}$, I need several input parameters. As I am interested in $\O(e^2)$ effects but work only at leading chiral order, I directly replace $F_0$ by the physical pion decay constant $F_\pi$. When calculating the fully inclusive decay rate, I take advantage of the cancellation of collinear singularities and send the lepton mass $\ml$ to zero, while I use the physical masses of the charged mesons \cite{Beringer2012}. In the calculation of the semi-inclusive decay rate with the photon energy cut $\Delta x$, I neglect terms that vanish in the limit $\ml\to0$ and evaluate the large logarithm $\ln\zl$ with the physical electron mass \cite{Beringer2012}.

In the NLO counterterm corrections, the low-energy constants $L_9^r$ and $L_{10}^r$ of the strong sector enter. The lattice determinations of these LECs have not yet reached `green status' in the FLAG report \cite{Aoki2013}. For $L_9^r$, I use the value of \cite{BijnensTalavera2002}, for $L_{10}^r$, I take the $\O(p^4)$ fit of \cite{Gonzalez-Alonso2008}, which is compatible with the available lattice determinations.

As for the case of the mass effects, I again use the estimates of \cite{Ananthanarayan2004,Moussallam1997} for the electromagnetic LECs with a 100\% error assigned to them.

The `leptonic' LECs $X_1^r$ and $X_6^r$ are unknown. $X_6^r$ contains the universal short-distance contribution \cite{Marciano1993}, which I split off following the treatment in \cite{Cirigliano2002}:
\begin{align}
	\begin{split}
		X_6^r(\mu) = \tilde X_6^r(\mu) + X_6^\mathrm{SD}, \quad e^2 X_6^\mathrm{SD} = 1 - S_\mathrm{EW}(M_\rho, M_Z) = - \frac{e^2}{4\pi^2} \ln\left( \frac{M_Z^2}{M_\rho^2} \right) ,
	\end{split}
\end{align}
such that $\tilde X_6^r$ is of the typical size of a LEC in \ChPT{}. I use the naïve dimensional estimate that those LECs are of the order $1/(4\pi)^2$. For the short-distance contribution, I take the value that includes leading logarithmic and QCD corrections \cite{Marciano1993}.

\begin{table}[H]
	\centering
	\begin{tabular}{c r r}
		\toprule
		$10^3 \cdot L_9^r(\mu)$ & $5.93 \pm 0.43\hphantom{0} \qquad$ & \cite{BijnensTalavera2002} \\
		$10^3 \cdot L_{10}^r(\mu)$ & $-5.22 \pm 0.06\hphantom{0} \qquad$ & \cite{Gonzalez-Alonso2008} \\[0.05cm]
		\hline \\[-0.35cm]
		$10^3 \cdot K_1^r(\mu)$ & $ -2.71 \pm 2.71\hphantom{0} \qquad$ & \cite{Ananthanarayan2004} \\
		$10^3 \cdot K_3^r(\mu)$ & $ 2.71 \pm 2.71\hphantom{0}  \qquad$ & \cite{Ananthanarayan2004} \\
		$10^3 \cdot K_5^r(\mu)$ & $ 11.59 \pm 11.59 \qquad$ & \cite{Ananthanarayan2004} \\
		$10^3 \cdot K_{12}^r(\mu)$ & $ -4.25 \pm 4.25\hphantom{0} \qquad$ & \cite{Moussallam1997}  \\[0.05cm]
		\hline \\[-0.35cm]
		$10^3 \cdot X_1^r(\mu)$ & $ 0 \pm 6.3\hphantom{00}  \qquad$ & \\
		$10^3 \cdot \tilde X_6^r(\mu)$ & $ 0 \pm 6.3\hphantom{00} \qquad$ & \\[0.05cm]
		\hline \\[-0.35cm]
		$S_\mathrm{EW}$ & $1.0232\qquad\quad\,$ & \cite{Marciano1993} \\[0.05cm]
		\hline \\[-0.35cm]
		$F_\pi$ & $(92.21 \pm 0.14)$~MeV & \cite{Beringer2012} \\
		\bottomrule
	\end{tabular}
	\caption{Input parameters for the evaluation of the photonic effects ($\mu = 770$~MeV).}
\end{table}

\subsubsection{Soft Photon Approximation vs.~Full Matrix Element}

\begin{figure}[H]
	\centering
	\large
	\scalebox{0.63}{
		\input{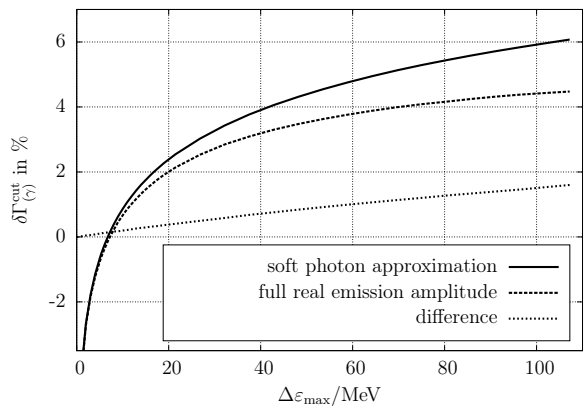}
		}
	\caption{Comparison of the $\O(e^2)$ photonic correction (virtual and real photons) to the semi-inclusive total decay rate as a function of the photon energy cut in $\Sigma_{\ell\nu\gamma}$, using the soft photon approximation vs.~the full radiative matrix element.}
	\label{plot:SPAvsHard}
\end{figure}

In a first step, I want to quantify the importance of considering the full (hard) matrix element for the radiative process instead of relying on the soft photon approximation. To this end, I compare the semi-inclusive total and differential decay rates (using the photon energy cut $\tilde x_\mathrm{max}$) with the decay rate, where the radiative process is just given by the SPA with a finite $\Delta \varepsilon$. The same energy cut in the two descriptions is obtained by setting
\begin{align}
	\begin{split}
		\tilde x_\mathrm{min} = \tilde x_\mathrm{max} \quad \Rightarrow \quad \Delta \varepsilon =  \frac{\sqrt{s_\ell}}{2} \tilde x_\mathrm{max} ( 1 - \zl).
	\end{split}
\end{align}
In this prescription, the photon energy cut is not constant but respects the bounds given by the phase space. The maximum photon energy is
\begin{align}
	\begin{split}
		\Delta\varepsilon_\mathrm{max} = \tilde x_\mathrm{max} \frac{(\mkp-2\mpip)^2 - \ml^2}{2(\mkp-2\mpip)} .
	\end{split}
\end{align}
I compare in the following the corrections to the total decay rate, defined by
\begin{align}
	\begin{split}
		\Gamma_{(\gamma)}^\mathrm{cut} = \Gamma^\mathrm{LO} \left( 1 + \delta \Gamma_{(\gamma)}^\mathrm{cut} \right) .
	\end{split}
\end{align}
In figure~\ref{plot:SPAvsHard}, the correction to the decay rate $\delta \Gamma_{(\gamma)}^\mathrm{cut}$ is shown as a function of the photon energy cut. The virtual corrections are evaluated using the central values of the input parameters. The soft photon approximation depends logarithmically on the energy cut (reflecting the IR divergence at low energies), whereas the correction using the full matrix element is somewhat smaller. Since I use a cut in the dilepton-photon rest frame, the result cannot be applied directly to the experiment, where an energy cut is present in the lab frame. However, I expect that the picture of the difference between full matrix element and soft photon approximation will look similar in the kaon centre-of-mass frame. In the relative form factor measurement of NA48/2, a 3~GeV photon energy cut was applied in the lab frame \cite{Batley2010}. This translates into a minimal detectable photon energy of 11.7~MeV in the centre-of-mass frame. For such a low photon energy, the soft approximation can be expected to still work well (the deviation in $\Sigma_{\ell\nu\gamma}$ is $\approx0.2\%$ of the total rate). However, the experimental cut is not sharp: at the outer edge of the calorimeter, the minimal detectable centre-of-mass photon energy is about 36.8~MeV and of course, only photons flying in the direction of the calorimeter can be detected. At larger photon energies, the error introduced by using a SPA is quite substantial (up to 1.6\% of the total rate for hard photons). This can be understood in terms of the collinear singularity: the SPA alone does not produce the correct dependence on the lepton mass, hence, the large logarithm does not cancel.

As explained before, the gauge invariant class of loop diagrams in figure~\ref{img:Kl4_mLoops} together with the corresponding counterterms has been neglected in the previous literature \cite{Cuplov2003, Cuplov2004}. To judge the influence of these diagrams, I compute the total inclusive decay rate, remove the cut ($\tilde x_\mathrm{max} = 1$) and sum the uncertainties due to the input parameters in quadrature. Using all the diagrams for the virtual corrections, I find
\begin{align}
	\begin{split}
		\delta \Gamma_{(\gamma)} = (4.53 \pm 0.66) \% ,
	\end{split}
\end{align}
whereas neglecting the mentioned class of diagrams results in
\begin{align}
	\begin{split}
		\delta \Gamma_{(\gamma)}^\mathrm{negl.} = (4.70 \pm 0.66) \% .
	\end{split}
\end{align}
The uncertainty is completely dominated by $X_1^r(\mu)$. Note that approximately half of the correction (2.32\%) is due to the short-distance enhancement.

\subsubsection{Comparison with Coulomb Factor $\times$ PHOTOS}

The Gamow-Sommerfeld (or Coulomb) factor is defined by
\begin{align}
	\begin{split}
		d\Gamma_\mathrm{Coulomb} &= d\Gamma \cdot \prod_{i < j} \frac{\omega_{ij}}{e^{\omega_{ij}}-1} ,
	\end{split}
\end{align}
where $i,j$ run over the three charged final state particles, $\pi^+$, $\pi^-$ and $\ell^+$, and where
\begin{align}
	\begin{split}
		\omega_{ij} := \frac{q_i q_j e^2}{2\beta_{ij}} , \quad \beta_{ij} := \sqrt{ 1 - \frac{4 m_i^2 m_j^2}{(s_{ij}-m_i^2-m_j^2)^2}}, \quad s_{ij} := (p_i + p_j)^2 .
	\end{split}
\end{align}
$q_{i,j}$ denote the charges of the particles in units of $e$.

The Coulomb factor is a semiclassical approximation of the final state interactions. However, it is non-perturbative and includes contributions to all orders in $e^2$. In $K_{e4}$, the factors involving the electron are negligible, the important contribution is the $\pi^+\pi^-$ interaction. An expansion of the Coulomb factor in $e^2$ gives
\begin{align}
	\begin{split}
		\frac{\omega_{\pi^+\pi^-}}{e^{\omega_{\pi^+\pi^-}}-1} = 1 + e^2 \frac{1 + \sigma_\pi^2(s)}{8\sigma_\pi(s)} + \O(e^4) .
	\end{split}
\end{align}
If one expands the triangle diagram~\ref{img:Kl4_NLOgLoop8} for $s$ near the threshold (i.e.~for small values of $\sigma_\pi$), exactly the same contribution to the correction of the decay rate is found, up to terms that are finite for $\sigma_\pi\to0$ (but contain e.g.~the IR divergence). The Coulomb factor is therefore an approximation of a part of the virtual corrections, resummed to all orders. It increases the fully inclusive total decay rate by 3.25\%, the $\O(e^2)$ part being responsible for 3.17\%.

The effect of PHOTOS can be described by a multiplicative factor on the decay rate, too,
\begin{align}
	\begin{split}
		d\Gamma_\mathrm{PHOTOS} = d\Gamma \cdot f_\mathrm{PHOTOS}(s,s_\ell,\cos\theta_\pi,\cos\theta_\ell,\phi) ,
	\end{split}
\end{align}
where I determine $f_\mathrm{PHOTOS}$ numerically through a simulation.

Note that PHOTOS assumes the virtual corrections to take such a value that the divergences cancel but that the fully inclusive total decay rate does not change \cite{Nanava2007}. The NA48/2 experiment however claims that PHOTOS has been used even in the determination of the form factor normalisation, i.e.~to take the effect of real photons on the total decay rate into account \cite{Batley2012}. The inclusion of PHOTOS increased the simulated decay rate by 0.69\%\footnote{B. Bloch-Devaux, private communication.}. I was not able to reproduce this number and suspect it to be only an effect due to finite resolution or statistical fluctuations. The results of my own simulations with a large statistics of $8 \cdot 10^{10}$ events are compatible with the assumption that PHOTOS does not change the fully inclusive total decay rate.

I compare now the results for the fully inclusive as well as for the semi-inclusive differential rate with a photon energy cut of $\Delta \varepsilon_\mathrm{max} = 40$~MeV in $\Sigma_{\ell\nu\gamma}$. I include only the $\O(e^2)$ contribution of the Coulomb factor.

\begin{figure}[H]
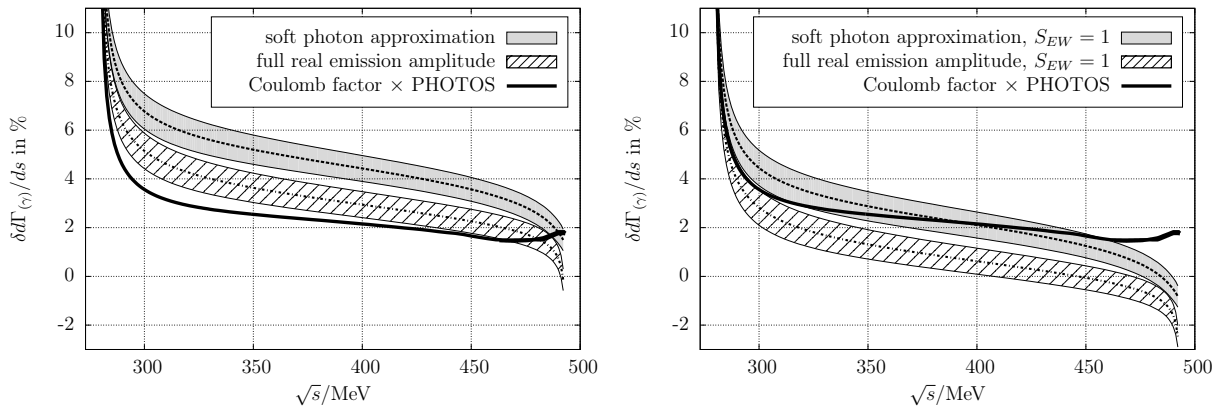

	\centering
	\large
	\scalebox{0.63}{
		\input{plots/SPAvsHardvsCoulPHDiffS.tex}
		\input{plots/SPAvsHardvsCoulPHDiffSNoSD.tex}
		}
	\caption{Comparison of the photonic corrections to the fully inclusive differential decay rate. The right plot excludes the short-distance enhancement factor.}
	\label{plot:SPAvsHardvsCoulPHDiffS}
\end{figure}

\begin{figure}[H]
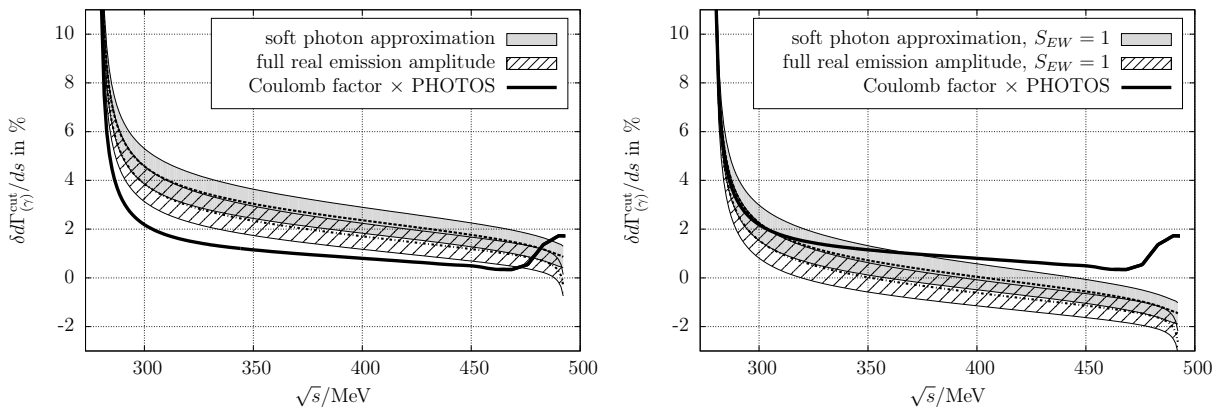

	\centering
	\large
	\scalebox{0.63}{
		\input{plots/SPAvsHardvsCoulPHDiffSCut.tex}
		\input{plots/SPAvsHardvsCoulPHDiffSCutNoSD.tex}
		}
	\caption{Comparison of the photonic corrections to the semi-inclusive total decay rate with a photon energy cut of $\Delta \varepsilon_\mathrm{max} = 40$~MeV in $\Sigma_{\ell\nu\gamma}$. The rise of the PHOTOS factor at large $s$ could be a numerical artefact, as the decay rate approaches zero in this phase space region.}
	\label{plot:SPAvsHardvsCoulPHDiffSCut}
\end{figure}

The plots in figures~\ref{plot:SPAvsHardvsCoulPHDiffS} and \ref{plot:SPAvsHardvsCoulPHDiffSCut} show the corrections to the differential decay rate. The divergence at the $\pi\pi$ threshold is the Coulomb singularity, reproduced in all descriptions. The rise of the PHOTOS factor at large values of $s$, however, could be a numerical artefact, because the differential decay rate drops to zero at the upper border of the phase space.

The comparison without the short-distance enhancement shows that the Coulomb factor $\times$ PHOTOS approach is relatively close to the soft photon approximation, which overestimates the radiative corrections. However, the short-distance factor has not been included in the experimental analysis, such that in total, the radiative corrections are underestimated.

Unfortunately, it is not possible to calculate the radiative corrections for a realistic setup with the experi\-mental cuts. Nevertheless, as NA48/2 determined the branching ratio in a fully inclusive measurement, it is possible to correct the normalisation of the form factors. For the relative values of the form factors, one has to assume that the Coulomb factor $\times$ PHOTOS approach is an acceptable description of the radiative corrections (a free normalisation factor corresponds to a free additive constant in the correction, hence the slopes of the corrections have to be compared).

I suggest to replace in a matching procedure the $\O(e^2)$ part of the Coulomb factor and the 0.69\% PHOTOS effect (or rather artefact) with the result of the here presented fixed order calculation, i.e.~to apply the following correction to the norm of the form factors $X\in\{F,G\}$:
\begin{align}
	\begin{split}
		| X | &= | X^\mathrm{exp} | \left( 1 + \frac{1}{2} \left( \delta\Gamma_\mathrm{Coulomb}^{e^2} + \delta\Gamma_\mathrm{PHOTOS} - \delta \Gamma_{(\gamma)} \right) \right) \\
			&= | X^\mathrm{exp} | \left( 0.9967 \pm 0.0033 \right) , \\
		\Rightarrow \delta |X| &= ( -0.33 \pm 0.33 ) \%.
	\end{split}
\end{align}
Note that replacing the systematic PHOTOS uncertainty with the above error increases the $0.62\%$ uncertainty of the NA48/2 norm measurement \cite{Batley2012} to 0.70\%.

The fact that the a posteriori correction is so small is at least partly accidental: as argued above, I have the strong suspicion that the estimate $\delta \Gamma_\mathrm{PHOTOS}=0.69\%$ is simply the outcome of statistical fluctuations. By chance, this number leads to a result close to the estimate obtained by Diamant-Berger in his analytic treatment of radiative corrections. For this reason, it has been considered so far as a reliable estimate\footnote{B. Bloch-Devaux, private communication.}.


\section{Discussion and Conclusion}

In the present work, I have computed the one-loop isospin breaking corrections to the $K_{\ell4}$ decay within \ChPT{} including leptons and photons. The corrections can be separated into mass effects and photonic effects. The mass effects for the $S$-wave are quite substantial but the result for the norm of the form factors suffers from large uncertainties, on the one hand due to the uncertainty in the LEC $L_4^r$, on the other hand due to higher order corrections. The mass effects for the $P$-waves are negligible.

For the photonic corrections, I have compared the fixed order calculation with the Coulomb $\times$ PHOTOS approach used in the experimental analysis of NA48/2. An a posteriori correction of the data is possible for the normalisation but not for the relative values of the form factors. The present calculation includes for the first time a treatment of the full radiative process and compares it with the soft photon approximation.

For possible forthcoming experiments on $K_{\ell4}$, I suggest that photonic corrections are applied in a Monte Carlo simulation that includes the exact matrix element. This can be done e.g.~with PHOTOS. The mass effects can be easily corrected a posteriori.

This work goes either beyond the isospin breaking treatments in previous literature or is complementary: I confirm the largest part of the amplitude calculation of \cite{Cuplov2003, Cuplov2004}, but correct their results by a neglected gauge invariant class of diagrams. I have included the full radiative process and shown that the soft photon approximation is not necessarily trustworthy and certainly not applicable for the fully inclusive decay.

I reproduce the NLO mass effect calculation for phases of the form factors done in \cite{Colangelo2009, Bernard2013}, but concentrate here on the absolute values of the form factors. As the NLO mass effect calculation suffers from large uncertainties, an extension of the dispersive framework of \cite{Bernard2013} to the norm of the form factors would be desirable.

To judge the reliability of the photonic corrections, one should ideally calculate them to higher chiral orders, which is however prohibitive (and would bring in many unknown low-energy constants). Here, I have assumed implicitly that the photonic corrections factorise and therefore modify the higher chiral orders with the same multiplicative correction as the lowest order. It is hard to judge if this assumption is justified: for this reason, I have attached a rather conservative estimate of the uncertainties to the photonic corrections presented here.

\section*{Acknowledgements}
\addcontentsline{toc}{section}{Acknowledgements}

I cordially thank my thesis advisor, Gilberto Colangelo, for his continuous support during the last years (in many respects and also at unearthly hours) and for having drawn my attention to the isospin breaking corrections to $K_{\ell4}$, which became a side-project to my thesis. I am very grateful to Brigitte Bloch-Devaux for her time and effort to provide me additional data and information and help me to understand diverse aspects of the experiment. I have enjoyed discussions with many people: Thomas Becher, Vincenzo Cirigliano, Vesna Cuplov, Jürg Gasser, Bastian Kubis, Stefan Lanz, Heiri Leutwyler, Lorenzo Mercolli, Emilie Passemar, Massimo Passera, Stefan Pislak, Christopher Smith, Peter Truöl and Zbigniew~W\c{a}s. I thank Brigitte Bloch-Devaux, Vincenzo Cirigliano, Gilberto Colangelo, Jürg Gasser, Serena Grädel and Emilie Passemar for corrections and many useful comments on the manuscript. I thank the Los Alamos National Laboratory, where part of this work was done, for the hospitality. The Albert Einstein Center for Fundamental Physics is supported by the ‘Innovations- und Kooperationsprojekt C--13’ of the ‘Schweizerische Universitätskonferenz SUK/CRUS’. This work was supported in part by the Swiss National Science Foundation.

\appendix
\numberwithin{equation}{section}


\section{Loop Functions}

\subsection{Scalar Functions}

I use the following conventions for the scalar loop functions:
\begin{align}
	\begin{split}
		A_0&(m^2) \\
			&:= \frac{1}{i} \int \frac{d^nq}{(2\pi)^n} \frac{1}{[ q^2 - m^2 ]} , \\
		B_0&(p^2, m_1^2, m_2^2) \\
			&:= \frac{1}{i} \int \frac{d^nq}{(2\pi)^n} \frac{1}{[ q^2 - m_1^2 ] [ (q+p)^2 - m_2^2 ]} , \\
		C_0&(p_1^2, (p_1-p_2)^2, p_2^2, m_1^2, m_2^2, m_3^2) \\
			&:= \frac{1}{i} \int \frac{d^nq}{(2\pi)^n} \frac{1}{[ q^2 - m_1^2 ] [ (q+p_1)^2 - m_2^2 ] [ (q+p_2)^2 - m_3^2 ]} , \\
		D_0&(p_1^2, (p_1-p_2)^2,(p_2-p_3)^2,p_3^2,p_2^2,(p_1-p_3)^2, m_1^2, m_2^2, m_3^2, m_4^2) \\
			&:= \frac{1}{i} \int \frac{d^nq}{(2\pi)^n} \frac{1}{[ q^2 - m_1^2 ] [ (q+p_1)^2 - m_2^2 ] [ (q+p_2)^2 - m_3^2 ] [ (q+p_3)^2 - m_4^2 ]} .
	\end{split}
\end{align}

The loop functions $A_0$ and $B_0$ are UV-divergent. The renormalised loop functions are defined in the $\overline{MS}$ scheme by
\begin{align}
	\begin{split}
		\label{eqn:RenormalisedLoopFunctions}
		A_0(m^2) &= -2 m^2 \lambda + \bar A_0(m^2) + \O(4-n) , \\
		B_0(p^2, m_1^2, m_2^2) &= -2\lambda + \bar B_0(p^2, m_1^2, m_2^2) + \O(4-n) ,
	\end{split}
\end{align}
where
\begin{align}
	\begin{split}
		\lambda = \frac{\mu^{n-4}}{16\pi^2} \left( \frac{1}{n-4} - \frac{1}{2} \left( \ln(4\pi) + 1 - \gamma_E \right) \right) .
	\end{split}
\end{align}
$\mu$ denotes the renormalisation scale.

The renormalised loop functions are given by \cite{Amoros2000}
\begin{align}
	\begin{split}
		\bar A_0(m^2) &= -\frac{m^2}{16\pi^2} \ln\left( \frac{m^2}{\mu^2} \right) , \\
		\bar B_0(p^2, m_1^2, m_2^2) &= -\frac{1}{16\pi^2} \frac{m_1^2 \ln\left(\frac{m_1^2}{\mu^2}\right) - m_2^2 \ln\left(\frac{m_2^2}{\mu^2}\right)}{m_1^2 - m_2^2} \\
			&+ \frac{1}{32\pi^2} \left( 2 + \left( -\frac{\Delta}{p^2} + \frac{\Sigma}{\Delta} \right) \ln\left( \frac{m_1^2}{m_2^2} \right) - \frac{\nu}{p^2} \ln\left( \frac{(p^2+\nu)^2 - \Delta^2}{(p^2-\nu)^2 - \Delta^2} \right) \right) ,
	\end{split}
\end{align}
where
\begin{align}
	\begin{split}
		\Delta &:= m_1^2 - m_2^2 , \\
		\Sigma &:= m_1^2 + m_2^2 , \\
		\nu &:= \sqrt{ (s - (m_1 + m_2)^2)(s-(m_1-m_2)^2) } = \lambda^{1/2}(s,m_1^2,m_2^2) .
	\end{split}
\end{align}

\subsection{Tensor-Coefficient Functions}

\label{sec:AppendixTensorCoefficientFunctions}

Although all the loop integrals can be expressed in terms of the basic scalar loop functions by means of a Passarino-Veltman reduction \cite{Hooft1979, Passarino1979}, this produces sometimes very long polynomial coefficients. I therefore also use the tensor coefficient functions. The tensor integrals that I use are defined by
{ \small
\begin{align}
	\begin{split}
		B^{\mu\nu}(p; m_1^2, m_2^2)
			&:= \frac{1}{i} \int \frac{d^nq}{(2\pi)^n} \frac{q^\mu q^\nu}{[ q^2 - m_1^2 ] [ (q+p)^2 - m_2^2 ]} , \\
		C^\mu(p_1,p_2; m_1^2, m_2^2, m_3^2)
			&:= \frac{1}{i} \int \frac{d^nq}{(2\pi)^n} \frac{q^\mu}{[ q^2 - m_1^2 ] [ (q+p_1)^2 - m_2^2 ] [ (q+p_2)^2 - m_3^2 ]} , \\
		C^{\mu\nu}(p_1,p_2; m_1^2, m_2^2, m_3^2)
			&:= \frac{1}{i} \int \frac{d^nq}{(2\pi)^n} \frac{q^\mu q^\nu}{[ q^2 - m_1^2 ] [ (q+p_1)^2 - m_2^2 ] [ (q+p_2)^2 - m_3^2 ]} , \\
		D^\mu(p_1,p_2,p_3; m_1^2, m_2^2, m_3^2,m_4^2)
			&:= \frac{1}{i} \int \frac{d^nq}{(2\pi)^n} \frac{q^\mu}{[ q^2 - m_1^2 ] [ (q+p_1)^2 - m_2^2 ] [ (q+p_2)^2 - m_3^2 ] [ (q+p_3)^2 - m_4^2 ]} , \\
		D^{\mu\nu}(p_1,p_2,p_3; m_1^2, m_2^2, m_3^2,m_4^2)
			&:= \frac{1}{i} \int \frac{d^nq}{(2\pi)^n} \frac{q^\mu q^\nu}{[ q^2 - m_1^2 ] [ (q+p_1)^2 - m_2^2 ] [ (q+p_2)^2 - m_3^2 ] [ (q+p_3)^2 - m_4^2 ]} .
	\end{split}
\end{align}
}
The tensor coefficients are then given by a Lorentz decomposition:
{ \small
\begin{align}
	\begin{split}
		B^{\mu\nu}(p; m_1^2, m_2^2) &= g^{\mu\nu} B_{00}(p^2, m_1^2, m_2^2) + p^\mu p^\nu B_{11}(p^2, m_1^2, m_2^2) , \\
		C^\mu(p_1, p_2; m_1^2, m_2^2, m_3^2) &= p_1^\mu C_1(p_1^2, (p_1-p_2)^2, p_2^2, m_1^2, m_2^2, m_3^2) \\
			& + p_2^\mu C_2(p_1^2, (p_1-p_2)^2, p_2^2, m_1^2, m_2^2, m_3^2) , \\
		C^{\mu\nu}(p_1, p_2; m_1^2, m_2^2, m_3^2) &= g^{\mu\nu} C_{00}(p_1^2, (p_1-p_2)^2, p_2^2, m_1^2, m_2^2, m_3^2) \\
			& + \sum_{i,j=1}^2 p_i^\mu p_j^\nu C_{ij}(p_1^2, (p_1-p_2)^2, p_2^2, m_1^2, m_2^2, m_3^2) , \\
		D^\mu(p_1, p_2, p_3; m_1^2, m_2^2, m_3^2, m_4^2) &= \sum_{i=1}^3 p_i^\mu D_i(p_1^2, (p_1-p_2)^2,(p_2-p_3)^2,p_3^2,p_2^2,(p_1-p_3)^2, m_1^2, m_2^2, m_3^2, m_4^2) , \\
		D^{\mu\nu}(p_1, p_2, p_3; m_1^2, m_2^2, m_3^2, m_4^2) &= g^{\mu\nu} D_{00}(p_1^2, (p_1-p_2)^2,(p_2-p_3)^2,p_3^2,p_2^2,(p_1-p_3)^2, m_1^2, m_2^2, m_3^2, m_4^2) \\
			& + \sum_{i,j=1}^3 p_i^\mu p_j^\nu D_{ij}(p_1^2, (p_1-p_2)^2,(p_2-p_3)^2,p_3^2,p_2^2,(p_1-p_3)^2, m_1^2, m_2^2, m_3^2, m_4^2) .
	\end{split}
\end{align}
}
Only some of those tensor coefficient functions are UV-divergent:
\begin{align}
	\begin{split}
		\label{eqn:RenormalisedTensorCoefficients}
		B_{00}(p^2, m_1^2, m_2^2) &= -\frac{\lambda}{2} \left( m_1^2 + m_2^2 - \frac{p^2}{3} \right) + \bar B_{00}(p^2, m_1^2, m_2^2) + \O(4-n) , \\
		B_{11}(p^2, m_1^2, m_2^2) &= -\frac{2}{3}\lambda + \bar B_{11}(p^2, m_1^2, m_2^2) + \O(4-n) , \\
		C_{00}(p_1^2, (p_1-p_2)^2, p_2^2, m_1^2, m_2^2, m_3^2) &= -\frac{\lambda}{2} + \bar C_{00}(p_1^2, (p_1-p_2)^2, p_2^2, m_1^2, m_2^2, m_3^2) + \O(4-n) .
	\end{split}
\end{align}

\subsection{Infrared Divergences in Loop Functions}

\label{sec:IRdivergentLoopFunctions}

The following explicit formulae are used to extract the IR divergence in the loop functions.

The derivative of the two-point function is IR-divergent:
\begin{align}
	\begin{split}
		\bar B_0^\prime(M^2, M^2, \mg^2) &= - \frac{1}{16\pi^2} \frac{1}{M^2} \left( 1 + \frac{1}{2} \ln\left( \frac{\mg^2}{M^2} \right) \right) + \O(\mg) , \\
		\bar B_0(M^2, M^2, \mg^2) &= \frac{1}{16\pi^2} \left( 1 - \ln\left( \frac{M^2}{\mu^2} \right) \right) + \O(\mg), \\
		\bar B_0(0, M^2, \mg^2) &= - \frac{1}{16\pi^2} \ln\left( \frac{M^2}{\mu^2} \right) + \O(\mg) .
	\end{split}
\end{align}
The IR-divergent three-point function is given by \cite{Beenakker1990}
\begin{align}
	\begin{split}
		C_0(m^2, s, M^2, \mg^2, m^2, M^2)
			&= \frac{1}{16\pi^2} \frac{x_s}{m M(1-x_s^2)} \Bigg( \ln x_s \left( -\frac{1}{2} \ln x_s + 2 \ln(1-x_s^2) + \ln\left( \frac{m M}{\mg^2} \right) \right) \\
				& - \frac{\pi^2}{6} + \dilog(x_s^2) + \frac{1}{2} \ln^2\left( \frac{m}{M} \right) + \dilog\left( 1 - x_s \frac{m}{M} \right) + \dilog\left( 1 - x_s \frac{M}{m} \right) \Bigg) \\
				& + \O(m_\gamma^2) ,
	\end{split}
\end{align}
where
\begin{align}
	\begin{split}
		x_s = -\frac{1 - \sqrt{1 - \frac{4 m M}{s - (m-M)^2}}}{1 + \sqrt{1 - \frac{4 m M}{s - (m-M)^2}}} .
	\end{split}
\end{align}



\section{Kinematics}

\subsection{Lorentz Frames and Transformations in $K_{\ell4}$}

\label{sec:LorentzTransformationsKl4}

Let us first look at the kaon rest frame $\Sigma_K$. From the relations
\begin{align}
	\begin{split}
		P &= p_1 + p_2 = \left(\sqrt{ s + \vec P^2}, \vec P \right) , \\
		L &= p_\ell + p_\nu = \left(\sqrt{ s_\ell + \vec P^2}, - \vec P \right) , \\
		p &= P + L = \left( \mkp, \vec 0 \right) ,
	\end{split}
\end{align}
one finds
\begin{align}
	\vec P^2 = \frac{\lambda_{K\ell}(s)}{4\mkp^2},
\end{align}
where $\lambda_{K\ell}(s) := \lambda(\mkp^2,s,s_\ell)$ and $\lambda(a,b,c):=a^2+b^2+c^2-2(ab+bc+ca)$.

I choose the $x$-axis along the dipion line of flight:
\begin{align}
	\begin{split}
		P &= \left( \frac{\mkp^2-s_\ell+s}{2\mkp}, \frac{\lambda_{K\ell}^{1/2}(s)}{2\mkp}, 0, 0 \right) , \\
		L &= \left( \frac{\mkp^2+s_\ell-s}{2\mkp}, - \frac{\lambda_{K\ell}^{1/2}(s)}{2\mkp}, 0, 0 \right) .
	\end{split}
\end{align}
In the dipion centre-of-mass frame $\Sigma_{2\pi}$, the boosted dipion four-momentum is
\begin{align}
	P^\prime = \Lambda_1^{-1} P = \left( \sqrt{s}, \vec 0 \right) .
\end{align}
$\Lambda_1$ is just a boost in the $x$-direction. Thus, I find
\begin{align}
	\Lambda_1 = \left(\begin{array}{cccc} \frac{\mkp^2+s-s_\ell}{2\mkp\sqrt{s}} & \frac{\lambda_{K\ell}^{1/2}(s)}{2\mkp\sqrt{s}} & 0 & 0 \\ \frac{\lambda_{K\ell}^{1/2}(s)}{2\mkp\sqrt{s}} & \frac{\mkp^2+s-s_\ell}{2\mkp\sqrt{s}} & 0 & 0 \\0 & 0 & 1 & 0 \\0 & 0 & 0 & 1\end{array}\right) .
\end{align}
Analogously, in the dilepton centre-of-mass frame $\Sigma_{\ell\nu}$, the boosted dilepton four-momentum is
\begin{align}
	L^\dprime = \Lambda_2^{-1} L = \left( \sqrt{s_\ell}, \vec 0 \right) .
\end{align}
$\Lambda_2$ is given by a rotation around the $x$-axis and a subsequent boost in the $x$-direction. I find
\begin{align}
	\label{eqn:LorentzTransformation2}
	\Lambda_2 = \left(\begin{array}{cccc} \frac{\mkp^2-s+s_\ell}{2\mkp\sqrt{s_\ell}} & -\frac{\lambda_{K\ell}^{1/2}(s)}{2\mkp\sqrt{s_\ell}} & 0 & 0 \\ -\frac{\lambda_{K\ell}^{1/2}(s)}{2\mkp\sqrt{s_\ell}} & \frac{\mkp^2-s+s_\ell}{2\mkp\sqrt{s_\ell}} & 0 & 0 \\0 & 0 & \cos\phi & \sin\phi \\0 & 0 & -\sin\phi & \cos\phi \end{array}\right) .
\end{align}

Let us determine the momenta of the four final-state particles in the kaon rest frame. In $\Sigma_{2\pi}$, the pion momenta
\begin{align}
	\begin{split}
		p_1^\prime &= \left( \sqrt{\mpip^2 + \vec p^2}, \vec p \right) , \\
		p_2^\prime &= \left( \sqrt{\mpip^2 + \vec p^2}, -\vec p \right)
	\end{split}
\end{align}
satisfy
\begin{align}
	\begin{split}
		P^\prime &= p_1^\prime + p_2^\prime = \left( \sqrt{s}, \vec 0 \right) .
	\end{split}
\end{align}
Therefore, we find
\begin{align}
	\vec p^2 = \frac{s}{4} - \mpip^2 ,
\end{align}
leading to
\begin{align}
	\begin{split}
		p_1^\prime &= \left( \frac{\sqrt{s}}{2}, \sqrt{\frac{s}{4} - \mpip^2} \cos\theta_\pi, \sqrt{\frac{s}{4} - \mpip^2} \sin\theta_\pi, 0 \right) , \\
		p_2^\prime &= \left( \frac{\sqrt{s}}{2}, -\sqrt{\frac{s}{4} - \mpip^2} \cos\theta_\pi, -\sqrt{\frac{s}{4} - \mpip^2} \sin\theta_\pi, 0 \right) .
	\end{split}
\end{align}
The pion momenta in $\Sigma_K$ are then given by
\begin{align}
	\begin{alignedat}{2}
		p_1 &= \Lambda_1 p_1^\prime = &&\Bigg( \frac{\mkp^2+s-s_\ell}{4\mkp} + \frac{\lambda_{K\ell}^{1/2}(s)}{4\mkp}\sigma_\pi(s)\cos\theta_\pi, \\
			& & &\quad \frac{\lambda_{K\ell}^{1/2}(s)}{4\mkp} + \frac{\mkp^2+s-s_\ell}{4\mkp}\sigma_\pi(s)\cos\theta_\pi, \sqrt{\frac{s}{4} - \mpip^2} \sin\theta_\pi, 0 \Bigg) , \\
		p_2 &= \Lambda_1 p_2^\prime = &&\Bigg( \frac{\mkp^2+s-s_\ell}{4\mkp} - \frac{\lambda_{K\ell}^{1/2}(s)}{4\mkp}\sigma_\pi(s)\cos\theta_\pi, \\
			& & &\quad \frac{\lambda_{K\ell}^{1/2}(s)}{4\mkp} - \frac{\mkp^2+s-s_\ell}{4\mkp}\sigma_\pi(s)\cos\theta_\pi, -\sqrt{\frac{s}{4} - \mpip^2} \sin\theta_\pi, 0 \Bigg) ,
	\end{alignedat}
\end{align}
where $\sigma_\pi(s) = \sqrt{1 - \frac{4\mpip^2}{s}}$.

Again, the analogous procedure for the dilepton system leads to the lepton momenta in the kaon system. In $\Sigma_{\ell\nu}$, the lepton momenta are
\begin{align}
	\begin{split}
		p_\ell^\dprime &= \left( \sqrt{\ml^2 + \vec p_\ell^2}, \vec p_\ell \right) , \quad p_\nu^\dprime = \left( |\vec p_\ell |, -\vec p_\ell \right) ,
	\end{split}
\end{align}
satisfying
\begin{align}
	L^\dprime = p_\ell^\dprime + p_\nu^\dprime = \left( \sqrt{s_\ell}, \vec 0 \right) ,
\end{align}
with the solution
\begin{align}
	\vec p_\ell^2 = \frac{\left(s_\ell-\ml^2\right)^2}{4s_\ell} ,
\end{align}
hence
\begin{align}
	\begin{split}
		p_\ell^\dprime &= \left( \frac{s_\ell+\ml^2}{2\sqrt{s_\ell}}, -\frac{s_\ell-\ml^2}{2\sqrt{s_\ell}} \cos\theta_\ell, \frac{s_\ell-\ml^2}{2\sqrt{s_\ell}} \sin\theta_\ell, 0 \right) , \\
		p_\nu^\dprime &= \left( \frac{s_\ell-\ml^2}{2\sqrt{s_\ell}}, \frac{s_\ell-\ml^2}{2\sqrt{s_\ell}} \cos\theta_\ell, -\frac{s_\ell-\ml^2}{2\sqrt{s_\ell}} \sin\theta_\ell, 0 \right) .
	\end{split}
\end{align}
I obtain the lepton momenta in $\Sigma_K$ by applying the Lorentz transformation $\Lambda_2$:
\begin{align}
	\begin{alignedat}{2}
		p_\ell &= \Lambda_2 p_\ell^\dprime = &&\Bigg( (1+z_\ell) \frac{\mkp^2-s+s_\ell}{4\mkp} + (1-z_\ell) \frac{\lambda_{K\ell}^{1/2}(s)}{4\mkp}\cos\theta_\ell, \\
			& & &\quad -(1+z_\ell)\frac{\lambda_{K\ell}^{1/2}(s)}{4\mkp} - (1-z_\ell) \frac{\mkp^2-s+s_\ell}{4\mkp}\cos\theta_\ell , \\
			& & &\quad \frac{s_\ell-\ml^2}{2\sqrt{s_\ell}} \sin\theta_\ell \cos\phi, -\frac{s_\ell-\ml^2}{2\sqrt{s_\ell}} \sin\theta_\ell \sin\phi \Bigg) , \\
		p_\nu &= \Lambda_2 p_\nu^\dprime = &&\Bigg( (1-z_\ell) \left( \frac{\mkp^2-s+s_\ell}{4\mkp} - \frac{\lambda_{K\ell}^{1/2}(s)}{4\mkp}\cos\theta_\ell \right), \\
			& & &\quad -(1-z_\ell) \left( \frac{\lambda_{K\ell}^{1/2}(s)}{4\mkp} - \frac{\mkp^2-s+s_\ell}{4\mkp}\cos\theta_\ell \right), \\
			& & &\quad -\frac{s_\ell-\ml^2}{2\sqrt{s_\ell}} \sin\theta_\ell \cos\phi, \frac{s_\ell-\ml^2}{2\sqrt{s_\ell}} \sin\theta_\ell \sin\phi \Bigg) ,
	\end{alignedat}
\end{align}
where $z_\ell = \ml^2/s_\ell$.

With these explicit expressions for the particle momenta, I calculate in the following all the Lorentz invariant products in terms of the five phase space variables.

The Lorentz invariant squares of the vectors (\ref{eqn:FourMomenta}) are given by
\begin{align}
	\begin{split}
		P^2 &= p_1^2 + 2p_1p_2 + p_2^2 = 2\mpip^2 + 2p_1p_2 = s , \\
		Q^2 &= p_1^2 - 2 p_1 p_2 + p_2^2 = 4\mpip^2 - s , \\
		L^2 &= p_\ell^2 + 2 p_\ell p_\nu + p_\nu^2 = m_\ell^2 + 2p_\ell p_\nu = s_\ell , \\
		N^2 &= p_\ell^2 - 2 p_\ell p_\nu + p_\nu^2 = 2m_\ell^2 - s_\ell.
	\end{split}
\end{align}
The remaining Lorentz invariant products are:
\begin{align}
	\begin{split}
		PQ ={}& p_1^2 - p_2^2 = 0 , \\
		PL ={}& \frac{1}{2}\left(p^2 - P^2 - L^2\right) = \frac{1}{2}\left(\mkp^2 - s - s_\ell \right) , \\
		PN ={}& \frac{1}{2}\left((p-2p_\nu)^2-P^2-N^2\right) = \frac{1}{2}z_\ell \left(\mkp^2 - s - s_\ell\right) + (1-z_\ell)X\cos\theta_\ell , \\
		QL ={}& Qp = \sigma_\pi X \cos\theta_\pi , \\
		QN ={}& z_\ell \sigma_\pi X \cos\theta_\pi + \sigma_\pi (1-z_\ell) \left\{ \frac{1}{2}\left(\mkp^2-s - s_\ell\right) \cos\theta_\pi \cos\theta_\ell - \sqrt{s s_\ell} \sin\theta_\pi \sin\theta_\ell \cos\phi \vphantom{\frac{1}{2}}\right\} , \\
		LN ={}& (p_\ell+p_\nu)(p_\ell-p_\nu) = m_\ell^2 , \\
		\< LNPQ \> :={}& \epsilon_{\mu\nu\rho\sigma} L^\mu N^\nu P^\rho Q^\sigma = -(1-z_\ell) \sigma_\pi X \sqrt{s_\ell s} \sin\theta_\pi \sin\theta_\ell \sin\phi .
	\end{split}
\end{align}

\subsection{Lorentz Frames and Transformations in $K_{\ell4\gamma}$}

\label{sec:LorentzTransformationsKl4g}

For the radiative process, I copy the results for the dipion subsystem from the $K_{\ell4}$ kinematics and therefore find the following expressions for the momenta in the kaon rest frame $\Sigma_K$:
\begin{align}
	\begin{split}
		P &= \left( \frac{\mkp^2-s_\ell+s}{2\mkp}, \frac{\lambda_{K\ell}^{1/2}(s)}{2\mkp}, 0, 0 \right) , \\
		L &= \left( \frac{\mkp^2+s_\ell-s}{2\mkp}, - \frac{\lambda_{K\ell}^{1/2}(s)}{2\mkp}, 0, 0 \right) .
	\end{split}
\end{align}
\begin{align}
	\begin{split}
		p_1 &= \Bigg( \frac{\mkp^2+s-s_\ell}{4\mkp} + \frac{\lambda_{K\ell}^{1/2}(s)}{4\mkp}\sigma_\pi(s)\cos\theta_\pi, \\
			&\qquad \frac{\lambda_{K\ell}^{1/2}(s)}{4\mkp} + \frac{\mkp^2+s-s_\ell}{4\mkp}\sigma_\pi(s)\cos\theta_\pi, \sqrt{\frac{s}{4} - \mpip^2} \sin\theta_\pi, 0 \Bigg) , \\
		p_2 &= \Bigg( \frac{\mkp^2+s-s_\ell}{4\mkp} - \frac{\lambda_{K\ell}^{1/2}(s)}{4\mkp}\sigma_\pi(s)\cos\theta_\pi, \\
			&\qquad \frac{\lambda_{K\ell}^{1/2}(s)}{4\mkp} - \frac{\mkp^2+s-s_\ell}{4\mkp}\sigma_\pi(s)\cos\theta_\pi, -\sqrt{\frac{s}{4} - \mpip^2} \sin\theta_\pi, 0 \Bigg) .
	\end{split}
\end{align}

We still need to determine the momenta of the photon and the two leptons. The photon and charged lepton momenta in $\Sigma_{\ell\nu\gamma}$ are given by
\begin{align}
	\begin{split}
		q^\dprime &= \Bigg( \frac{\sqrt{s_\ell}}{2} x, - \frac{\sqrt{s_\ell}}{2}\sqrt{x^2 - 4 \zg} \cos\theta_\gamma, \frac{\sqrt{s_\ell}}{2}\sqrt{x^2 - 4 \zg} \sin\theta_\gamma, 0 \Bigg) , \\
		p_\ell^\dprime &= \Bigg( \frac{\sqrt{s_\ell}}{2} y, \begin{aligned}[t]
			& \frac{\sqrt{s_\ell}}{2} \sqrt{y^2 - 4 \zl} \left( \sin\theta_\gamma \sin\theta_{\ell\gamma} \cos\phi_\ell - \cos\theta_\gamma \cos\theta_{\ell\gamma} \right) , \\
			& \frac{\sqrt{s_\ell}}{2} \sqrt{y^2 - 4 \zl} \left( \cos\theta_\gamma \sin\theta_{\ell\gamma} \cos\phi_\ell + \sin\theta_\gamma \cos\theta_{\ell\gamma} \right) , \\
			& \frac{\sqrt{s_\ell}}{2} \sqrt{y^2 - 4 \zl} \sin\theta_{\ell\gamma} \sin\phi_\ell \Bigg) ,
			\end{aligned}
	\end{split}
\end{align}
where $\theta_{\ell\gamma}$ denotes the angle between photon and lepton in $\Sigma_{\ell\nu\gamma}$:
\begin{align}
	\begin{split}
		\cos\theta_{\ell\gamma} = \frac{x(y-2) + 2(1-y + \zl +\zg)}{\sqrt{x^2 - 4 \zg} \sqrt{y^2 - 4 \zl}} .
	\end{split}
\end{align}
The neutrino momentum is then easily found by $p_\nu^\dprime = L^\dprime - q^\dprime - p_\ell^\dprime$.

The momenta in the kaon rest frame $\Sigma_K$ are given by
\begin{align}
	\begin{split}
		q &= \Lambda_2 q^\dprime , \quad p_\ell = \Lambda_2 p_\ell^\dprime , \quad p_\nu = \Lambda_2 p_\nu^\dprime ,
	\end{split}
\end{align}
where $\Lambda_2$ is defined in (\ref{eqn:LorentzTransformation2}). I do not state here the expressions explicitly, as they are rather long. I use them to calculate in the following all the Lorentz invariant products in terms of the eight phase space variables.

The Lorentz invariant squares of the vectors (\ref{eqn:FourMomentaKl4g}) are
\begin{align}
	\begin{split}
		P^2 &= p_1^2 + 2p_1p_2 + p_2^2 = 2\mpip^2 + 2p_1p_2 = s , \\
		Q^2 &= p_1^2 - 2 p_1 p_2 + p_2^2 = 4\mpip^2 - s , \\
		L^2 &= (p_\ell+q)^2 + 2 (p_\ell+q) p_\nu + p_\nu^2 = \sg + 2(p_\ell+q) p_\nu = s_\ell , \\
		N^2 &= (p_\ell+q)^2 - 2 (p_\ell+q) p_\nu + p_\nu^2 = 2 \sg - s_\ell = s_\ell( 2x + 2y - 3) .
	\end{split}
\end{align}

The remaining Lorentz invariant products involving the vectors (\ref{eqn:FourMomentaKl4g}) are given by:
\begin{align}
	\begin{split}
		PQ ={}& 0 , \quad PL = \frac{1}{2}\left(\mkp^2 - s - s_\ell \right) , \quad QL = \sigma_\pi X \cos\theta_\pi , \quad LN = s_\ell(x+y-1) , \\
		PN ={}& (x+y-1) \frac{1}{2} \left(\mkp^2 - s - s_\ell\right) + X \Big( \sqrt{x^2 - 4 \zg} \cos\theta_\gamma \\
			&+ \sqrt{y^2 - 4 \zl} \left( \cos\theta_{\ell\gamma} \cos\theta_\gamma - \sin\theta_{\ell\gamma} \sin\theta_\gamma \cos\phi_\ell \right) \Big) , \\
		QN ={}& (x+y-1) \sigma_\pi X \cos\theta_\pi \\
			&+ \sigma_\pi \begin{aligned}[t] & \bigg\{ \frac{1}{2}\left(\mkp^2-s - s_\ell\right) \cos\theta_\pi \Big( \sqrt{x^2-4\zg} \cos\theta_\gamma \\
				&\qquad + \sqrt{y^2-4\zl} \left( \cos\theta_{\ell\gamma}\cos\theta_\gamma - \sin\theta_{\ell\gamma}\sin\theta_\gamma\cos\phi_\ell \right) \Big) \\
				& - \sqrt{s s_\ell} \sin\theta_\pi \begin{aligned}[t] &\bigg[ \cos\phi \Big( \sqrt{x^2 - 4 \zg} \sin\theta_\gamma \\
					& \qquad + \sqrt{y^2 - 4\zl} \left( \cos\theta_{\ell\gamma} \sin\theta_\gamma + \sin\theta_{\ell\gamma} \cos\theta_\gamma \cos\phi_\ell \right) \Big) \\
				& + \sin\phi \sqrt{y^2 - 4 \zl} \sin\theta_{\ell\gamma} \sin\phi_\ell \bigg] \bigg\} ,
				\end{aligned} 
			\end{aligned} \\
		\< LNPQ \> :={}& \epsilon_{\mu\nu\rho\sigma} L^\mu N^\nu P^\rho Q^\sigma \\
			={}& - \sigma_\pi X \sqrt{s s_\ell} \sin\theta_\pi 
				\begin{aligned}[t]
					&\Big( \sqrt{x^2 - 4 \zg} \sin\phi \sin\theta_\gamma \\
					&+ \sqrt{y^2 - 4 \zl}
					\begin{aligned}[t]
						& \big( \sin\phi \left( \cos\theta_{\ell\gamma} \sin\theta_\gamma + \sin\theta_{\ell\gamma}\cos\theta_\gamma \cos\phi_\ell \right) \\
						& - \cos\phi \sin\theta_{\ell\gamma} \sin\phi_\ell \big) \Big) .
					\end{aligned}
				\end{aligned}
	\end{split}
\end{align}
In addition, we need the Lorentz invariant products involving $q$:
\begin{align}
	\begin{split}
		Pq &= \frac{x}{4} \left( \mkp^2 - s - s_\ell \right) + \frac{X}{2} \sqrt{x^2 - 4 \zg} \cos\theta_\gamma , \\
		Qq &= \frac{\sigma_\pi}{2}
				\begin{aligned}[t]
					&\bigg[ x X \cos\theta_\pi + \sqrt{x^2 - 4 \zg} 
						\begin{aligned}[t]
							&\bigg( \cos\theta_\pi  \frac{1}{2} (\mkp^2-s-s_\ell) \cos\theta_\gamma  - \sin\theta_\pi \sqrt{s s_\ell} \sin\theta_\gamma \cos\phi \bigg) \bigg] ,
						\end{aligned} \\
				\end{aligned} \\
		Lq &= \frac{s_\ell}{2} x , \\
		Nq &= \frac{s_\ell}{2} \left( x + 2( y - 1 + \zg - \zl ) \right) , \\
		\< LNPq \> &= \frac{1}{2} X s_\ell \sqrt{x^2 - 4\zg} \sqrt{y^2 - 4 \zl} \sin\theta_{\ell\gamma} \sin\theta_\gamma \sin\phi_\ell , \\
		\< LNQq \> &= \frac{1}{2} \sigma_\pi s_\ell \sqrt{x^2-4\zg} \sqrt{y^2-4\zl} \sin\theta_{\ell\gamma} \\
			&\quad \cdot \begin{aligned}[t] 
				& \bigg( \frac{1}{2} ( \mkp^2 - s - s_\ell) \cos\theta_\pi \sin\theta_\gamma \sin\phi_\ell  - \sqrt{s s_\ell} \sin\theta_\pi \left( \sin\phi \cos\phi_\ell - \cos\phi \sin\phi_\ell \cos\theta_\gamma \right) \bigg) ,
				\end{aligned} \\
		\< LPQq \> &= - \frac{1}{2} \sigma_\pi X \sqrt{s s_\ell} \sin\theta_\pi \sqrt{x^2 - 4\zg} \sin\theta_\gamma \sin\phi , \\
		\< NPQq \> &= \frac{1}{2} \sigma_\pi \sqrt{s s_\ell} \\
				& \cdot 
				\begin{aligned}[t]
					&\bigg\{ \sqrt{x^2 - 4\zg} \sqrt{y^2-4\zl} \sin\theta_{\ell\gamma}  \bigg(  - \sqrt{s s_\ell} \cos\theta_\pi \sin\theta_\gamma \sin\phi_\ell \\
					& \quad  + \frac{1}{2}(\mkp^2-s-s_\ell) \sin\theta_\pi \left( \sin\phi \cos\phi_\ell - \cos\phi \sin\phi_\ell \cos\theta_\gamma \right) \bigg) \\
					& + X \sin\theta_\pi
					\begin{aligned}[t]
						&\bigg( x \sqrt{y^2-4\zl}  \big( - \sin\theta_{\ell\gamma} \left( \cos\phi \sin\phi_\ell - \sin\phi \cos\phi_\ell \cos\theta_\gamma \right)  + \cos\theta_{\ell\gamma} \sin\phi \sin\theta_\gamma \big) \\
						& - (y-1) \sqrt{x^2-4\zg} \sin\phi \sin\theta_\gamma \bigg) \bigg\} .
					\end{aligned}
				\end{aligned}
	\end{split}
\end{align}


\section{Decay Rates}

\label{sec:DecayRates}

\subsection{Decay Rate for $K_{\ell4}$}

\subsubsection{Isospin Limit}

\label{sec:DecayRateIsospinLimit}

The partial decay rate for the $K_{\ell4}$ decay is given by
\begin{align}
	d\Gamma = \frac{1}{2\mkp(2\pi)^8} \sum_\mathrm{spins} | \mathcal{T} |^2 \delta^{(4)}(p - P - L) \frac{d^3 p_1}{2p_1^0} \frac{d^3 p_2}{2p_2^0} \frac{d^3 p_\ell}{2p_\ell^0} \frac{d^3 p_\nu}{2p_\nu^0}.
\end{align}
The kinematics of the decay is described by the 5 variables $s$, $s_\ell$, $\theta_\pi$, $\theta_\ell$ and $\phi$. The remaining 7 integrals can be performed explicitly \cite{Cabibbo1965}. Let us review the reduction of the partial decay rate to the five-dimensional phase space integral.

The spin summed square of the matrix element
\begin{align}
	\begin{split}
		\mathcal{T} &= \frac{G_F}{\sqrt{2}} V_{us}^* \bar u(p_\nu) \gamma_\mu (1-\gamma^5)v(p_\ell) \; \big\< \pi^+(p_1) \pi^-(p_2) \big| \bar s \gamma^\mu(1-\gamma^5) u \big| K^+(p) \big\> \\
			&= \frac{G_F}{\sqrt{2}} V_{us}^* \; \mathcal{L}_\mu \mathcal{H}^\mu ,
	\end{split}
\end{align}
where $\mathcal{L}_\mu :=  \bar u(p_\nu) \gamma_\mu (1-\gamma^5)v(p_\ell)$ and $\mathcal{H}^\mu := \big\< \pi^+(p_1) \pi^-(p_2) \big| \bar s \gamma^\mu(1-\gamma^5) u \big| K^+(p) \big\>$, can be written as
\begin{align}
	\begin{split}
		\sum_\mathrm{spins} | \mathcal{T} |^2 &= \frac{G_F^2 |V_{us}|^2}{2} \mathcal{H}^\mu {\mathcal{H}^*}^\nu \sum_\mathrm{spins} \mathcal{L}_\mu \mathcal{L}_\nu^* .
	\end{split}
\end{align}
The spin sum can be performed with standard trace techniques:
\begin{align}
	\begin{split}
		4 \mathcal{L}_{\mu\nu} := \sum_\mathrm{spins} \mathcal{L}_\mu \mathcal{L}_\nu^* &= \sum_\mathrm{spins} \bar u(p_\nu) \gamma_\mu (1-\gamma^5)v(p_\ell) \bar v(p_\ell) \gamma_\nu (1-\gamma^5)u(p_\nu) \\
			&= \tr\left[ \slashed p_\nu \gamma_\mu (1-\gamma^5) (\slashed p_\ell - m_\ell) \gamma_\nu (1-\gamma^5) \right] \\
			&= -2 g_{\mu\nu} ( L^2 - N^2 ) + 4 (L_\mu L_\nu - N_\mu N_\nu) + 4 i \epsilon_{\mu\nu\rho\sigma} L^\rho N^\sigma \\
			&= 4 \left( g_{\mu\nu} ( m_\ell^2 - s_\ell ) + L_\mu L_\nu - N_\mu N_\nu  + i \epsilon_{\mu\nu\rho\sigma} L^\rho N^\sigma \right) .
	\end{split}
\end{align}
After the contraction with the hadronic matrix element, expressed in terms of the form factors,
\begin{align}
	\mathcal{H}^\mu &= -\frac{H}{\mkp^3} \epsilon_{\mu\nu\rho\sigma} L^\nu P^\rho Q^\sigma  + i \frac{1}{\mkp} \left( P_\mu F + Q_\mu G + L_\mu R \right) ,
\end{align}
all the scalar products can be expressed in terms of the five phase space variables $s$, $s_\ell$, $\theta_\pi$, $\theta_\ell$ and $\phi$.

Let us now consider the phase space measure:
\begin{align}
	\begin{split}
		dI :={}& \delta^{(4)}(p-p_1-p_2-p_\ell-p_\nu) \frac{d^3 p_1}{2p_1^0} \frac{d^3 p_2}{2p_2^0} \frac{d^3 p_\ell}{2p_\ell^0} \frac{d^3 p_\nu}{2p_\nu^0} \\
		={}& \delta^{(4)}(p-p_1-p_2-p_\ell-p_\nu) \frac{d^3 p_1}{2p_1^0} \frac{d^3 p_2}{2p_2^0} \frac{d^3 p_\ell}{2p_\ell^0} \frac{d^3 p_\nu}{2p_\nu^0} \\
		&\qquad\cdot \delta^{(4)}(p_1 + p_2 - P) \delta^{(4)}(p_\ell+p_\nu - L) d^4P d^4L \; \theta(P^0) \theta(L^0) \\
		={}& ds \, ds_\ell \, \delta^{(4)}(p-P-L) \, d^4P \delta(s - P^2) \theta(P^0)  d^4L \delta(s_\ell - L^2) \theta(L^0) \\
		&\qquad\cdot \delta^{(4)}(p_1+p_2-P)  \frac{d^3 p_1}{2p_1^0} \frac{d^3 p_2}{2p_2^0} \, \delta^{(4)}(p_\ell+p_\nu - L) \frac{d^3 p_\ell}{2p_\ell^0} \frac{d^3 p_\nu}{2p_\nu^0} .
	\end{split}
\end{align}
The phase space integral can be split into three separately Lorentz invariant pieces:
\begin{align}
	\begin{split}
		dI ={}& dI_1 dI_2 dI_3 , \\
		dI_1 :={}& ds \, ds_\ell \, \delta^{(4)}(p-P-L) \, d^4P \delta(s - P^2) \theta(P^0)  d^4L \delta(s_\ell - L^2) \theta(L^0) , \\
		dI_2 :={}& \delta^{(4)}(p_1+p_2-P)  \frac{d^3 p_1}{2p_1^0} \frac{d^3 p_2}{2p_2^0} , \\
		dI_3 :={}& \delta^{(4)}(p_\ell+p_\nu - L) \frac{d^3 p_\ell}{2p_\ell^0} \frac{d^3 p_\nu}{2p_\nu^0} .
	\end{split}
\end{align}
Each of these three pieces can be evaluated in a convenient frame. For $dI_1$, I choose the kaon rest frame:
\begin{align}
	\begin{split}
		dI_1 &= ds \, ds_\ell \, \delta^{(3)}(\vec p - \vec P - \vec L) \delta\left(p^0 - \sqrt{\vec P^2 + s} - \sqrt{\vec L^2 + s_\ell} \right) \frac{d^3P}{2\sqrt{\vec P^2 + s}} \frac{d^3L}{2\sqrt{\vec L^2 + s_\ell}} \\
		&= ds \, ds_\ell \, \delta^{(3)}(\vec P + \vec L) \delta\left(\mkp - \sqrt{\vec P^2 + s} - \sqrt{\vec L^2 + s_\ell} \right) \frac{d^3P}{2\sqrt{\vec P^2 + s}} \frac{d^3L}{2\sqrt{\vec L^2 + s_\ell}} \\
		&= ds \, ds_\ell \, \delta\left(\mkp - \sqrt{\vec P^2 + s} - \sqrt{\vec P^2 + s_\ell} \right) \frac{d^3P}{2\sqrt{\vec P^2 + s}} \frac{1}{2\sqrt{\vec P^2 + s_\ell}} \\
		&= \pi ds \, ds_\ell \, \delta\left(\mkp - \sqrt{\vec P^2 + s} - \sqrt{\vec P^2 + s_\ell} \right) \frac{\vec P^2}{\sqrt{\vec P^2 + s} \sqrt{\vec P^2 + s_\ell}} d|\vec P| \\
		&= \pi ds \, ds_\ell \, \delta\left( |\vec P| - \frac{\lambda^{1/2}(\mkp^2,s,s_\ell)}{2 \mkp} \right) \frac{|\vec P|}{\sqrt{\vec P^2 + s}+\sqrt{\vec P^2 + s_\ell}}  d|\vec P| \\
		&= \pi ds \, ds_\ell \, \frac{\lambda^{1/2}_{K\ell}(s)}{2\mkp^2} = \pi ds \, ds_\ell \, \frac{X}{\mkp^2} .
	\end{split}
\end{align}
I have used that the integrand depends on $\vec P$ only through $\vec P^2$.

The second piece is evaluated in the dipion frame:
\begin{align}
	\begin{split}
		dI_2 &= \delta^{(3)}(\vec p_1 + \vec p_2 - \vec P) \delta\left(\sqrt{\vec p_1^2 + \mpip^2} + \sqrt{\vec p_2^2 + \mpip^2} - P^0\right) \frac{d^3 p_1}{2\sqrt{\vec p_1^2 + \mpip^2}} \frac{d^3 p_2}{2\sqrt{\vec p_2^2 + \mpip^2}} \\
			&= \delta^{(3)}(\vec p_1 + \vec p_2) \delta\left(\sqrt{\vec p_1^2 + \mpip^2} + \sqrt{\vec p_2^2 + \mpip^2} - \sqrt{s}\right) \frac{d^3 p_1}{2\sqrt{\vec p_1^2 + \mpip^2}} \frac{d^3 p_2}{2\sqrt{\vec p_2^2 + \mpip^2}} \\
			&= \delta\left(2\sqrt{\vec p_1^2 + \mpip^2} - \sqrt{s}\right) \frac{d^3 p_1}{4 (\vec p_1^2 + \mpip^2)} = \delta\left( |\vec p_1| - \sqrt{\frac{s}{4} - \mpip^2} \right) \frac{\pi}{4} \sigma_\pi(s) d\cos\theta_\pi d |\vec p_1| \\
			&= \frac{\pi}{4} \sigma_\pi(s) d\cos\theta_\pi ,
	\end{split}
\end{align}
and the third piece analogously in the dilepton frame:
\begin{align}
	\begin{split}
		dI_3 &= \delta^{(3)}(\vec p_\ell + \vec p_\nu - \vec L) \delta\left(\sqrt{\vec p_\ell^2 + \ml^2} + |\vec p_\nu| - L^0\right) \frac{d^3 p_\ell}{2\sqrt{\vec p_\ell^2 + \ml^2}} \frac{d^3 p_\nu}{2|\vec p_\nu|} \\
			&= \delta^{(3)}(\vec p_\ell + \vec p_\nu) \delta\left(\sqrt{\vec p_\ell^2 + \ml^2} + |\vec p_\nu| - \sqrt{s_\ell}\right) \frac{d^3 p_\ell}{2\sqrt{\vec p_\ell^2 + \ml^2}} \frac{d^3 p_\nu}{2|\vec p_\nu|} \\
			&= \delta\left(\sqrt{\vec p_\ell^2 + \ml^2} + |\vec p_\ell| - \sqrt{s_\ell}\right) \frac{d^3 p_\ell}{4|\vec p_\ell|\sqrt{\vec p_\ell^2 + \ml^2}} = \delta\left( |\vec p_\ell| - \frac{s_\ell-\ml^2}{2\sqrt{s_\ell}} \right) \frac{1}{8}(1 - z_\ell) d\cos\theta_\ell d\phi d |\vec p_\ell| \\
			&= \frac{1}{8} (1-z_\ell) d\cos\theta_\ell d\phi .
	\end{split}
\end{align}
Putting the three pieces together, I find
\begin{align}
	\begin{split}
		dI = \frac{\lambda^{1/2}_{K\ell}(s)}{\mkp^2} \frac{\pi^2}{64} (1-z_\ell) \sigma_\pi(s) \, ds \, ds_\ell \, d\cos\theta_\pi \, d\cos\theta_\ell \, d\phi ,
	\end{split}
\end{align}
and for the differential decay rate
\begin{align}
	\begin{split}
		d\Gamma &= \frac{1}{2^{15}\pi^6} \frac{\lambda^{1/2}_{K\ell}(s)}{\mkp^3} (1-z_\ell) \sigma_\pi(s)  \sum_\mathrm{spins} | \mathcal{T} |^2 \, ds \, ds_\ell \, d\cos\theta_\pi \, d\cos\theta_\ell \, d\phi \\
		&= G_F^2 |V_{us}|^2 \frac{(1-z_\ell) \sigma_\pi(s)X}{2^{13}\pi^6\mkp^3} \mathcal{H}^\mu {\mathcal{H}^*}^\nu \mathcal{L}_{\mu\nu} \, ds \, ds_\ell \, d\cos\theta_\pi \, d\cos\theta_\ell \, d\phi  \\
		&=: G_F^2 |V_{us}|^2 \frac{(1-z_\ell) \sigma_\pi(s)X}{2^{13}\pi^6\mkp^5} J_5(s,s_\ell,\theta_\pi,\theta_\ell,\phi) \, ds \, ds_\ell \, d\cos\theta_\pi \, d\cos\theta_\ell \, d\phi .
	\end{split}
\end{align}
A rather tedious calculation yields (in accordance with \cite{Bijnens1994})
\begin{align}
	\begin{split}
		J_5(s,s_\ell,\theta_\pi,\theta_\ell,\phi) &= \mkp^2 \mathcal{H}^\mu {\mathcal{H}^*}^\nu \mathcal{L}_{\mu\nu} \\
		&= 2(1-z_\ell) \begin{aligned}[t]
			& \bigg[ I_1 + I_2 \cos(2\theta_\ell) + I_3 \sin^2(\theta_\ell) \cos(2\phi) + I_4 \sin(2\theta_\ell) \cos(\phi) \\
			& + I_5 \sin(\theta_\ell) \cos(\phi) + I_6 \cos(\theta_\ell) + I_7 \sin(\theta_\ell) \sin(\phi) + I_8 \sin(2\theta_\ell) \sin(\phi) \\
			& + I_9 \sin^2(\theta_\ell) \sin(2\phi) \bigg] , \end{aligned}
	\end{split}
\end{align}
where
\begin{align}
	\begin{split}
		\label{eqn:DecayRateFormFactorsIsoLimit}
		I_1 :={}& \frac{1}{4} \left((1+z_\ell) |F_1|^2+\frac{1}{2} (3+z_\ell) \sin^2(\theta_\pi) \left(|F_2|^2+|F_3|^2\right)+2 z_\ell |F_4|^2\right) , \\
		I_2 :={}& -\frac{1}{4}(1-z_\ell) \left(|F_1|^2-\frac{1}{2} \sin^2(\theta_\pi)\left(|F_2|^2+|F_3|^2\right)\right) , \\
		I_3 :={}& -\frac{1}{4} (1-z_\ell) \sin^2(\theta_\pi) \left(|F_2|^2-|F_3|^2\right) , \\
		I_4 :={}& \frac{1}{2} (1-z_\ell) \sin(\theta_\pi) \Re\left(F_1^*F_2\right) , \\
		I_5 :={}& -\sin(\theta_\pi) \left(\Re\left(F_1^* F_3\right)+z_\ell \Re\left(F_2^*F_4\right)\right) , \\
		I_6 :={}& z_\ell \Re\left(F_1^*F_4\right)-\sin^2(\theta_\pi) \Re\left(F_2^*F_3\right) , \\
		I_7 :={}& \sin(\theta_\pi) \left(z_\ell \Im\left(F_3^*F_4\right)-\Im\left(F_1^*F_2\right)\right) , \\
		I_8 :={}& \frac{1}{2} (1-z_\ell) \sin(\theta_\pi) \Im\left(F_1^*F_3\right) , \\
		I_9 :={}& -\frac{1}{2} (1-z_\ell) \sin^2(\theta_\pi) \Im\left(F_2^*F_3\right) .
	\end{split}
\end{align}

\subsubsection{Broken Isospin}

\label{sec:DecayRateIsospinBroken}

In the case of broken isospin, the Lorentz structure of the $K_{\ell4}$ matrix element is modified by the presence of the additional tensorial form factor. The expression for the spin sum has to be adapted. This is, however, the only necessary modification. The phase space is still parametrised by the same five kinematic variables.

The $T$-matrix element is given by (see also (\ref{eqn:TMatrixBrokenIsospin}))
\begin{align}
	\begin{split}
		\mathcal{T} &= \frac{G_F}{\sqrt{2}} V_{us}^* \left( \bar u(p_\nu) \gamma_\mu (1-\gamma^5)v(p_\ell) \mathcal{H}^\mu + \bar u(p_\nu) \sigma_{\mu\nu}(1+\gamma^5) v(p_\ell) \mathcal{T}^{\mu\nu} \right), \\
		\mathcal{H}^\mu &= \mathcal{V}^\mu - \mathcal{A}^\mu , \quad \mathcal{T}^{\mu\nu} = \frac{1}{\mkp^2} p_1^\mu p_2^\nu T .
	\end{split}
\end{align}

Let us calculate the spin sum of the squared $T$-matrix:
\begin{align}
	\begin{split}
		\sum_\mathrm{spins} | \mathcal{T} |^2 = \frac{G_F^2 |V_{us}|^2}{2} &\Bigg( \mathcal{H}^\mu \mathcal{H^*}^\nu\sum_\mathrm{spins} \mathcal{L}_\mu \mathcal{L}_\nu^* + \mathcal{T}^{\mu\nu} \mathcal{T^*}^{\rho\sigma} \sum_\mathrm{spins} \mathcal{\hat L}_{\mu\nu} \mathcal{\hat L^*}_{\rho\sigma} + 2 \Re \bigg[ \mathcal{H}^\mu \mathcal{T^*}^{\rho\sigma} \sum_\mathrm{spins} \mathcal{L}_\mu \mathcal{\hat L^*}_{\rho\sigma} \bigg]  \Bigg) ,
	\end{split}
\end{align}
where again $\mathcal{L}_\mu =  \bar u(p_\nu) \gamma_\mu (1-\gamma^5)v(p_\ell)$ and $\mathcal{\hat L}_{\mu\nu} := \bar u(p_\nu) \sigma_{\mu\nu}(1+\gamma^5) v(p_\ell)$.

The differential decay rate is given by
\begin{align}
	\begin{split}
		d\Gamma &= \frac{1}{2^{15}\pi^6} \frac{\lambda^{1/2}_{K\ell}(s)}{\mkp^3} (1-z_\ell) \sigma_\pi(s)  \sum_\mathrm{spins} | \mathcal{T} |^2 \, ds \, ds_\ell \, d\cos\theta_\pi \, d\cos\theta_\ell \, d\phi \\
		&=: G_F^2 |V_{us}|^2 \frac{(1-z_\ell) \sigma_\pi(s)X}{2^{13}\pi^6\mkp^5} J_5(s,s_\ell,\theta_\pi,\theta_\ell,\phi) \, ds \, ds_\ell \, d\cos\theta_\pi \, d\cos\theta_\ell \, d\phi ,
	\end{split}
\end{align}
where now
\begin{align}
	\begin{split}
		J_5 :={}& J_5^{V-A} + J_5^T + J_5^\mathrm{int} , \\
		J_5^{V-A} :={}&  \frac{\mkp^2}{4} \mathcal{H}^\mu \mathcal{H^*}^\nu \sum_\mathrm{spins} \mathcal{L}_\mu \mathcal{L}_\nu^* , \\
		J_5^{T} :={}&  \frac{\mkp^2}{4} \mathcal{T}^{\mu\nu} \mathcal{T^*}^{\rho\sigma} \sum_\mathrm{spins} \mathcal{\hat L}_{\mu\nu} \mathcal{\hat L^*}_{\rho\sigma} , \\
		J_5^\mathrm{int} :={}&  \frac{\mkp^2}{2} \Re \bigg[ \mathcal{H}^\mu \mathcal{T^*}^{\rho\sigma} \sum_\mathrm{spins} \mathcal{L}_\mu \mathcal{\hat L^*}_{\rho\sigma} \bigg] .
	\end{split}
\end{align}
$J_5^{V-A}$ agrees with $J_5$ in the isospin limit, but with the form factors $F_1$, \ldots, $F_4$ replaced by the isospin corrected ones. $J_5^T$ is due to the tensorial form factor only, $J_5^\mathrm{int}$ is the interference of the tensorial and the $V-A$ part.

$J_5$ can still be written in the form
\begin{align}
	\begin{split}
		J_5(s,s_\ell,\theta_\pi,\theta_\ell,\phi)
		&= 2(1-z_\ell) \begin{aligned}[t]
			&\bigg[ I_1 + I_2 \cos(2\theta_\ell) + I_3 \sin^2(\theta_\ell) \cos(2\phi) + I_4 \sin(2\theta_\ell) \cos(\phi) \\
			& + I_5 \sin(\theta_\ell) \cos(\phi) + I_6 \cos(\theta_\ell) + I_7 \sin(\theta_\ell) \sin(\phi) + I_8 \sin(2\theta_\ell) \sin(\phi) \\
			& + I_9 \sin^2(\theta_\ell) \sin(2\phi) \bigg] , \end{aligned}
	\end{split}
\end{align}
where $I_i = I_i^{V-A} + I_i^T + I_i^\mathrm{int}$. $I_i^{V-A}$ correspond to the functions $I_i$ in the isospin limit (\ref{eqn:DecayRateFormFactorsIsoLimit}). The additional pieces are given by
\begin{align}
	\begin{split}
		\label{eqn:DecayRateFormFactorsIsoBrokenTensorial}
		I_1^T ={}& \frac{1}{4} z_\ell \left((1+z_\ell)+\sin^2(\theta_\pi) \left( (1+3z_\ell) \frac{X^2}{s s_\ell} - \frac{1}{2}(1-z_\ell) \right) \right) |F_5|^2 , \\
		I_2^T ={}& \frac{1}{4} z_\ell (1-z_\ell) \left( 1 - \sin^2(\theta_\pi) \left( \frac{X^2}{s s_\ell} + \frac{3}{2} \right) \right) |F_5|^2 , \\
		I_3^T ={}& \frac{1}{4} z_\ell (1-z_\ell) \sin^2(\theta_\pi) |F_5|^2 , \\
		I_4^T ={}& -\frac{1}{4} z_\ell (1-z_\ell) \sin(2\theta_\pi) \frac{PL}{\sqrt{s s_\ell}} |F_5|^2 , \\
		I_5^T ={}& -\frac{1}{2} z_\ell^2 \sin(2\theta_\pi) \frac{X}{\sqrt{s s_\ell}} |F_5|^2 , \\
		I_6^T ={}& -z_\ell^2 \sin^2(\theta_\pi) \frac{PL \; X}{s s_\ell} |F_5|^2 , \\
		I_7^T ={}& I_8^T = I_9^T = 0
	\end{split}
\end{align}
and
\begin{align}
	\begin{split}
		\label{eqn:DecayRateFormFactorsIsoBrokenInterference}
		I_1^\mathrm{int} ={}& z_\ell \bigg( -\cos(\theta_\pi) \Re(F_1^* F_5) - \frac{PL}{\sqrt{s s_\ell}} \sin^2(\theta_\pi) \Re(F_2^* F_5) - \frac{X}{\sqrt{s s_\ell}} \sin^2(\theta_\pi) \Re(F_3^* F_5) \bigg)  , \\
		I_2^\mathrm{int} ={}& I_3^\mathrm{int} = I_4^\mathrm{int} = 0 , \\
		I_5^\mathrm{int} ={}& z_\ell \bigg( \frac{X}{\sqrt{s s_\ell}} \sin(\theta_\pi) \Re(F_1^* F_5) + \sin(\theta_\pi) \cos(\theta_\pi) \Re(F_3^* F_5) - \frac{PL}{\sqrt{s s_\ell}} \sin(\theta_\pi) \Re(F_4^* F_5) \bigg)  , \\
		I_6^\mathrm{int} ={}& z_\ell \bigg( \frac{X}{\sqrt{s s_\ell}} \sin^2(\theta_\pi) \Re(F_2^* F_5) + \frac{PL}{\sqrt{s s_\ell}} \sin^2(\theta_\pi) \Re(F_3^* F_5) - \cos(\theta_\pi) \Re(F_4^* F_5) \bigg) , \\
		I_7^\mathrm{int} ={}& z_\ell \bigg( \frac{PL}{\sqrt{s s_\ell}} \sin(\theta_\pi) \Im(F_1^* F_5) - \sin(\theta_\pi)\cos(\theta_\pi) \Im(F_2^* F_5) + \frac{X}{\sqrt{s s_\ell}} \sin(\theta_\pi) \Im(F_4^* F_5) \bigg) , \\
		I_8^\mathrm{int} ={}& I_9^\mathrm{int} = 0 .
	\end{split}
\end{align}
These results agree with \cite{Cuplov2004} apart from the different normalisation of $F_5$.

\subsection{Decay Rate for $K_{\ell4\gamma}$}

\label{sec:RadiativeDecayRate}

The partial decay rate for the $K_{\ell4\gamma}$ decay is given by
\begin{align}
	d\Gamma_\gamma = \frac{1}{2\mkp(2\pi)^{11}} \sum_{\substack{\mathrm{spins} \\ \mathrm{polar.}}} | \mathcal{T}_\gamma |^2 \delta^{(4)}(p - P - L) \frac{d^3 p_1}{2p_1^0} \frac{d^3 p_2}{2p_2^0} \frac{d^3 p_\ell}{2p_\ell^0} \frac{d^3 p_\nu}{2p_\nu^0} \frac{d^3 q}{2q^0}.
\end{align}
The kinematics of the decay is described by the 8 variables $s$, $s_\ell$, $\theta_\pi$, $\theta_\gamma$, $\phi$, $x$, $y$ and $\phi_\ell$. The remaining 7 integrals can be performed explicitly. The reduction of the partial decay rate to the eight-dimensional phase space integral is performed in the following.

The spin summed square of the matrix element
\begin{align}
	\begin{split}
		\mathcal{T}_\gamma &= - \frac{G_F}{\sqrt{2}} e V_{us}^* \epsilon_\mu(q)^* \bigg[ \mathcal{H}^{\mu\nu} \; \mathcal{L}_\nu +  \mathcal{H}_\nu \; \mathcal{\tilde L}^{\mu\nu} \bigg] ,
	\end{split}
\end{align}
where
\begin{align}
	\begin{split}
		\mathcal{L}_\nu &:=  \bar u(p_\nu) \gamma_\nu (1-\gamma^5)v(p_\ell) , \\
		\mathcal{\tilde L}^{\mu\nu} &:= \frac{1}{2 p_\ell q} \bar u(p_\nu) \gamma^\nu (1-\gamma^5)(m_\ell - \slashed p_\ell - \slashed q) \gamma^\mu v(p_\ell) ,
	\end{split}
\end{align}
can be written as
\begin{align}
	\begin{split}
		\sum_{\substack{\mathrm{spins}\\ \mathrm{polar.}}} | \mathcal{T}_\gamma |^2 &= \frac{e^2 G_F^2 |V_{us}|^2}{2} \sum_\mathrm{polar.} \epsilon_\mu(q)^* \epsilon_\rho(q)
			\begin{aligned}[t]
				&\Bigg[ \mathcal{H}_\nu \mathcal{H}^*_\sigma \sum_\mathrm{spins} \mathcal{\tilde L}^{\mu\nu} \mathcal{\tilde L}^{*\rho\sigma} + \mathcal{H}^{\mu\nu} \mathcal{H}^{*\rho\sigma} \sum_\mathrm{spins} \mathcal{L}_\nu \mathcal{L}^*_\sigma \\
				&+ 2 \Re\bigg( \mathcal{H}^{\mu\nu} \mathcal{H}^{*\sigma} \sum_\mathrm{spins} \mathcal{L}_\nu \mathcal{\tilde L}^{*\rho}{}_{\sigma} \bigg) \Bigg] .
			\end{aligned}
	\end{split}
\end{align}
All the spin sums can be performed with standard trace techniques. As I give the photon an artificial small mass $m_\gamma$, I have to use the polarisation sum formula for a massive vector boson:
\begin{align}
	\begin{split}
		\sum_\mathrm{polar.}  \epsilon_\mu(q)^* \epsilon_\rho(q) &= - g_{\mu\rho} + \frac{q_\mu q_\rho}{m_\gamma^2} .
	\end{split}
\end{align}
Using the Ward identity, I find that the second term in the polarisation sum formula does only contribute at~$\O(m_\gamma^2)$:
\begin{align}
	\begin{split}
		& \frac{q_\mu q_\rho}{m_\gamma^2}
			\begin{aligned}[t]
				\Bigg[ \mathcal{H}_\nu \mathcal{H}^*_\sigma & \sum_\mathrm{spins} \mathcal{\tilde L}^{\mu\nu} \mathcal{\tilde L}^{*\rho\sigma} + \mathcal{H}^{\mu\nu} \mathcal{H}^{*\rho\sigma} \sum_\mathrm{spins} \mathcal{L}_\nu \mathcal{L}^*_\sigma + 2 \Re\bigg( \mathcal{H}^{\mu\nu} \mathcal{H}^{*\sigma} \sum_\mathrm{spins} \mathcal{L}_\nu \mathcal{\tilde L}^{*\rho}{}_{\sigma} \bigg) \Bigg]
			\end{aligned} \\
			&= \frac{1}{m_\gamma^2} \Re \Bigg[ \mathcal{H}^\nu \mathcal{H}^{*\sigma} \sum_\mathrm{spins} \left( q^\mu q^\rho \mathcal{\tilde L}_{\mu\nu} \mathcal{\tilde L}_{\rho\sigma}^* + \mathcal{L}_\nu \mathcal{L}_\sigma^* + 2 q^\rho \mathcal{L}_\nu \mathcal{\tilde L}_{\rho\sigma}^* \right) \Bigg] \\
			&= \frac{4 m_\gamma^2}{( \hat L q + \hat N q )^2} \Re \Bigg[ \mathcal{H}^\nu \mathcal{H}^{*\sigma} \left( g_{\nu\sigma} \frac{\hat N^2 - \hat L^2}{2} + \hat L_\nu \hat L_\sigma - \hat N_\nu \hat N_\sigma + i \epsilon_{\nu\sigma\alpha\beta} \hat L^\alpha \hat N^\beta \right) \Bigg].
	\end{split}
\end{align}
I therefore find the following results for the spin and polarisation sums:
\begin{align}
	\begin{split}
		\sum_{\substack{\mathrm{spins} \\ \mathrm{polar.}}}  \epsilon_\mu(q)^* \epsilon_\rho(q) \mathcal{\tilde L}^{\mu\nu} \mathcal{\tilde L}^{*\rho\sigma} &= \frac{8}{Lq + Nq}
			\begin{aligned}[t]
				&\bigg( g^{\nu\sigma} (Nq - Lq) + q^\nu L^\sigma + q^\sigma L^\nu - q^\nu N^\sigma - q^\sigma N^\nu \\
				&+ i \epsilon^{\nu\sigma\alpha\beta} L_\alpha q_\beta - i \epsilon^{\nu\sigma\alpha\beta} N_\alpha q_\beta \bigg) \\
			\end{aligned} \\
			& - \frac{16 \ml^2}{(Lq + Nq)^2} \cdot \Big(g^{\nu\sigma} \frac{N^2-L^2}{2} + L^\nu L^\sigma - N^\nu N^\sigma + i \epsilon^{\nu\sigma\alpha\beta} L_\alpha N_\beta \Big) + \O(m_\gamma^2) ,
	\end{split}
\end{align}
\begin{align}
	\begin{split}
		\sum_{\substack{\mathrm{spins} \\ \mathrm{polar.}}}  \epsilon_\mu(q)^* \epsilon_\rho(q) \mathcal{L}_\nu \mathcal{L}_\sigma^* &= - 4 g_{\mu\rho} \bigg( g_{\nu\sigma} \frac{\hat N^2 - \hat L^2}{2} + \hat L_\nu \hat L_\sigma - \hat N_\nu \hat N_\sigma + i \epsilon_{\nu\sigma\alpha\beta} \hat L^\alpha \hat N^\beta \bigg) + \O(m_\gamma^2) , \\
	\end{split}
\end{align}
\begin{align}
	\begin{split}
		\sum_{\substack{\mathrm{spins} \\ \mathrm{polar.}}}  \epsilon_\mu(q)^* \epsilon_\rho(q) \mathcal{L}_\nu \mathcal{\tilde L}^{*\rho}{}_\sigma &= \frac{4}{Lq + Nq} \Bigg[ L_\mu L_\nu L_\sigma - N_\mu N_\nu N_\sigma + N_\mu L_\nu L_\sigma - L_\mu N_\nu N_\sigma \\
			& - q_\mu L_\nu L_\sigma + q_\mu N_\nu N_\sigma - q_\nu L_\mu L_\sigma + q_\nu N_\mu N_\sigma + q_\sigma L_\mu N_\nu - q_\sigma L_\nu N_\mu \\
			& + g_{\mu\nu} \left( \frac{N^2-L^2}{2} q_\sigma - Nq \, L_\sigma + Lq \, N_\sigma \right) + g_{\mu\sigma} \left( \frac{L^2-N^2}{2} q_\nu - Lq \, L_\nu + Nq \, N_\nu \right) \\
			& + g_{\nu\sigma} \left( \frac{N^2-L^2}{2} (L_\mu + N_\mu - q_\mu) + Lq \, L_\mu - Nq \, N_\mu \right) - i g_{\nu\sigma} \epsilon_{\mu\alpha\beta\gamma} L^\alpha N^\beta q^\gamma \\
			& + (L_\sigma - N_\sigma) \frac{i}{2} \epsilon_{\mu\nu\alpha\beta} (L^\alpha + N^\alpha) q^\beta + (L_\nu - N_\nu) \frac{i}{2} \epsilon_{\mu\sigma\alpha\beta} (L^\alpha + N^\alpha) q^\beta \\
			& + (L_\mu + N_\mu) \frac{i}{2} \epsilon_{\nu\sigma\alpha\beta} (-L^\alpha + N^\alpha) q^\beta + (L_\mu + N_\mu - q_\mu) i \epsilon_{\nu\sigma\alpha\beta} L^\alpha N^\beta \\
			& + \frac{i}{2}\epsilon_{\mu\nu\sigma\alpha} (L^\alpha - N^\alpha) (Lq + Nq)  \Bigg] + \O(m_\gamma^2) .
	\end{split}
\end{align}

I perform the contraction with the hadronic part and express all the scalar products in terms of the eight phase space variables. Neglecting the contribution form the anomalous sector, one can express the hadronic matrix elements in terms of the following form factors:
\begin{align}
	\begin{split}
		\mathcal{H}^\mu &= \frac{i}{\mkp} \left( P^\mu F + Q^\mu G + L^\mu R \right) , \\
		\mathcal{H}^{\mu\nu} &= \frac{i}{\mkp} g^{\mu\nu} \Pi + \frac{i}{\mkp^2}\left( P^\mu \Pi_0^\nu + Q^\mu \Pi_1^\nu + L^\mu \Pi_2^\nu \right) , \\
		\Pi_i^\nu &= \frac{1}{\mkp} \left( P^\nu \Pi_{i0} + Q^\nu \Pi_{i1} + L^\nu \Pi_{i2} + q^\nu \Pi_{i3}  \right) .
	\end{split}
\end{align}
The $K_{\ell4}$ form factors $F$, $G$, $R$ depend on scalar products of $P$, $Q$ and $L$, hence, they can be expressed as functions of $s$, $s_\ell$ and $\theta_\pi$. The $K_{\ell4\gamma}$ form factors $\Pi$ and $\Pi_{ij}$ depend on the scalar products of $P$, $Q$, $L$ and $q$. They are therefore functions of the six phase space variables $s$, $s_\ell$, $\theta_\pi$, $\theta_\gamma$, $\phi$ and $x$.

I consider now the phase space measure:
\begin{align}
	\begin{split}
		dI_\gamma :={}& \delta^{(4)}(p-p_1-p_2-p_\ell-p_\nu-q) \frac{d^3 p_1}{2p_1^0} \frac{d^3 p_2}{2p_2^0} \frac{d^3 p_\ell}{2p_\ell^0} \frac{d^3 p_\nu}{2p_\nu^0} \frac{d^3q}{2q^0} \\
		={}& \delta^{(4)}(p-p_1-p_2-p_\ell-p_\nu-q) \frac{d^3 p_1}{2p_1^0} \frac{d^3 p_2}{2p_2^0} \frac{d^3 p_\ell}{2p_\ell^0} \frac{d^3 p_\nu}{2p_\nu^0}\frac{d^3q}{2q^0} \\
		&\qquad \cdot \delta^{(4)}(p_1 + p_2 - P) \delta^{(4)}(p_\ell+p_\nu+q - L) d^4P d^4L \; \theta(P^0) \theta(L^0) \\
		={}& ds \, ds_\ell \, \delta^{(4)}(p-P-L) \, d^4P \delta(s - P^2) \theta(P^0)  d^4L \delta(s_\ell - L^2) \theta(L^0) \\
		&\qquad \cdot \delta^{(4)}(p_1+p_2-P)  \frac{d^3 p_1}{2p_1^0} \frac{d^3 p_2}{2p_2^0} \, \delta^{(4)}(p_\ell+p_\nu+q - L) \frac{d^3 p_\ell}{2p_\ell^0} \frac{d^3 p_\nu}{2p_\nu^0} \frac{d^3q}{2q^0} .
	\end{split}
\end{align}
The phase space integral can again be split into three separately Lorentz invariant pieces:
\begin{align}
	\begin{split}
		dI_\gamma ={}& dI_1^\gamma dI_2^\gamma dI_3^\gamma , \\
		dI_1^\gamma :={}& ds \, ds_\ell \, \delta^{(4)}(p-P-L) \, d^4P \delta(s - P^2) \theta(P^0)  d^4L \delta(s_\ell - L^2) \theta(L^0) , \\
		dI_2^\gamma :={}& \delta^{(4)}(p_1+p_2-P)  \frac{d^3 p_1}{2p_1^0} \frac{d^3 p_2}{2p_2^0} , \\
		dI_3^\gamma :={}& \delta^{(4)}(p_\ell+p_\nu+q - L) \frac{d^3 p_\ell}{2p_\ell^0} \frac{d^3 p_\nu}{2p_\nu^0} \frac{d^3q}{2q^0} .
	\end{split}
\end{align}
Each of these three pieces can be evaluated in a convenient frame. $dI_1^\gamma$ and $dI_2^\gamma$ can be evaluated in complete analogy to $K_{\ell4}$, i.e.~in the kaon and dipion rest frames:
\begin{align}
	\begin{split}
		dI_1^\gamma &=  \pi ds \, ds_\ell \, \frac{\lambda^{1/2}_{K\ell}(s)}{2\mkp^2} = \pi ds \, ds_\ell \, \frac{X}{\mkp^2} , \quad dI_2^\gamma = \frac{\pi}{4} \sigma_\pi(s) d\cos\theta_\pi .
	\end{split}
\end{align}

The third piece represents now a three body decay. I first perform the neutrino momentum integrals in the three body rest frame:
\begin{align}
	\begin{split}
		dI_3^\gamma &= \delta^{(3)}(\vec p_\ell + \vec p_\nu + \vec q - \vec L) \delta\left(\sqrt{\vec p_\ell^2 + \ml^2} + |\vec p_\nu| + \sqrt{\vec q^2 + \mg^2} - L^0\right) \frac{d^3 p_\ell}{2\sqrt{\vec p_\ell^2 + \ml^2}} \frac{d^3 p_\nu}{2|\vec p_\nu|} \frac{d^3q}{2\sqrt{\vec q^2 + \mg^2}} \\
			&= \delta\left(\sqrt{\vec p_\ell^2 + \ml^2} + |\vec p_\ell + \vec q| + \sqrt{\vec q^2 + \mg^2} - \sqrt{s_\ell} \right) \frac{d^3 p_\ell}{2\sqrt{\vec p_\ell^2 + \ml^2}} \frac{1}{2|\vec p_\ell + \vec q|} \frac{d^3q}{2\sqrt{\vec q^2 + \mg^2}} \\
			&= \delta\left(\sqrt{|\vec p_\ell|^2 + \ml^2} + \sqrt{ |\vec p_\ell|^2 + |\vec q|^2 + 2 |\vec p_\ell| |\vec q| \cos\theta_{\ell\gamma} } + \sqrt{|\vec q|^2 + \mg^2} - \sqrt{s_\ell} \right) \\
				&\quad \cdot \frac{|\vec p_\ell|^2 d |\vec p_\ell| d\cos\theta_{\ell\gamma} d\phi_\ell |\vec q|^2 d|\vec q| d\cos\theta_\gamma d\phi}{8\sqrt{|\vec p_\ell|^2 + \ml^2} \sqrt{ |\vec p_\ell|^2 + |\vec q|^2 + 2 |\vec p_\ell| |\vec q| \cos\theta_{\ell\gamma} } \sqrt{|\vec q|^2 + \mg^2}} \\
			&= \frac{|\vec p_\ell| |\vec q|}{8\sqrt{|\vec p_\ell|^2 + \mg^2}\sqrt{|\vec q|^2 + \mg^2}} d|\vec p_\ell| d|\vec q| d\phi_\ell d\cos\theta_\gamma d\phi \\
			&= \frac{1}{8} d p_\ell^0 dq^0 d\phi_\ell d\cos\theta_\gamma d\phi = \frac{s_\ell}{32} dx dy d\phi_\ell d\cos\theta_\gamma d\phi ,
	\end{split}
\end{align}
where I have used the angle $\theta_{\ell\gamma}$ between the photon and the lepton.

Putting the three pieces together, I find
\begin{align}
	\begin{split}
		dI_\gamma &= \frac{\lambda^{1/2}_{K\ell}(s)}{\mkp^2} \frac{\pi^2}{256} \sigma_\pi(s) s_\ell \, ds \, ds_\ell \, d\cos\theta_\pi \, d\cos\theta_\gamma \, d\phi \, dx \, dy \, d\phi_\ell , 
	\end{split}
\end{align}
and for the differential decay rate
\begin{align}
	\begin{split}
		d\Gamma_\gamma &= \frac{1}{2\mkp(2\pi)^{11}} \sum_{\substack{\mathrm{spins} \\ \mathrm{polar.}}} | \mathcal{T}_\gamma |^2 dI_\gamma \\
			&= \frac{1}{2^{20}\pi^9} \frac{\lambda^{1/2}_{K\ell}(s)}{\mkp^3} \sigma_\pi(s) s_\ell  \sum_{\substack{\mathrm{spins} \\ \mathrm{polar.}}} | \mathcal{T}_\gamma |^2 \, ds \, ds_\ell \, d\cos\theta_\pi \, d\cos\theta_\gamma \, d\phi \, dx \, dy \, d\phi_\ell \\
			&= G_F^2 |V_{us}|^2 e^2 \frac{s_\ell \, \sigma_\pi(s) X}{2^{20}\pi^9 \mkp^7} J_8 \, ds \, ds_\ell \, d\cos\theta_\pi \, d\cos\theta_\gamma \, d\phi \, dx \, dy \, d\phi_\ell ,
	\end{split}
\end{align}
where
\begin{align}
	\begin{split}
		J_8 &= \mkp^4 \sum_\mathrm{polar.} \epsilon_\mu(q)^* \epsilon_\rho(q)
			\begin{aligned}[t]
				&\Bigg[ \mathcal{H}_\nu \mathcal{H}^*_\sigma \sum_\mathrm{spins} \mathcal{\tilde L}^{\mu\nu} \mathcal{\tilde L}^{*\rho\sigma} + \mathcal{H}^{\mu\nu} \mathcal{H}^{*\rho\sigma} \sum_\mathrm{spins} \mathcal{L}_\nu \mathcal{L}^*_\sigma + 2 \Re\bigg( \mathcal{H}^{\mu\nu} \mathcal{H}^{*\sigma} \sum_\mathrm{spins} \mathcal{L}_\nu \mathcal{\tilde L}^{*\rho}{}_{\sigma} \bigg) \Bigg] .
			\end{aligned}
	\end{split}
\end{align}


\section{\ChPT{} with Photons and Leptons}

\label{sec:AppendixChPT}

In order to settle the conventions, I collect here the most important formulae needed to define \ChPT{} with photons and leptons \cite{Weinberg1968, GasserLeutwyler1984, GasserLeutwyler1985, Urech1995,Knecht2000}.

We consider $SU(3)$ \ChPT{}, where the Goldstone bosons are collected in the $SU(3)$ matrix
\begin{align}
	\begin{split}
		U = \exp\left( \frac{i \sqrt{2}}{F_0} \phi \right) ,
	\end{split}
\end{align}
with
\begin{align}
	\begin{split}
		\phi &= \sum_{a=1}^8 \lambda_a \phi_a = \left( \begin{matrix}
				\pi^0 \left( \frac{1}{\sqrt{2}} + \frac{\epsilon}{\sqrt{6}} \right) + \eta \left( \frac{1}{\sqrt{6}} - \frac{\epsilon}{\sqrt{2}} \right) & \pi^+ & K^+ \\
				\pi^- & \pi^0 \left(  \frac{\epsilon}{\sqrt{6}} - \frac{1}{\sqrt{2}} \right) + \eta \left( \frac{1}{\sqrt{6}} + \frac{\epsilon}{\sqrt{2}} \right) & K^0 \\
				K^- & \bar K^0 & - \eta \sqrt{\frac{2}{3}} - \pi^0 \sqrt{\frac{2}{3}} \epsilon
			 \end{matrix} \right) .
	\end{split}
\end{align}
At leading order, the Lagrangian is given by\footnote{I denote by $\< \cdot \>$ the flavour trace.}
\begin{align}
	\begin{split}
		\mathcal{L}_\mathrm{eff}^\mathrm{LO} &= \mathcal{L}_{p^2} + \mathcal{L}_{e^2} + \mathcal{L}_\mathrm{QED}, \\
		\mathcal{L}_{p^2} &= \frac{F_0^2}{4} \< D_\mu U D^\mu U^\dagger + \chi U^\dagger + U \chi^\dagger \> , \\
		\mathcal{L}_{e^2} &= e^2 F_0^4 Z \< U Q U^\dagger Q \> , \\
		\mathcal{L}_\mathrm{QED} &= - \frac{1}{4} F_{\mu\nu} F^{\mu\nu} + \sum_{\ell} \left[ \bar \ell (i \slashed \p + e \slashed A - m_\ell) \ell + \bar \nu_{\ell L} i \slashed \p \nu_{\ell L} \right] ,
	\end{split}
\end{align}
where
\begin{align}
	\begin{split}
		D_\mu U &= \p_\mu U - i r_\mu U + i U l_\mu , \\
		\chi &= 2 B_0 (s + i p) , \quad r_\mu = v_\mu + a_\mu , \quad l_\mu = v_\mu - a_\mu , \\
		F_{\mu\nu} &= \p_\mu A_\nu - \p_\nu A_\mu , \\
		\nu_{\ell L} &= \frac{1-\gamma_5}{2} \nu_\ell .
	\end{split}
\end{align}
The external fields are fixed by
\begin{align}
	\begin{split}
		s + i p &= \mathcal{M} = \mathrm{diag}( m_u, m_d, m_s ) , \\
		r_\mu &= -e A_\mu Q , \\
		l_\mu &= -e A_\mu Q  +  \sum_{\ell} \left( \bar \ell \gamma_\mu \nu_{\ell L} Q_L^w + \bar \nu_{\ell L} \gamma_\mu \ell Q_L^{w\dagger} \right) , \\
		Q &= \frac{1}{3} \mathrm{diag}( 2, -1, -1 ) , \\
		Q_L^w &= -2 \sqrt{2} G_F T, \quad T = \left( \begin{matrix} 0 & V_{ud} & V_{us} \\ 0 & 0 & 0 \\ 0 & 0 & 0 \end{matrix} \right) .
	\end{split}
\end{align}

By expanding $\mathcal{L}_\mathrm{eff}^\mathrm{LO}$ in the meson fields, we can extract the mass terms. At leading order, I find:
\begin{align}
	\begin{split}
		\mpio^2 &= 2 B_0 \hat m , \\
		\mpip^2 &= 2 B_0 \hat m + 2 e^2 Z F_0^2 , \\
		\mko^2 &= B_0\left( m_s + \hat m + \frac{2\epsilon}{\sqrt{3}}( m_s - \hat m ) \right) , \\
		\mkp^2 &= B_0\left( m_s + \hat m - \frac{2\epsilon}{\sqrt{3}}( m_s - \hat m ) \right) + 2 e^2 Z F_0^2 , \\
		\meta^2 &= \frac{4}{3} B_0 \left( m_s + \frac{\hat m}{2} \right) .
	\end{split}
\end{align}
At this order, the masses obey the Gell-Mann -- Okubo relation:
\begin{align}
	\begin{split}
		2 \mkp^2 + 2 \mko^2 - 2 \mpip^2 + \mpio^2 = 3 \meta^2.
	\end{split}
\end{align}
Let us define
\begin{align}
	\label{eqn:LOMassDifferences}
	\begin{split}
		\Delta_\pi &:= \mpip^2 - \mpio^2 = 2 e^2 Z F_0^2 , \\
		\Delta_K &:= \mkp^2 - \mko^2 = 2 e^2 Z F_0^2 + B_0 (m_u - m_d) .
	\end{split}
\end{align}

The next-to-leading-order Lagrangian is given by
\begin{align}
	\begin{split}
		\mathcal{L}_\mathrm{eff}^\mathrm{NLO} &= \mathcal{L}_\mathrm{eff}^\mathrm{LO} + \mathcal{L}_{p^4} + \mathcal{L}_{e^2 p^2} + \mathcal{L}_\mathrm{lept} + \mathcal{L}_\gamma,
	\end{split}
\end{align}
where
\begin{align}
	\begin{split}
		\mathcal{L}_{p^4} &= L_1 \< D_\mu U D^\mu U^\dagger \> \<D_\nu U D^\nu U^\dagger \> + L_2 \< D_\mu U D_\nu U^\dagger \> \< D^\mu U D^\nu U^\dagger \> \\
			&+ L_3 \< D_\mu U D^\mu U^\dagger D_\nu U D^\nu U^\dagger \> + L_4 \< D_\mu U D^\mu U^\dagger \> \< \chi U^\dagger + U \chi^\dagger \> \\
			&+ L_5 \< D_\mu U D^\mu U^\dagger ( \chi U^\dagger + U \chi^\dagger ) \> + L_6 \< \chi U^\dagger + U \chi^\dagger \>^2 + L_7 \< \chi U^\dagger - U \chi^\dagger \>^2 \\
			&+ L_8 \< U \chi^\dagger U \chi^\dagger + \chi U^\dagger \chi U^\dagger \> - i L_9  \< F_R^{\mu\nu} D_\mu U D_\nu U^\dagger + F_L^{\mu\nu} D_\mu U^\dagger D_\nu U \> + L_{10} \< U F_L^{\mu\nu} U^\dagger F^R_{\mu\nu} \> \\
			&+ H_1 \< F_R^{\mu\nu} F^R_{\mu\nu} + F_L^{\mu\nu} F^L_{\mu\nu} \> + H_2 \< \chi \chi^\dagger \> , \\
	\end{split}
\end{align}
\begin{align}
	\begin{split}
		\mathcal{L}_{e^2 p^2} &= e^2 F_0^2 \begin{aligned}[t]
			&\bigg\{ K_1 \< Q Q \> \< D_\mu U D^\mu U^\dagger \>  + K_2 \< Q U^\dagger Q U \> \< D_\mu U D^\mu U^\dagger \> \\
			& + K_3 \left( \< Q U^\dagger D_\mu U \> \< Q U^\dagger D^\mu U \> + \< Q U D_\mu U^\dagger \> \< Q U D^\mu U^\dagger \> \right) \\
			& + K_4 \< Q U^\dagger D_\mu U \> \< Q U D^\mu U^\dagger \>  + K_5 \< Q Q ( D_\mu U^\dagger D^\mu U + D_\mu U D^\mu U^\dagger) \> \\
			& + K_6 \< U Q U^\dagger Q D_\mu U D^\mu U^\dagger + U^\dagger Q U Q D_\mu U^\dagger D^\mu U \> + K_7 \< Q Q \> \< \chi U^\dagger + U \chi^\dagger \> \\
			& + K_8 \< Q U^\dagger Q U \> \< \chi U^\dagger + U \chi^\dagger \>  + K_9 \< Q Q ( U^\dagger \chi + \chi^\dagger U + \chi U^\dagger + U \chi^\dagger ) \> \\
			& + K_{10} \< Q U^\dagger Q \chi + Q U Q \chi^\dagger + Q U^\dagger Q U \chi^\dagger U + Q U Q U^\dagger \chi U^\dagger \> \\
			& - K_{11} \< Q U^\dagger Q \chi + Q U Q \chi^\dagger - Q U^\dagger Q U \chi^\dagger U - Q U Q U^\dagger \chi U^\dagger \> \\
			& + i K_{12} \< \big[ [ l_\mu , Q ] , Q \big] D^\mu U^\dagger U + \big[ [ r_\mu , Q ] , Q \big] D^\mu U U^\dagger \> \\
			& - K_{13} \< [ l_\mu , Q ] U^\dagger [ r^\mu , Q ] U \> + 2 K_{14} \< l_\mu [ l^\mu , Q ] Q + r_\mu [ r^\mu , Q ] Q \> \bigg\} ,
			\end{aligned} \\
	\end{split}
\end{align}
\begin{align}
	\begin{split}
		\mathcal{L}_\mathrm{lept} &= e^2 \sum_\ell \begin{aligned}[t]
			&\bigg\{ F_0^2 \begin{aligned}[t]
				&\Big[ X_1 \bar \ell \gamma_\mu \nu_{\ell L} i \< D^\mu U Q_L^w U^\dagger Q - D^\mu U^\dagger Q U Q_L^w \> \\
				& - X_2 \bar \ell \gamma_\mu \nu_{\ell L} i \< D^\mu U Q_L^w U^\dagger Q + D^\mu U^\dagger Q U Q_L^w \> \\
				& + X_3 m_\ell \bar \ell \nu_{\ell L} \< Q_L^w U^\dagger Q U \>  + X_4 \bar \ell \gamma_\mu \nu_{\ell L} \< Q_L^w l^\mu Q - Q_L^w Q l^\mu \> \\
				& + X_5 \bar \ell \gamma_\mu \nu_{\ell L} \< Q_L^w U^\dagger r^\mu Q U - Q_L^w U^\dagger Q r^\mu U \> + h.c. \Big]  + X_6 \bar \ell ( i \slashed \p + e \slashed A) \ell + X_7 m_\ell \bar \ell \ell \bigg\} ,
				\end{aligned}
			\end{aligned}
	\end{split}
\end{align}
\begin{align}
	\begin{split}
		\mathcal{L}_\gamma &= e^2 X_8 F_{\mu\nu} F^{\mu\nu} .
	\end{split}
\end{align}

The low-energy constants (LECs) are UV-divergent. Their finite part is defined by
\begin{align}
	\begin{split}
		\label{eqn:RenormalisedLECs}
		L_i &= \Gamma_i \lambda + L_i^r(\mu) , \\
		H_i &= \Delta_i \lambda + H_i^r(\mu) , \\
		K_i &= \Sigma_i \lambda + K_i^r(\mu) , \\
		X_i &= \Xi_i \lambda + X_i^r(\mu) ,
	\end{split}
\end{align}
where
\begin{align}
	\begin{split}
		\label{eqn:UVDivergenceLambda}
		\lambda = \frac{\mu^{n-4}}{16\pi^2} \left( \frac{1}{n-4} - \frac{1}{2} \left( \ln(4\pi) + 1 - \gamma_E \right) \right) .
	\end{split}
\end{align}
The coefficients $\Gamma_i$, $\Delta_i$, $\Sigma_i$ and $\Xi_i$ can be found in \cite{GasserLeutwyler1985, Urech1995, Knecht2000}.


\section{Feynman Diagrams}

\label{sec:AppendixDiagrams}

\subsection{Mass Effects}

\label{sec:AppendixDiagramsMassEffects}

\subsubsection{Loop Diagrams}
The meson loop diagrams contribute as follows to the form factors $F$ and $G$:
\begin{align}
	\begin{split}
		\delta F^\mathrm{NLO}_\mathrm{tadpole} &= \frac{1}{12 F_0^2} \left[ A_0(\mpio^2) + 4 A_0(\mpip^2) + 8 A_0(\mko^2) + 8 A_0(\mkp^2) + 9 A_0(\meta^2) \right] , \\
		\delta G^\mathrm{NLO}_\mathrm{tadpole} &= \frac{1}{4 F_0^2} \left[ A_0(\mpio^2) + 4 A_0(\mpip^2) + 4 A_0(\mkp^2) + A_0(\meta^2) \right] ,
	\end{split}
\end{align}
\begin{align}
	\begin{split}
		\delta F^\mathrm{NLO}_\text{$s$-loop} &=  \frac{1}{F_0^2} \begin{aligned}[t]
				& \bigg[ 3(s-\mpio^2) B_0(s,\mpio^2,\mpio^2)  + 3( s + 4 \Delta_\pi) B_0(s,\mpip^2,\mpip^2) \\
				& + \left( \frac{3}{2}s + \Delta_K - \Delta_\pi \right) B_0(s,\mko^2,\mko^2) + 3(4 \Delta_\pi + s) B_0(s,\mkp^2,\mkp^2) \\
				& + 3 \mpio^2 B_0(s,\meta^2,\meta^2) - 2 A_0(\mpio^2) - 2 A_0(\mpip^2) - A_0(\mko^2) - 2 A_0(\mkp^2) \\
				& + 2 \sqrt{3} \epsilon \begin{aligned}[t]
					&\Big( 3(s-\mpio^2) B_0(s,\mpio^2,\mpio^2) + \frac{2}{3}(\mko^2 - \mpio^2) B_0(s, \mko^2, \mko^2) \\
					& + ( 4 \mpio^2 - 3s) B_0(s,\meta^2,\mpio^2) - \mpio^2 B_0(s,\meta^2,\meta^2) - A_0(\mpio^2) + A_0(\meta^2) \Big) \bigg] ,
					\end{aligned}
			\end{aligned} \\
		\delta G^\mathrm{NLO}_\text{$s$-loop} &= \frac{1}{6 F_0^2} \begin{aligned}[t]
			& \bigg[ (s - 4 \mkp^2) B_0(s,\mkp^2,\mkp^2) - \frac{1}{2}(s - 4 \mko^2) B_0(s,\mko^2,\mko^2) \\
			& + (s - 4 \mpip^2) B_0(s,\mpip^2,\mpip^2) - 2 A_0(\mkp^2) + A_0(\mko^2) - 2 A_0(\mpip^2) \\
			& + \frac{2 \mko^2 - 4 \mkp^2 - 4 \mpip^2 + s}{16\pi^2} \bigg] ,
			\end{aligned}
	\end{split}
\end{align}
{ \small
\begin{align}
	\begin{split}
		\delta F^\mathrm{NLO}_\text{$t$-loop} &= \frac{1}{6 F_0^2} \Bigg[ \frac{1}{4 t^2} \begin{aligned}[t]
				& \bigg( \mkp^2 \left(2 t-6\meta^2\right)+6 \meta^2 \mpip^2+3 \meta^2 t \\
				& +6 \mko^2 \left(\mkp^2-\mpip^2-t\right)-3\mpio^2 t-2 \mpip^2 t \bigg)  \left(\mko^2-\meta^2\right) B_0\left(0,\meta^2,\mko^2\right) \end{aligned} \\
			& + \frac{1}{4 t^2} \begin{aligned}[t] 
				& \bigg(\mko^2 \left(2\mkp^2 \left(6 \meta^2-t\right)-3 \meta^2 \left(4 \mpip^2+3 t\right)+t \left(3 \mpio^2+2\mpip^2-12 t\right)\right) \\
				& + \left(\meta^2-t\right) \left(\mkp^2 \left(2 t-6 \meta^2\right)+3 \meta^2\left(2 \mpip^2+t\right)-t \left(3 \mpio^2+2 \mpip^2\right)\right) \\
				& + 6 \mko^4\left(-\mkp^2+\mpip^2+t\right)\bigg) B_0\left(t,\meta^2,\mko^2\right) \end{aligned} \\
			& + \frac{1}{2t^2} \begin{aligned}[t]
				& \bigg(\mko^2 \left(\mkp^2-\mpip^2\right)-\mkp^2\left(\mpio^2+2 t\right) \\
				& + \mpio^2 \mpip^2+3 \mpio^2 t+2 \mpip^2 t \bigg) \left(\mko^2-\mpio^2\right) B_0\left(0,\mko^2,\mpio^2\right) \end{aligned} \\
			& + \frac{1}{2t^2} \begin{aligned}[t] 
				& \bigg(\mko^2\left(2 \mkp^2 \left(\mpio^2+t\right)-\mpio^2 \left(2 \mpip^2+3 t\right)+t \left(3 t-2\mpip^2\right)\right) \\
				& + \mko^4 \left(\mpip^2-\mkp^2\right) - \left(\mpio^4+\mpio^2 t-2 t^2\right) \left(\mkp^2-\mpip^2-3 t\right)\bigg) B_0\left(t,\mko^2,\mpio^2\right) \end{aligned} \\
			& + \frac{1}{t^2} \begin{aligned}[t]
				& \bigg(\mkp^4-2 \mkp^2\left(\mpip^2+t\right)  + 3 \mpio^2 t+\mpip^4-\mpip^2 t\bigg) \left(\mkp^2-\mpip^2\right) B_0\left(0,\mkp^2,\mpip^2\right) \end{aligned} \\
			& + \frac{1}{t^2} \begin{aligned}[t]
				& \bigg(-\mkp^6+\mkp^4\left(3 \mpip^2+2 t\right)-\mkp^2 \left(t \left(3 \mpio^2+t\right)+3 \mpip^4\right) \\
				& + \mpip^2 t \left(3\mpio^2-5 t\right)+3 \mpio^2 t^2+\mpip^6-2 \mpip^4 t \bigg) B_0\left(t,\mkp^2,\mpip^2\right) \end{aligned} \\
			& - \frac{3 \left(\mkp^2-\mpip^2+t\right)}{2 t} A_0\left(\meta^2\right) + \frac{\left(-\mkp^2+\mpip^2+3t\right)}{2 t} A_0\left(\mpio^2\right) \\
			& + \frac{\left(\mpip^2-\mkp^2\right)}{t} A_0\left(\mpip^2\right) - A_0\left(\mkp^2\right) \\
			& + \frac{\left(\mkp^2-\mpip^2-3 t\right)\left(3 \meta^2+4 \mko^2+2 \mkp^2+\mpio^2+2 \mpip^2-2 t\right)}{64 \pi ^2 t}  \Bigg] \\
			& + \frac{1}{6 F_0^2} \sqrt{3} \epsilon \Bigg[ \frac{1}{9 t^2} \bigg(\mko^4-\mko^2 \left(2 \mpio^2+t\right)+\mpio^4-2\mpio^2 t\bigg) (\mko^2-\mpio^2) B_0(0,\meta^2,\mko^2)  \\
			& + \frac{1}{9 t^2} \begin{aligned}[t]
				& \bigg(-\mko^6+\mko^4 \left(3\mpio^2+13 t\right)-\mko^2 \left(3 \mpio^4+14 \mpio^2 t+57 t^2\right) \\
				& + \mpio^6+\mpio^4 t+3\mpio^2 t^2+27 t^3 \bigg) B_0(t,\meta^2,\mko^2) \end{aligned} \\
			& - \frac{1}{t^2} \bigg(\mko^4-\mko^2 \left(2 \mpio^2+t\right)+\mpio^4+2\mpio^2 t \bigg) (\mko^2-\mpio^2) B_0(0,\mko^2,\mpio^2) \\
			& + \frac{1}{t^2} \begin{aligned}[t]
				& \bigg(\mko^6-\mko^4 \left(3\mpio^2+t\right)+\mko^2 \left(3 \mpio^4+2 \mpio^2 t+t^2\right) \\
				& - \left(\mpio^2-t\right)^2\left(\mpio^2+3 t\right)\bigg) B_0(t,\mko^2,\mpio^2) \end{aligned} \\
			& - \frac{\left(\mko^2-\mpio^2+t\right)}{t} A_0(\meta^2) + \frac{\left(\mko^2-\mpio^2+t\right)}{t} A_0(\mpio^2) \\
			& + \frac{\left(\mko^2-\mpio^2\right) \left(\mko^2-\mpio^2-3t\right)}{24 \pi ^2 t} \Bigg] , \\
	\end{split}
\end{align}
\begin{align}
	\begin{split}
		\delta G^\mathrm{NLO}_\text{$t$-loop} &= \frac{1}{6 F_0^2} \Bigg[ \frac{1}{4 t^2} \begin{aligned}[t]
			& \bigg( \mkp^2 \left(6 \meta^2-2t\right)-6 \meta^2 \mpip^2+3 \meta^2 t-6 \mko^2 \left(\mkp^2-\mpip^2\right) \\
			& + 3 \mpio^2 t+2\mpip^2 t-6 t^2 \bigg) \left(\mko^2-\meta^2\right) B_0\left(0,\meta^2,\mko^2\right) \end{aligned} \\
		& + \frac{1}{4 t^2} \begin{aligned}[t]
			& \bigg( -\mko^2 \left(2 \mkp^2\left(6 \meta^2-t\right)+3 \meta^2 \left(t-4 \mpip^2\right)+t \left(3 \mpio^2+2\mpip^2\right)\right) \\
			& + \left(\meta^2-t\right) \left(\mkp^2 \left(6 \meta^2-2 t\right)+\meta^2 \left(3t-6 \mpip^2\right)+t \left(3 \mpio^2+2 \mpip^2-6 t\right)\right) \\
			& + 6 \mko^4\left(\mkp^2-\mpip^2\right) \bigg) B_0\left(t,\meta^2,\mko^2\right) \end{aligned} \\
		& - \frac{1}{2t^2} \begin{aligned}[t] 
			& \bigg( \mko^2 \left(\mkp^2-\mpip^2+t\right)-\mkp^2\left(\mpio^2+2 t\right) \\
			& + \mpio^2 \mpip^2+2 \mpio^2 t+2 \mpip^2 t+t^2 \bigg) \left(\mko^2-\mpio^2\right)B_0\left(0,\mko^2,\mpio^2\right) \end{aligned} \\
		& + \frac{1}{2t^2} \begin{aligned}[t]
			& \bigg( \mko^2\left(-2 \mkp^2 \left(\mpio^2+t\right)+\mpio^2 \left(2 \mpip^2+t\right)+t \left(2 \mpip^2-5t\right)\right) \\
			& - \left(\mpio^2-t\right) \left(\mkp^2 \left(-\left(\mpio^2+2 t\right)\right)+\mpio^2\left(\mpip^2+2 t\right)+t \left(2 \mpip^2+7 t\right)\right) \\
			& + \mko^4 \left(\mkp^2-\mpip^2+t\right) \bigg) B_0\left(t,\mko^2,\mpio^2\right) \end{aligned} \\
		& - \frac{1}{t^2} \begin{aligned}[t]
			& \bigg( \mkp^4-\mkp^2\left(2 \mpip^2+t\right)+t \left(3 \mpio^2+t\right) \\
			& + \mpip^4-2 \mpip^2t \bigg) \left(\mkp^2-\mpip^2\right) B_0\left(0,\mkp^2,\mpip^2\right) \end{aligned} \\
		& + \frac{1}{t^2} \begin{aligned}[t]
			& \bigg( \mkp^6-\mkp^4 \left(3 \mpip^2+t\right)+\mkp^2 \left(t \left(3\mpio^2-t\right)+3 \mpip^4-2 \mpip^2 t\right) \\
			& + 3 \mpip^2 t \left(t-\mpio^2\right)+t^2 \left(t-3\mpio^2\right)-\mpip^6+3 \mpip^4 t \bigg) B_0\left(t,\mkp^2,\mpip^2\right) \end{aligned} \\
		& - \frac{3\left(-\mkp^2+\mpip^2+t\right)}{2 t} A_0\left(\meta^2\right) + \frac{\left(\mkp^2-\mpip^2-5 t\right)}{2 t} A_0\left(\mpio^2\right) \\
		& + \frac{\left(\mkp^2-\mpip^2-2t\right)}{t} A_0\left(\mpip^2\right) + A_0\left(\mkp^2\right) \\
		& - \frac{\left(\mkp^2-\mpip^2+t\right) \left(3 \meta^2+4\mko^2+2 \mkp^2+\mpio^2+2 \mpip^2-2 t\right)}{64 \pi ^2 t} \Bigg] \\
		& + \frac{1}{6 F_0^2} \sqrt{3} \epsilon \Bigg[ \frac{1}{9 t^2} \bigg( \mko^4-2 \mko^2 \mpio^2+\mpio^4-3 \mpio^2 t-3 t^2 \bigg) \left(\mpio^2-\mko^2\right) B_0\left(0,\meta^2,\mko^2\right) \\
		& + \frac{1}{9 t^2} \begin{aligned}[t]
			& \bigg( \mko^6-3 \mko^4 \left(\mpio^2+4 t\right)+3 \mko^2 \left(\mpio^4+4 \mpio^2 t+5 t^2\right) \\
			& - \mpio^6+3 \mpio^2 t^2-18 t^3 \bigg) B_0\left(t,\meta^2,\mko^2\right) \end{aligned} \\
		& + \frac{1}{t^2} \bigg( \mko^4-2 \mko^2 \mpio^2+\mpio^4+\mpio^2 t+t^2 \bigg) \left(\mko^2-\mpio^2\right) B_0\left(0,\mko^2,\mpio^2\right) \\
		& + \frac{1}{t^2} \bigg( -\mko^6+3 \mko^4 \mpio^2+\mko^2 \left(t^2-3 \mpio^4\right)+\left(\mpio^2-t\right)^2 \left(\mpio^2+2 t\right)  \bigg) B_0\left(t,\mko^2,\mpio^2\right) \\
		& - \frac{\left(-\mko^2+\mpio^2+t\right)}{t} A_0\left(\meta^2\right) + \frac{\left(-\mko^2+\mpio^2+t\right)}{t} A_0\left(\mpio^2\right) \\
		& - \frac{\left(\mko^2-\mpio^2\right) \left(\mko^2-\mpio^2+t\right)}{24 \pi ^2 t} \Bigg] , 
	\end{split}
\end{align}
}
\begin{align}
	\begin{split}
		\delta F^\mathrm{NLO}_\text{$u$-loop} &= \delta G^\mathrm{NLO}_\text{$u$-loop} \\
			&= \frac{1}{2F_0^2} \begin{aligned}[t]
				& \bigg[ B_0(u,\mkp^2,\mpip^2) ( \mkp^2 + 3 \mpip^2 - 2 \mpio^2 - u) + \frac{1}{3} A_0(\mkp^2) + \frac{1}{3} A_0(\mpip^2) \bigg] .
				\end{aligned}
	\end{split}
\end{align}

\subsubsection{Counterterms}

The counterterm contribution to the form factors is given by:
\begin{align}
	\begin{split}
		\delta F^\mathrm{NLO}_\mathrm{ct} &= \frac{1}{F_0^2} \begin{aligned}[t]
				& \Bigg[ 32(s-2\mpip^2) L_1 + 8(\mkp^2+s-s_\ell) L_2 + 4(\mkp^2 - 3 \mpip^2 + 2 s -t) L_3 \\
				& + 8\bigg(2 \mko^2 + 5 \mpio^2 - \frac{4 \sqrt{3} \epsilon}{3} (\mko^2-\mpio^2)\bigg) L_4 \\
				& + 4(\mkp^2 + 2 \mpip^2 - 3 \Delta_\pi) L_5 + 2 s_\ell L_9 \Bigg]  + \frac{2}{9} e^2 \left( 84 K_2 + 37 K_6 \right) ,
				\end{aligned} \\
		\delta G^\mathrm{NLO}_\mathrm{ct} &= \frac{1}{F_0^2} \begin{aligned}[t]
				& \Bigg[ 8(t-u) L_2 - 4(\mkp^2 + \mpip^2 - t) L_3  + 8\bigg(2 \mko^2 + \mpio^2 - \frac{4 \sqrt{3} \epsilon}{3} (\mko^2 - \mpio^2) \bigg) L_4 \\
				& + 4(\mkp^2 + 2 \mpip^2 - 3 \Delta_\pi) L_5 + 2 s_\ell L_9 \Bigg] + \frac{2}{9} e^2 \left( 12 K_2 + 18 K_4 + 25 K_6 \right) .
				\end{aligned}
	\end{split}
\end{align}

\subsubsection{External Leg Corrections}
Let us first consider the pion self-energy: it is given by
\begin{align}
	\begin{split}
		\Sigma_{\pi^+}(p^2) = i ( \mathcal{D}_{\pi^+}^\mathrm{loop} + \mathcal{D}_{\pi^+}^\mathrm{ct} ) ,
	\end{split}
\end{align}
where $p$ denotes the external pion momentum.

The value of the tadpole diagram is
\begin{align}
	\begin{split}
		\mathcal{D}_{\pi^+}^\mathrm{loop} &= \frac{i}{6 F_0^2} \bigg[  p^2 \left( A_0(\mkp^2) + A_0(\mko^2) + 2 A_0(\mpip^2) + 2 A_0(\mpio^2) \right) \\
			& - \mpip^2 \left( A_0(\mkp^2) + A_0(\mko^2) + A_0(\meta^2) + 2 A_0(\mpip^2) - A_0(\mpio^2) \right) \bigg] \\
			& - \frac{i}{3} e^2 Z \left( 6 A_0(\mkp^2) - A_0(\meta^2) + 12 A_0(\mpip^2) + 3 A_0(\mpio^2) \right) ,
	\end{split}
\end{align}
and the counterterm is given by
\begin{align}
	\begin{split}
		\mathcal{D}_{\pi^+}^\mathrm{ct} &= p^2 \Bigg[ \frac{8 i}{F_0^2} \bigg( (2\mkp^2 - 2 \mpip^2 + 3 \mpio^2) L_4 + \mpio^2 L_5 + \frac{4 \sqrt{3} \epsilon}{3} (\mkp^2-\mpip^2) L_4 \bigg) + \frac{4i}{9} e^2 (6 K_2 + 5 K_6) \Bigg] \\
			& + \frac{16 i}{F_0^2} \begin{aligned}[t]
				& \bigg( (-2 \mpio^2 \mkp^2 + 3 \mpip^4 - 4 \mpio^2 \mpip^2) L_6 + \mpip^2( \mpip^2 - 2 \mpio^2) L_8 \\
				& - \frac{4 \sqrt{3} \epsilon}{3} \mpip^2(\mkp^2-\mpip^2) L_6 \bigg) - \frac{4i}{9} e^2 \left( 3(6\mkp^2+5\mpip^2) K_8 + 23 \mpip^2 K_{10} \right) . \end{aligned}
	\end{split}
\end{align}
Since the full propagator is
\begin{align}
	\begin{split}
		\frac{i}{p^2 - \mpip^2 - \Sigma_{\pi^+}(p^2)} = \frac{i Z_{\pi^+}}{p^2 - M_{\pi^+,\mathrm{ph}}^2} + \mathrm{regular} ,
	\end{split}
\end{align}
the field strength renormalisation $Z_{\pi^+}$ can be computed as
\begin{align}
	\begin{split}
		Z_{\pi^+} &= 1 + \Sigma_{\pi^+}^\prime(M_{\pi^+,\mathrm{ph}}^2) + h.o. =  1 + \Sigma_{\pi^+}^\prime(\mpip^2) + h.o. ,
	\end{split}
\end{align}
where $h.o.$ denotes higher order terms.

The physical mass, i.e.~the position of the pole is given by
\begin{align}
	\begin{split}
		M_{\pi^+,\mathrm{ph}}^2 = \mpip^2 + \delta \mpip^2 , \quad \delta \mpip^2 = \Sigma_{\pi^+}(M_{\pi^+,\mathrm{ph}}^2) = \Sigma_{\pi^+}(\mpip^2) + h.o.
	\end{split}
\end{align}

I find the following expression for the field strength renormalisation of the pion:
\begin{align}
	\begin{split}
		Z_{\pi^+} &= 1 - \frac{1}{F_0^2} \begin{aligned}[t] 
			& \bigg( \frac{1}{6} \left( A_0(\mko^2) + A_0(\mkp^2) + 2 A_0(\mpip^2) + 2 A_0(\mpio^2) \right) \\
			& + 8(2 \mkp^2 - 2 \mpip^2 + 3 \mpio^2) L_4 + 8 \mpio^2 L_5 \\
			& + \frac{32\sqrt{3} \epsilon}{3} (\mkp^2-\mpip^2) L_4 \bigg) - \frac{4}{9} e^2 \left( 6 K_2 + 5 K_6 \right) . \end{aligned}
	\end{split}
\end{align}

We still have to compute the external leg correction for the kaon. The values of the two self-energy diagrams for a charged kaon are given by
{ \small
\begin{align}
	\begin{split}
		\mathcal{D}_{K^+}^\mathrm{loop} &= p^2 \begin{aligned}[t] 
				&\Bigg[ \frac{i}{12F_0^2} \left( 2 A_0(\mko^2) + 4 A_0(\mkp^2) + 3 A_0(\meta^2) + 2 A_0(\mpip^2) + A_0(\mpio^2) \right) \\
				& - \frac{i \sqrt{3} \epsilon}{6 F_0^2} \left( A_0(\meta^2) - A_0(\mpio^2) \right) \Bigg] \end{aligned} \\
			& - \frac{i \mkp^2}{12 F_0^2} \begin{aligned}[t] 
				& \bigg( 2 A_0(\mko^2) + 4 A_0(\mkp^2) - A_0(\meta^2) + 2 A_0(\mpip^2) + A_0(\mpio^2) \\
				& + 2 \sqrt{3}\epsilon \left( A_0(\meta^2) - A_0(\mpio^2) \right) \bigg)  - \frac{2i}{3} e^2 Z \left( 6 A_0(\mkp^2) + A_0(\meta^2) + 3 A_0(\mpip^2) \right) , \end{aligned}
	\end{split}
\end{align}
\begin{align}
	\begin{split}
		\mathcal{D}_{K^+}^\mathrm{ct} &= p^2 \Bigg[ \frac{8 i}{F_0^2} \bigg( (2\mkp^2 - 2 \mpip^2 + 3 \mpio^2) L_4 + (\mkp^2 - \mpip^2 + \mpio^2) L_5 \\
			& + \frac{4 \sqrt{3} \epsilon}{3} (\mkp^2-\mpip^2) L_4 \bigg) + \frac{4i}{9} e^2 (6 K_2 + 5 K_6) \Bigg] \\
			& + \frac{16 i}{F_0^2} \bigg( (\mkp^2(4 \mpip^2 - 2 \mkp^2 - 5\mpio^2) + \mpip^2 \Delta_\pi) L_6 \\
			& - \mkp^2( \mkp^2 - 2 \Delta_\pi) L_8 - \frac{4 \sqrt{3} \epsilon}{3} \mkp^2(\mkp^2-\mpip^2) L_6 \bigg) \\
			& - \frac{4i}{9} e^2 \left( 3(8 \mkp^2 + 3 \mpip^2) K_8 + (20\mkp^2 + 3 \mpip^2) K_{10} \right) .
	\end{split}
\end{align}
}
The field strength renormalisation of the kaon is given by
\begin{align}
	\begin{split}
		Z_{K^+} &= 1 \begin{aligned}[t]
			&- \frac{1}{F_0^2} \begin{aligned}[t] 
				& \bigg( \frac{1}{12} \left( 2 A_0(\mko^2) + 4 A_0(\mkp^2) + 3 A_0(\meta^2) + 2 A_0(\mpip^2) + A_0(\mpio^2) \right) \\
				& + 8(2 \mkp^2 - 2 \mpip^2 + 3 \mpio^2) L_4 + 8 ( \mkp^2 - \Delta_\pi) L_5 \\
				& + \frac{32\sqrt{3} \epsilon}{3} (\mkp^2-\mpip^2) L_4 - \frac{\sqrt{3}\epsilon}{6} \left(A_0(\meta^2) - A_0(\mpio^2)\right) \bigg) \end{aligned} \\
			&- \frac{4}{9} e^2 \left( 6 K_2 + 5 K_6 \right) . \end{aligned}
	\end{split}
\end{align}

The contribution of the field strength renormalisation to the amplitude consists of the LO tree diagrams multiplied by a factor of $\sqrt{Z_i}$ for every external particle $i$. Therefore, the NLO external leg corrections to the form factors are given by
\begin{align}
	\begin{split}
		\delta F_Z^\mathrm{NLO} &= \delta G_Z^\mathrm{NLO} =  Z_{\pi^+} \sqrt{Z_{K^+}} - 1 \\
			&= \begin{aligned}[t]
				&- \frac{1}{F_0^2} \begin{aligned}[t]
					& \bigg( \frac{1}{24} \left( 6 A_0(\mko^2) + 8 A_0(\mkp^2) + 3 A_0(\meta^2) + 10 A_0(\mpip^2) + 9 A_0(\mpio^2) \right) \\
					& + 12(2\mkp^2 - 2 \mpip^2 + 3 \mpio^2) L_4 + 4(\mkp^2 - \mpip^2 + 3 \mpio^2) L_5 \\
					& - \frac{\sqrt{3} \epsilon}{12} \left( A_0(\meta^2) - A_0(\mpio^2) \right) + 16\sqrt{3} \epsilon (\mkp^2-\mpip^2) L_4 \bigg)
					\end{aligned} \\
				& - \frac{2}{3} e^2 ( 6 K_2 + 5 K_6 ) .
				\end{aligned}
	\end{split}
\end{align}

\subsection{Photonic Effects}

\label{sec:AppendixDiagramsPhotonicEffects}

\subsubsection{Loop Diagrams}

Here, I give the explicit expressions for the contributions of the loop diagrams shown in figures~\ref{img:Kl4_gLoops} and \ref{img:Kl4_mLoops} to the form factors $F$ and $G$.

The first four diagrams only contain bulb topologies. Their contribution, expressed in terms of the scalar loop function $B_0$, is given by
\begin{align}
	\begin{split}
		\delta F^\mathrm{NLO}_{\gamma-\mathrm{loop},a} &= \frac{4}{3} e^2 \left( B_0(0,\mkp^2,\mg^2) - 4 B_0(\mkp^2,\mkp^2,\mg^2) \right), \\
		\delta G^\mathrm{NLO}_{\gamma-\mathrm{loop},a} &= 0 , \\
		\delta F^\mathrm{NLO}_{\gamma-\mathrm{loop},b} &= \delta G^\mathrm{NLO}_{\gamma-\mathrm{loop},b} = -\delta F^\mathrm{NLO}_{\gamma-\mathrm{loop},c} = \delta G^\mathrm{NLO}_{\gamma-\mathrm{loop},c} =  \frac{2}{3} e^2 \left( B_0(0,\mpip^2,\mg^2) - 4 B_0(\mpip^2, \mpip^2, \mg^2) \right) , \\
		\delta F^\mathrm{NLO}_{\gamma-\mathrm{loop},d} &= \delta G^\mathrm{NLO}_{\gamma-\mathrm{loop},d} = 0 ,
	\end{split}
\end{align}
hence, in total
\begin{align}
	\begin{split}
		\delta F^\mathrm{NLO}_{\gamma-\mathrm{loop},a-d} &= \frac{4}{3} e^2 \left( B_0(0,\mkp^2,\mg^2) - 4 B_0(\mkp^2,\mkp^2,\mg^2) \right), \\
		\delta G^\mathrm{NLO}_{\gamma-\mathrm{loop},a-d} &= \frac{4}{3} e^2 \left( B_0(0,\mpip^2,\mg^2) - 4 B_0(\mpip^2, \mpip^2, \mg^2) \right) .
	\end{split}
\end{align}

The next six diagrams consist of triangle topologies. My results agree with \cite{Cuplov2004} up to the contribution of the additional term in the massive gauge boson propagator (which cancels in the end), though I choose to employ Passarino-Veltman reduction techniques to write everything in terms of the basic scalar loop functions $A_0$, $B_0$ and $C_0$, even if this results in longer expressions.
{
\small
\begin{align}
	\begin{split}
		\delta F^\mathrm{NLO}_{\gamma-\mathrm{loop},e} &= \frac{1}{2} e^2 \begin{aligned}[t]
			&\Bigg( 4(\mkp^2+\mpip^2-t) C_0(\mpip^2,t, \mkp^2,\mg^2,\mpip^2,\mkp^2) \\
			& + \begin{aligned}[t]
				& \bigg(3 \mkp^4 - 5 \mpip^4 - 6 \mkp^2 \mpip^2 - 2 t(3\mkp^2-\mpip^2) + 3 t^2 \bigg) \frac{2 B_0(\mpip^2,\mpip^2,\mg^2)}{\lambda(t,\mpip^2,\mkp^2)} \end{aligned} \\
			& + \bigg(\mkp^4+2\mkp^2(\mpip^2-3t) + 5(\mpip^2-t)^2 \bigg) \frac{2 B_0(\mkp^2,\mkp^2,\mg^2)}{\lambda(t,\mpip^2,\mkp^2)} \\
			& - \begin{aligned}[t]
				& \bigg( (\mkp^2-\mpip^2)^3 + t(\mkp^4 - 3 \mpip^4 + 2\mkp^2\mpip^2) \\
				& - t^2(13\mkp^2 + 7 \mpip^2) + 11 t^3 \bigg) \frac{B_0(t,\mpip^2,\mkp^2)}{t \lambda(t,\mpip^2,\mkp^2)} - B_0(0,\mpip^2,\mg^2) \end{aligned} \\
			& - 2 B_0(0,\mkp^2,\mg^2) + \frac{\mkp^2-\mpip^2}{t} B_0(0,\mpip^2,\mkp^2) \Bigg)  - e^2 \frac{A_0(\mg^2)}{\mg^2} , \end{aligned}
	\end{split}
\end{align}
\begin{align}
	\begin{split}
		\delta G^\mathrm{NLO}_{\gamma-\mathrm{loop},e} &= \frac{1}{2} e^2 \begin{aligned}[t]
			&\Bigg( 4(\mkp^2+\mpip^2-t) C_0(\mpip^2,t, \mkp^2,\mg^2,\mpip^2,\mkp^2) \\
			& + \bigg( 3 \mkp^4 + 2 \mkp^2(\mpip^2 - 3t) + 3(\mpip^2-t)^2 \bigg) \frac{2 B_0(\mpip^2,\mpip^2,\mg^2)}{\lambda(t,\mpip^2,\mkp^2)} \\
			& -  \bigg( 3\mkp^4+2\mkp^2(3\mpip^2-t) - (\mpip^2-t)^2 \bigg) \frac{2B_0(\mkp^2,\mkp^2,\mg^2)}{\lambda(t,\mpip^2,\mkp^2)} \\
			& + \begin{aligned}[t]
				& \bigg( (\mkp^2-\mpip^2)^3 + t(\mkp^4 - 3 \mpip^4 + 2\mkp^2\mpip^2) \\
				& + 3t^2(\mkp^2 + 3 \mpip^2) - 5 t^3 \bigg) \frac{B_0(t,\mpip^2,\mkp^2)}{t \lambda(t,\mpip^2,\mkp^2)} \end{aligned} \\
			& - B_0(0,\mpip^2,\mg^2) - \frac{\mkp^2-\mpip^2}{t} B_0(0,\mpip^2,\mkp^2) \Bigg) - e^2 \frac{A_0(\mg^2)}{\mg^2} , \end{aligned}
	\end{split}
\end{align}
\begin{align}
	\begin{split}
		\delta F^\mathrm{NLO}_{\gamma-\mathrm{loop},f} &= \delta G^\mathrm{NLO}_{\gamma-\mathrm{loop},f} \\
			&= - e^2 \begin{aligned}[t]
			& \bigg( 2(\mkp^2+\mpip^2-u) C_0(\mpip^2,u,\mkp^2,\mg^2,\mpip^2,\mkp^2) \\
			& + B_0(\mpip^2,\mpip^2,\mg^2) + B_0(\mkp^2,\mkp^2,\mg^2) - B_0(u,\mpip^2,\mkp^2)  \bigg)  + e^2 \frac{A_0(\mg^2)}{\mg^2} , \end{aligned}
	\end{split}
\end{align}
\begin{align}
	\begin{split}
		\delta F^\mathrm{NLO}_{\gamma-\mathrm{loop},g} &= e^2 \begin{aligned}[t]
			& \Bigg( 2(\mkp^2+\ml^2-\tl) C_0(\ml^2,\tl,\mkp^2,\mg^2,\ml^2,\mkp^2) \\
			& + \bigg( 3 \mkp^4 - \mkp^2 (5\ml^2 + 6 \tl) - \ml^2 \tl - 2 \ml^4 + 3 \tl^2 \bigg) \frac{2 B_0(\ml^2,\ml^2,\mg^2)}{3 \lambda(\tl,\ml^2,\mkp^2)} \\
			& + \bigg( \mkp^4 - 8 \mkp^2 \tl  + 7(\ml^2-\tl)^2 \bigg) \frac{B_0(\mkp^2,\mkp^2,\mg^2)}{3 \lambda(\tl,\ml^2,\mkp^2)} \\
			& + \bigg( \mkp^2(3\tl - 2 \ml^2) + \ml^2 \tl + 2 \ml^4 - 3 \tl^2 \bigg) \frac{2 B_0(\tl,\ml^2,\mkp^2)}{3 \lambda(\tl,\ml^2,\mkp^2)} \\
			& - \frac{1}{3} B_0(0,\mkp^2,\mg^2) \Bigg)  - e^2 \frac{A_0(\mg^2)}{\mg^2} , \end{aligned}
	\end{split}
\end{align}
\begin{align}
	\begin{split}
		\delta G^\mathrm{NLO}_{\gamma-\mathrm{loop},g} &= e^2 \begin{aligned}[t]
			& \Bigg( 2(\mkp^2+\ml^2-\tl) C_0(\ml^2,\tl,\mkp^2,\mg^2,\ml^2,\mkp^2) \\
			& + \bigg( \mkp^4 - \mkp^2 (\ml^2 + 2 \tl) + \tl(\tl-\ml^2) \bigg) \frac{2 B_0(\ml^2,\ml^2,\mg^2)}{\lambda(\tl,\ml^2,\mkp^2)} \\
			& - \bigg( \mkp^4  - (\ml^2-\tl)^2 \bigg) \frac{B_0(\mkp^2,\mkp^2,\mg^2)}{\lambda(\tl,\ml^2,\mkp^2)} \\
			& + \bigg( \mkp^2 + \ml^2 - \tl \bigg) \frac{2 \tl B_0(\tl,\ml^2,\mkp^2)}{\lambda(\tl,\ml^2,\mkp^2)} \Bigg)  - e^2 \frac{A_0(\mg^2)}{\mg^2} , \end{aligned}
	\end{split}
\end{align}
\begin{align}
	\begin{split}
		\delta F^\mathrm{NLO}_{\gamma-\mathrm{loop},h} &= e^2 \begin{aligned}[t]
			& \Bigg( 2(2\mpip^2-s) C_0(\mpip^2,s,\mpip^2,\mg^2,\mpip^2,\mpip^2) \\
			& + 2 B_0(\mpip^2,\mpip^2,\mg^2) - B_0(s,\mpip^2,\mpip^2) \Bigg)  - e^2 \frac{A_0(\mg^2)}{\mg^2} , \end{aligned}
	\end{split}
\end{align}
\begin{align}
	\begin{split}
		\delta G^\mathrm{NLO}_{\gamma-\mathrm{loop},h} &= e^2 \begin{aligned}[t]
			& \Bigg( 2(2\mpip^2-s) C_0(\mpip^2,s,\mpip^2,\mg^2,\mpip^2,\mpip^2) \\
			& + \frac{2(8\mpip^2-3s)}{4\mpip^2-s} B_0(\mpip^2,\mpip^2,\mg^2) \\
			& - \frac{4(2\mpip^2-s)}{4\mpip^2-s} B_0(s,\mpip^2,\mpip^2) - B_0(0,\mpip^2,\mg^2) \Bigg)  - e^2 \frac{A_0(\mg^2)}{\mg^2} , \end{aligned}
	\end{split}
\end{align}
\begin{align}
	\begin{split}
		\delta F^\mathrm{NLO}_{\gamma-\mathrm{loop},i} &= \delta G^\mathrm{NLO}_{\gamma-\mathrm{loop},i} \\
			&= e^2 \begin{aligned}[t]
			& \Bigg( -2(\mpip^2 + \ml^2 - s_{1\ell} ) C_0(\ml^2, s_{1\ell}, \mpip^2, \mg^2, \ml^2, \mpip^2) \\
			& + \bigg( 2 \ml^4 + \ml^2(s_{1\ell} + 5 \mpip^2) - 3(\mpip^2-s_{1\ell})^2  \bigg) \frac{2 B_0(\ml^2, \ml^2, \mg^2)}{3\lambda(s_{1\ell}, \ml^2, \mpip^2)} \\
			& + \bigg( \mpip^4 + 7 \ml^4 - 2 s_{1\ell}( 4 \mpip^2 + 7 \ml^2 ) + 7 s_{1\ell}^2 \bigg) \frac{B_0(\mpip^2, \mpip^2, \mg^2)}{3\lambda(s_{1\ell}, \ml^2, \mpip^2)} \\
			& - \bigg( 4 \ml^4 - 2 \ml^2(2 \mpip^2 - s_{1\ell}) - 6 s_{1\ell}(s_{1\ell}-\mpip^2) \bigg) \frac{B_0(s_{1\ell}, \ml^2, \mpip^2)}{3\lambda(s_{1\ell}, \ml^2, \mpip^2)} \\
			& + \frac{1}{3} B_0(0, \mpip^2, \mg^2) \Bigg) +  e^2 \frac{A_0(\mg^2)}{\mg^2} , \end{aligned}
	\end{split}
\end{align}
\begin{align}
	\begin{split}
		\delta F^\mathrm{NLO}_{\gamma-\mathrm{loop},j} &= e^2 \begin{aligned}[t]
			& \Bigg( 2 (\mpip^2 + \ml^2 - s_{2\ell}) C_0(\ml^2, s_{2\ell}, \mpip^2, \mg^2, \ml^2, \mpip^2) \\
			& + \bigg( 4 \ml^4 + \ml^2(\mpip^2 - 7 s_{2\ell}) + 3 (\mpip^2-s_{2\ell})^2 \bigg) \frac{2 B_0(\ml^2, \ml^2, \mg^2)}{3 \lambda(s_{2\ell}, \ml^2, \mpip^2)} \\
			& + \bigg( \ml^4 - 5 \mpip^4 - 12 \ml^2 \mpip^2 - 2s_{2\ell}(\ml^2 - 2 \mpip^2) + s_{2\ell}^2 \bigg) \frac{B_0(\mpip^2, \mpip^2, \mg^2)}{3\lambda(s_{2\ell}, \ml^2, \mpip^2)} \\
			& - \bigg( 8 \ml^4  - 2 \ml^2(7s_{2\ell} + 4\mpip^2) - 6 s_{2\ell}(\mpip^2-s_{2\ell}) \bigg) \frac{B_0(s_{2\ell}, \ml^2, \mpip^2)}{3\lambda(s_{2\ell}, \ml^2, \mpip^2)} \\
			& + \frac{1}{6} B_0(0, \mpip^2, \mg^2) \Bigg) - e^2 \frac{A_0(\mg^2)}{\mg^2} , \end{aligned}
	\end{split}
\end{align}
\begin{align}
	\begin{split}
		\delta G^\mathrm{NLO}_{\gamma-\mathrm{loop},j} &= e^2 \begin{aligned}[t]
			& \Bigg( 2 (\mpip^2 + \ml^2 - s_{2\ell}) C_0(\ml^2, s_{2\ell}, \mpip^2, \mg^2, \ml^2, \mpip^2) \\
			& + \bigg( 2 \ml^4 - \ml^2(\mpip^2 + 5 s_{2\ell}) + 3 (\mpip^2-s_{2\ell})^2 \bigg) \frac{2 B_0(\ml^2, \ml^2, \mg^2)}{3 \lambda(s_{2\ell}, \ml^2, \mpip^2)} \\
			& + \bigg( 5 \ml^4 - 2 \ml^2 ( 5 s_{2\ell} + 6 \mpip^2) + (s_{2\ell} - \mpip^2)(5 s_{2\ell} + \mpip^2) \bigg) \frac{B_0(\mpip^2, \mpip^2, \mg^2)}{3\lambda(s_{2\ell}, \ml^2, \mpip^2)} \\
			& - \bigg( 4 \ml^4  - 2 \ml^2(5s_{2\ell} + 2\mpip^2) - 6 s_{2\ell}(\mpip^2-s_{2\ell}) \bigg) \frac{B_0(s_{2\ell}, \ml^2, \mpip^2)}{3\lambda(s_{2\ell}, \ml^2, \mpip^2)} \\
			& - \frac{1}{6} B_0(0, \mpip^2, \mg^2) \Bigg) - e^2 \frac{A_0(\mg^2)}{\mg^2} . \end{aligned}
	\end{split}
\end{align}
}

The remaining eight diagrams in this first set are loop corrections to the diagram~\ref{img:Kl4_LO2}. Here, the Passarino-Veltman reduction \cite{Hooft1979, Passarino1979} produces too lengthy expressions, hence, I use the tensor-coefficient functions (see appendix~\ref{sec:AppendixTensorCoefficientFunctions}):
\begin{align}
	\small
	\begin{split}
		\delta F^\mathrm{NLO}_{\gamma-\mathrm{loop},k} &= \delta G^\mathrm{NLO}_{\gamma-\mathrm{loop},k} = -\frac{4}{3} e^2 B_0(s_\ell, \mkp^2, \mg^2) + \frac{4}{3} e^2 \frac{B_{00}(s_\ell, \mkp^2, \mg^2)}{\mg^2} ,
	\end{split}
\end{align}
\begin{align}
	\small
	\begin{split}
		\delta F^\mathrm{NLO}_{\gamma-\mathrm{loop},l} &= e^2 \begin{aligned}[t]
			& \Bigg( \frac{1}{3} B_0(s_\ell, \mkp^2, \mg^2) + \frac{1}{3} B_0(\mkp^2, \mkp^2, \mg^2) - \frac{1}{12} B_0(0, \mkp^2, \mg^2) \\
			& - (s + \nu) C_0(s_\ell, s, \mkp^2, \mg^2, \mkp^2, \mkp^2) - \nu C_1(s_\ell, s, \mkp^2, \mg^2, \mkp^2, \mkp^2) \\
			& - \frac{s + 3\nu}{2} C_2(s_\ell, s, \mkp^2, \mg^2, \mkp^2, \mkp^2)  - \frac{\nu}{2} C_{12}(s_\ell, s, \mkp^2, \mg^2, \mkp^2, \mkp^2) \\
			& - \frac{\nu}{2} C_{22}(s_\ell, s, \mkp^2, \mg^2, \mkp^2, \mkp^2)  \Bigg) - \frac{1}{3} e^2 \frac{B_{00}(s_\ell, \mkp^2, \mg^2)}{\mg^2} ,
			\end{aligned}
	\end{split}
\end{align}
\begin{align}
	\small
	\begin{split}
		\delta G^\mathrm{NLO}_{\gamma-\mathrm{loop},l} &= e^2 C_{00}(s_\ell, s, \mkp^2, \mg^2, \mkp^2, \mkp^2) - e^2 \frac{B_{00}(s_\ell, \mkp^2, \mg^2)}{\mg^2} ,
	\end{split}
\end{align}
\begin{align}
	\small
	\begin{split}
		\delta F^\mathrm{NLO}_{\gamma-\mathrm{loop},m} &= \delta G^\mathrm{NLO}_{\gamma-\mathrm{loop},m} \\
			&= -e^2 \begin{aligned}[t]
				& \Bigg( \frac{1}{3} B_0(s_\ell, \mkp^2, \mg^2) + \frac{1}{3} B_0(\mpip^2, \mpip^2, \mg^2) - \frac{1}{12} B_0(0, \mpip^2, \mg^2) \\
				& + (\mkp^2 + \mpip^2 - u) C_0(\mpip^2, u, s_\ell, \mg^2, \mpip^2, \mkp^2) \\
				& + \frac{\mkp^2 + \mpip^2 - u}{2} C_1(\mpip^2, u, s_\ell, \mg^2, \mpip^2, \mkp^2) \Bigg)  + \frac{1}{3} e^2 \frac{B_{00}(s_\ell, \mkp^2, \mg^2)}{\mg^2} , \end{aligned}
	\end{split}
\end{align}
\begin{align}
	\small
	\begin{split}
		\delta F^\mathrm{NLO}_{\gamma-\mathrm{loop},n} &= e^2 \begin{aligned}[t]
			& \Bigg( - \frac{2}{3} B_0(\mpip^2, \mpip^2, \mg^2) + \frac{1}{3} B_0(s_\ell, \mkp^2, \mg^2) + \frac{1}{6} B_0(0, \mpip^2, \mg^2) \\
			& + (s_\ell + \mpip^2 - u) C_0(\mpip^2, t, s_\ell, \mg^2, \mpip^2, \mkp^2) \\
			& - \frac{\mkp^2 + 5 \mpip^2 - 3s - t}{2} C_1(\mpip^2, t, s_\ell, \mg^2, \mpip^2, \mkp^2) \\
			& + (s_\ell + \mpip^2 - u) C_2(\mpip^2, t, s_\ell, \mg^2, \mpip^2, \mkp^2) \\
			& + C_{00}(\mpip^2, t, s_\ell, \mg^2, \mpip^2, \mkp^2) + \frac{s - 2\mpip^2}{2} C_{11}(\mpip^2, t, s_\ell, \mg^2, \mpip^2, \mkp^2) \\
			& + \frac{s_\ell + \mpip^2 - u}{2} C_{12}(\mpip^2, t, s_\ell, \mg^2, \mpip^2, \mkp^2) \Bigg) - \frac{4}{3} e^2 \frac{B_{00}(s_\ell, \mkp^2, \mg^2)}{\mg^2} , \end{aligned}
	\end{split}
\end{align}
\begin{align}
	\small
	\begin{split}
		\delta G^\mathrm{NLO}_{\gamma-\mathrm{loop},n} &= - \delta F^\mathrm{NLO}_{\gamma-\mathrm{loop},n} + 2 e^2 C_{00}(\mpip^2, t, s_\ell, \mg^2, \mpip^2, \mkp^2) - 2 e^2 \frac{B_{00}(s_\ell, \mkp^2, \mg^2)}{\mg^2} ,
	\end{split}
\end{align}
\begin{align}
	\small
	\begin{split}
		\delta F^\mathrm{NLO}_{\gamma-\mathrm{loop},o} &= \delta G^\mathrm{NLO}_{\gamma-\mathrm{loop},o} \\
			&= \frac{4}{3} e^2 \frac{1}{\ml^2-s_\ell} \begin{aligned}[t]
				& \Bigg( \ml^2 B_0(\ml^2, \ml^2, \mg^2) - s_\ell B_0(s_\ell, \mkp^2, \mg^2) \\
				& + \ml^2 (\mkp^2 - s_\ell) C_0(\ml^2, 0, s_\ell, \mg^2, \ml^2, \mkp^2) \Bigg)  - \frac{4}{3} e^2 \frac{B_{00}(s_\ell, \mkp^2, \mg^2)}{\mg^2} , \end{aligned}
	\end{split}
\end{align}
\begin{align}
	\small
	\begin{split}
		\delta F^\mathrm{NLO}_{\gamma-\mathrm{loop},p} &= e^2 \begin{aligned}[t]
			& \Bigg( \frac{1}{12} B_0(0, \mkp^2, \mg^2) - \frac{1}{3} B_0(\mkp^2, \mkp^2, \mg^2) - \frac{1}{3} B_0(s_\ell, \mkp^2, \mg^2) \\
			& + (\nu + s) C_0(\mkp^2, s, s_\ell, \mg^2, \mkp^2, \mkp^2) + \frac{1}{2} (\nu + s) C_1(\mkp^2, s, s_\ell, \mg^2, \mkp^2, \mkp^2) \\
			& + \nu C_{1+2}(\mkp^2, s, s_\ell, \mg^2, \mkp^2, \mkp^2) + \frac{1}{2} \nu C_{11+12}(\mkp^2, s, s_\ell, \mg^2, \mkp^2, \mkp^2) \\
			& + \frac{1}{3} \ml^2 C_{1+2}(\ml^2, 0, s_\ell, \mg^2, \ml^2, \mkp^2) + \frac{1}{3} \ml^2 C_2(\mkp^2, \tl, \ml^2, \mg^2, \mkp^2, \ml^2) \\
			& - \ml^2 (\nu + s) D_{2+3}(\mkp^2, \tl, 0, s_\ell, \ml^2, s, \mg^2, \mkp^2, \ml^2, \mkp^2) \\
			& - \ml^2 \nu D_{12+13}(\mkp^2, \tl, 0, s_\ell, \ml^2, s, \mg^2, \mkp^2, \ml^2, \mkp^2) \\
			& + \ml^2 (s_{1\ell} - s_{2\ell}) D_{22+23}(\mkp^2, \tl, 0, s_\ell, \ml^2, s, \mg^2, \mkp^2, \ml^2, \mkp^2) \\
			& - \ml^2 \nu D_{23+33}(\mkp^2, \tl, 0, s_\ell, \ml^2, s, \mg^2, \mkp^2, \ml^2, \mkp^2) \Bigg) + \frac{1}{3} e^2 \frac{B_{00}(s_\ell, \mkp^2, \mg^2)}{\mg^2} , \end{aligned}
	\end{split}
\end{align}
where I use the abbreviation
\begin{align}
	\begin{split}
		C_{i+j}(X) &:= C_i(X) + C_j(X) , \\
		D_{i+j}(X) &:= D_i(X) + D_j(X) .
	\end{split}
\end{align}

\begin{align}
	\small
	\begin{split}
		\delta G^\mathrm{NLO}_{\gamma-\mathrm{loop},p} &= - e^2 \begin{aligned}[t]
			& \Bigg( C_{00}(\mkp^2, s, s_\ell, \mg^2, \mkp^2, \mkp^2) \\
			& + 2 \ml^2 D_{00}(\ml^2, \tl, s, s_\ell, \mkp^2, 0, \mg^2, \ml^2, \mkp^2, \mkp^2) \Bigg)  + e^2 \frac{B_{00}(s_\ell, \mkp^2, \mg^2)}{\mg^2} , \end{aligned}
	\end{split}
\end{align}
\begin{align}
	\small
	\begin{split}
		\delta F^\mathrm{NLO}_{\gamma-\mathrm{loop},q} &= \delta G^\mathrm{NLO}_{\gamma-\mathrm{loop},q} \\
			&= e^2 \begin{aligned}[t]
			& \Bigg( - \frac{1}{12} B_0(0, \mpip^2, \mg^2) + \frac{1}{3} B_0(\mpip^2, \mpip^2, \mg^2) + \frac{1}{3} B_0(s_\ell, \mkp^2, \mg^2) \\
			& + (\mkp^2 + \mpip^2 - u) C_0(\mpip^2, u, s_\ell, \mg^2, \mpip^2, \mkp^2) \\
			& + \frac{1}{2} (\mkp^2 + \mpip^2 - u) C_1(\mpip^2, u, s_\ell, \mg^2, \mpip^2, \mkp^2) \\
			& - \frac{1}{3} \ml^2 C_{1+2}(\ml^2, 0, s_\ell, \mg^2, \ml^2, \mkp^2) - \frac{1}{3} \ml^2 C_1(\ml^2, s_{1\ell}, \mpip^2, \mg^2, \ml^2, \mpip^2) \\
			& - \ml^2 (\mkp^2 + \mpip^2 - u) D_{2+3}(\mpip^2, s_{1\ell}, 0, s_\ell, \ml^2, u, \mg^2, \mpip^2, \ml^2, \mkp^2) \Bigg) \end{aligned} \\
			&  - \frac{1}{3} e^2 \frac{B_{00}(s_\ell, \mkp^2, \mg^2)}{\mg^2} ,
	\end{split}
\end{align}
\begin{align}
	\small
	\begin{split}
		\delta F^\mathrm{NLO}_{\gamma-\mathrm{loop},r} &= e^2 \begin{aligned}[t]
			& \Bigg(- \frac{1}{6} B_0(0, \mpip^2, \mg^2) + \frac{2}{3} B_0(\mpip^2, \mpip^2, \mg^2) - \frac{1}{3} B_0(s_\ell, \mkp^2, \mg^2) \\
			& - (\mpip^2 + s_\ell - u) C_0(\mpip^2, t, s_\ell, \mg^2, \mpip^2, \mkp^2) \\
			& + \frac{1}{2} (3 \mpip^2 - 2 s - s_\ell + u) C_1(\mpip^2, t, s_\ell, \mg^2, \mpip^2, \mkp^2) \\
			& - (\mpip^2 + s_\ell - u) C_2(\mpip^2, t, s_\ell, \mg^2, \mpip^2, \mkp^2) - C_{00}(\mpip^2, t, s_\ell, \mg^2, \mpip^2, \mkp^2) \\
			& + \frac{1}{2}(2\mpip^2 - s) C_{11}(\mpip^2, t, s_\ell, \mg^2, \mpip^2, \mkp^2) \\
			& - \frac{1}{2} (\mpip^2 + s_\ell - u) C_{12}(\mpip^2, t, s_\ell, \mg^2, \mpip^2, \mkp^2) \\
			& + \frac{1}{3} \ml^2 C_{1+2}(\ml^2, 0, s_\ell, \mg^2, \ml^2, \mkp^2) - \frac{2}{3} \ml^2 C_1(\ml^2, s_{2\ell}, \mpip^2, \mg^2, \ml^2, \mpip^2) \\
			& + \ml^2 (\mpip^2 + s_\ell - u) D_1(s_\ell, t, s_{2\ell}, \ml^2, \mpip^2, 0, \mg^2, \mkp^2, \mpip^2, \ml^2) \\
			& + \ml^2 (\mpip^2 + s_\ell - u) D_3(s_\ell, t, s_{2\ell}, \ml^2, \mpip^2, 0, \mg^2, \mkp^2, \mpip^2, \ml^2) \\
			& + \ml^2 (\mpip^2 + s_\ell - u) D_{11}(s_\ell, t, s_{2\ell}, \ml^2, \mpip^2, 0, \mg^2, \mkp^2, \mpip^2, \ml^2) \\
			& + \ml^2 (s - 2 \mpip^2) D_{12}(s_\ell, t, s_{2\ell}, \ml^2, \mpip^2, 0, \mg^2, \mkp^2, \mpip^2, \ml^2) \\
			& + \ml^2 (\ml^2 + 2 \mpip^2 - s_{1\ell} + s_\ell - u) D_{13}(s_\ell, t, s_{2\ell}, \ml^2, \mpip^2, 0, \mg^2, \mkp^2, \mpip^2, \ml^2) \\
			& + \ml^2 (s - 2 \mpip^2) D_{23}(s_\ell, t, s_{2\ell}, \ml^2, \mpip^2, 0, \mg^2, \mkp^2, \mpip^2, \ml^2) \\
			& + \ml^2 (\ml^2 + \mpip^2 - s_{1\ell}) D_{33}(s_\ell, t, s_{2\ell}, \ml^2, \mpip^2, 0, \mg^2, \mkp^2, \mpip^2, \ml^2) \Bigg) \end{aligned} \\
			& + \frac{4}{3} e^2 \frac{B_{00}(s_\ell, \mkp^2, \mg^2)}{\mg^2} ,
	\end{split}
\end{align}
\begin{align}
	\small
	\begin{split}
		\delta G^\mathrm{NLO}_{\gamma-\mathrm{loop},r} &= - \delta F^\mathrm{NLO}_{\gamma-\mathrm{loop},r} - 2 e^2 C_{00}(\mpip^2, t, s_\ell, \mg^2, \mpip^2, \mkp^2) + 2 e^2 \frac{B_{00}(s_\ell, \mkp^2, \mg^2)}{\mg^2} .
	\end{split}
\end{align}
I use the notation $\nu = t - u$, $\lambda_{K\ell}(s) = \lambda(\mkp^2, s, s_\ell)$, $\lambda_{\pi\ell}(s) = \lambda(\mpip^2, s, s_\ell)$.

Next, I give the explicit expressions for the diagrams of the second set, shown in figure~\ref{img:Kl4_mLoops}. These diagrams contain a photon pole in the $s$-channel and mesonic loops.
\begin{align}
	\begin{split}
		\delta F^\mathrm{NLO}_{\gamma-\mathrm{pole},a} &= 0 , \\
		\delta G^\mathrm{NLO}_{\gamma-\mathrm{pole},a} &= -e^2 \frac{1}{3s} \begin{aligned}[t]
			& \Bigg( 2( s - 4\mkp^2) B_0(s, \mkp^2, \mkp^2) + (s - 4\mpip^2) B_0(s,\mpip^2, \mpip^2) \\
			& - 4 A_0(\mkp^2) - 2 A_0(\mpip^2) - \frac{4\mkp^2 + 2\mpip^2 -s}{8\pi^2}  \Bigg) , \end{aligned}
	\end{split}
\end{align}
\begin{align}
	\begin{split}
		\delta F^\mathrm{NLO}_{\gamma-\mathrm{pole},b} &= -\delta F^\mathrm{NLO}_{\gamma-\mathrm{pole},c} = -e^2 \frac{s - 6 \mpip^2}{144 \pi^2 \mg^2} , \\
		\delta G^\mathrm{NLO}_{\gamma-\mathrm{pole},b} &= -\delta G^\mathrm{NLO}_{\gamma-\mathrm{pole},c} =  -e^2 \frac{1}{6s} \begin{aligned}[t]
			& \Bigg( ( s - 4\mkp^2) B_0(s, \mkp^2, \mkp^2) + 2(s - 4\mpip^2) B_0(s,\mpip^2, \mpip^2) \\
			& - 2 A_0(\mkp^2) - 4 A_0(\mpip^2) - \frac{2\mkp^2 + 4\mpip^2 -s}{8\pi^2}  \Bigg) , \end{aligned}
	\end{split}
\end{align}
\begin{align}
	\begin{split}
		\delta F^\mathrm{NLO}_{\gamma-\mathrm{pole},d} &= 0 , \\
		\delta G^\mathrm{NLO}_{\gamma-\mathrm{pole},d} &= -e^2 \frac{1}{3s} \bigg( A_0(\mpio^2) + 8 A_0(\mpip^2) + 2 A_0(\mko^2) + 16 A_0(\mkp^2) + 3 A_0(\meta^2) \bigg) ,
	\end{split}
\end{align}
\begin{align}
	\begin{split}
		\delta F^\mathrm{NLO}_{\gamma-\mathrm{pole},e} &= 0 , \\
		\delta G^\mathrm{NLO}_{\gamma-\mathrm{pole},e} &= e^2 \frac{1}{3s} \bigg( A_0(\mpio^2) + 2 A_0(\mpip^2) + 2 A_0(\mko^2) + 4 A_0(\mkp^2) + 3 A_0(\meta^2) \bigg) ,
	\end{split}
\end{align}
\begin{align}
	\begin{split}
		\delta F^\mathrm{NLO}_{\gamma-\mathrm{pole},f} &= \delta F^\mathrm{NLO}_{\gamma-\mathrm{pole},g} = 0 , \\
		\delta G^\mathrm{NLO}_{\gamma-\mathrm{pole},f} &= -\delta G^\mathrm{NLO}_{\gamma-\mathrm{pole},g} = -e^2 \frac{1}{3s} \bigg( 2 A_0(\mpio^2) + 8 A_0(\mpip^2) + A_0(\mko^2) + 4 A_0(\mkp^2) \bigg) .
	\end{split}
\end{align}
In the sum of these diagrams, the contribution to $F$ vanishes:
\begin{align}
	\begin{split}
		\delta F^\mathrm{NLO}_{\gamma-\mathrm{pole}} &= 0 , \\
		\delta G^\mathrm{NLO}_{\gamma-\mathrm{pole}} &= - e^2 \frac{1}{3s} \begin{aligned}[t]
			& \Bigg( 2 ( s - 4\mkp^2) B_0(s, \mkp^2, \mkp^2) + (s - 4\mpip^2) B_0(s,\mpip^2, \mpip^2) \\
			& + 8 A_0(\mkp^2) + 4 A_0(\mpip^2) - \frac{4\mkp^2 + 2\mpip^2 -s}{8\pi^2}  \Bigg) . \end{aligned}
	\end{split}
\end{align}

\subsubsection{Counterterms}

The individual contributions of the counterterm diagrams, shown in figure~\ref{img:Kl4_gCT}, are given by
\begin{align}
	\begin{split}
		\delta F^\mathrm{NLO}_{\gamma-\mathrm{ct},a} &= \frac{2}{9} e^2 \left( 12 K_1 + 19 K_5 + 9 K_{12} - 30 X_1 \right) , \\
		\delta G^\mathrm{NLO}_{\gamma-\mathrm{ct},a} &= \frac{2}{9} e^2 \left( 12 K_1 + 36 K_3 + 7 K_5 + 9 K_{12} + 6 X_1 \right) , \\
		\delta F^\mathrm{NLO}_{\gamma-\mathrm{ct},b} &= - e^2 \frac{4(t-u)}{s} \left( L_9 + L_{10} \right) , \\
		\delta G^\mathrm{NLO}_{\gamma-\mathrm{ct},b} &= - e^2 \frac{4}{s} \begin{aligned}[t]
													& \bigg( (\mkp^2 + s - s_\ell) L_9 + (\mkp^2 - s - s_\ell) L_{10}  + 4( 2 \mkp^2 + \mpip^2) L_4 + 4 \mkp^2 L_5  \bigg) , \end{aligned} \\
		\delta F^\mathrm{NLO}_{\gamma-\mathrm{ct},c} &= \delta F^\mathrm{NLO}_{\gamma-\mathrm{ct},d} = 0 , \\
		\delta G^\mathrm{NLO}_{\gamma-\mathrm{ct},c} &= -\delta G^\mathrm{NLO}_{\gamma-\mathrm{ct},d} = - e^2 \frac{4}{s} \left( 4(2\mkp^2 + \mpip^2) L_4 + 4 \mpip^2 L_5 + s L_9 \right) , \\
		\delta F^\mathrm{NLO}_{\gamma-\mathrm{ct},e} &= 0 , \\
		\delta G^\mathrm{NLO}_{\gamma-\mathrm{ct},e} &= e^2 \frac{16}{s} \left( (2\mkp^2 + \mpip^2) L_4 + \mkp^2 L_5 \right) .
	\end{split}
\end{align}
Their sum is
\begin{align}
	\begin{split}
		\delta F^\mathrm{NLO}_{\gamma-\mathrm{ct}} &= \frac{2}{9} e^2 \left( 12 K_1 + 19 K_5 + 9 K_{12} - 30 X_1 \right) - e^2 \frac{4(t-u)}{s}(L_9 + L_{10}) , \\
		\delta G^\mathrm{NLO}_{\gamma-\mathrm{ct}} &= \frac{2}{9} e^2 \left( 12 K_1 + 36 K_3 + 7 K_5 + 9 K_{12} + 6 X_1 \right) - e^2 \frac{4}{s} \left( (\mkp^2 + s - s_\ell) L_9 + (\mkp^2 - s - s_\ell) L_{10} \right) .
	\end{split}
\end{align}

\subsubsection{External Leg Corrections}

\label{sec:AppendixExternalLegCorrectionsPhotonicEffects}
I first compute the external leg corrections for the mesons (figures~\ref{img:Kl4_mSEgLoop} and \ref{img:Kl4_mSECT}). The field strength renormalisation of a charged meson is related to the self-energy by
\begin{align}
	\begin{split}
		Z_{m^+}^\gamma &= 1 + \Sigma_{m^+}^{\gamma\prime}(M_{m^+,\mathrm{ph}}^2) + h.o. =  1 + \Sigma_{m^+}^{\gamma\prime}(M_{m^+}^2) + h.o. , \\
		\Sigma_{m^+}^\gamma(p^2) &= i ( \mathcal{D}_{m^+}^{\gamma-\mathrm{loop}} + \mathcal{D}_{m^+}^{\gamma-\mathrm{ct}} ) , \quad \Sigma_{m^+}^{\gamma\prime}(p^2) = \frac{d}{d p^2} \Sigma_{m^+}^\gamma(p^2) ,
	\end{split}
\end{align}
where $p$ denotes the meson momentum and $h.o.$ stands for higher order terms.

I find the following field strength renormalisations:
\begin{align}
	\begin{split}
		Z_{\pi^+}^\gamma &= 1 + e^2 \left( \frac{A_0(\mg^2)}{\mg^2} + 2 B_0(\mpip^2, \mpip^2, \mg^2) + 4 \mpip^2 B_0^\prime(\mpip^2, \mpip^2, \mg^2) \right) - \frac{4}{9} e^2 \left( 6 K_1 + 5 K_5 \right) , \\
		Z_{K^+}^\gamma &= 1 + e^2 \left( \frac{A_0(\mg^2)}{\mg^2} + 2 B_0(\mkp^2, \mkp^2, \mg^2) + 4 \mkp^2 B_0^\prime(\mkp^2, \mkp^2, \mg^2) \right) - \frac{4}{9} e^2 \left( 6 K_1 + 5 K_5 \right) .
	\end{split}
\end{align}

Finally, we need the field strength renormalisation of the lepton. The two diagrams~\ref{img:Kl4_lSEgLoop} and \ref{img:Kl4_lSECT} contribute to the self-energy:
\begin{align}
	\begin{split}
		Z_\ell^\gamma &= 1 + \Sigma_\ell^{\gamma\prime}(\ml) + h.o. , \\
		\Sigma_\ell^\gamma(\slashed p) &=  i ( \mathcal{D}_\ell^{\gamma-\mathrm{loop}} + \mathcal{D}_\ell^{\gamma-\mathrm{ct}} ) , \quad \Sigma_\ell^{\gamma\prime}(\slashed p) = \frac{d}{d \slashed p} \Sigma_\ell^\gamma(\slashed p) .
	\end{split}
\end{align}
Up to terms that vanish for $\mg\to0$, the lepton self-energy is given by
\begin{align}
	\begin{split}
		Z_\ell^\gamma &= 1 + e^2 \left( \frac{3 A_0(\mg^2)}{\mg^2} - \frac{3 A_0(\ml^2)}{\ml^2} - X_6 - \frac{3}{16\pi^2}  \right) .
	\end{split}
\end{align}
The contribution of the field strength renormalisation to the form factors is therefore
\begin{align}
	\begin{split}
		\delta F_{\gamma-Z}^\mathrm{NLO} &= \delta G_{\gamma-Z}^\mathrm{NLO} =  Z_{\pi^+}^\gamma \sqrt{Z_{K^+}^\gamma Z_\ell^\gamma} - 1 \\
			&= e^2 \begin{aligned}[t]
				& \bigg( B_0(\mkp^2, \mkp^2, \mg^2) + 2 B_0(\mpip^2, \mpip^2, \mg^2) \\
				& - \frac{A_0(\mkp^2)}{\mkp^2} - \frac{2 A_0(\mpip^2)}{\mpip^2} - \frac{3 A_0(\ml^2)}{2 \ml^2} + \frac{6 A_0(\mg^2)}{\mg^2}  - 4 K_1 - \frac{10}{3} K_5 - \frac{1}{2} X_6 - \frac{15}{32\pi^2} \bigg) .
				\end{aligned}
	\end{split}
\end{align}
The mass renormalisation of the lepton is given by
\begin{align}
	\label{eqn:LeptonMassRenormalisation}
	\begin{split}
		\ml^\mathrm{ph} = \ml^\mathrm{NLO} + h.o., \quad \ml^\mathrm{NLO} &= \ml^\mathrm{bare} + \delta \ml = \ml^\mathrm{bare} + \Sigma_\ell^\gamma(\slashed p = \ml) \\
			&= e^2 \ml \left( \frac{1}{16\pi^2} - \frac{3 A_0(\ml^2)}{\ml^2} - X_6 - X_7 \right) .
	\end{split}
\end{align}

\clearpage

\addcontentsline{toc}{section}{References}
\bibliographystyle{my-physrev}
\bibliography{Literature}

\end{document}